\documentclass[sigconf, nonacm]{acmart}

\newcommand\vldbdoi{10.14778/3457390.3457395}

\newcommand\vldbavailabilityurl{}
\newcommand\vldbpagestyle{empty} 

\usepackage{graphics}
\usepackage{subcaption}
\usepackage{enumerate}
\usepackage{balance}
\usepackage[linesnumbered,ruled,vlined]{algorithm2e}
\usepackage[a-1b]{pdfx}

\newtheoremstyle{MyStyle}%
  {1.1pt} % Space above
  {1.1pt} % Space below
  {} % Body font
  {} % Indent amount
  {\bfseries} % Theorem head font
  {.} % Punctuation after theorem head
  {.5em} % Space after theorem head
  {} % Theorem head spec (can be left empty, meaning `normal')
  
\newcommand\eqnlabel{\addtocounter{equation}{1}\tag{\theequation}}

\newcommand{\myop}{\textit{op}}
\theoremstyle{MyStyle} \newtheorem{MyDefinition}{Definition}[section]
\theoremstyle{MyStyle} \newtheorem{MyLemma}{Lemma}[section]
\theoremstyle{MyStyle} 
\theoremstyle{MyStyle} \newtheorem{MyExample}{Example}[section]

\begin{document}
\title{Symmetric Continuous Subgraph Matching with Bidirectional Dynamic Programming}

%%
%% The "author" command and its associated commands are used to define the authors and their affiliations.
\author{Seunghwan Min}
\affiliation{%
  \institution{Seoul National University}
%   \city{Seoul}
%   \country{Korea}
}
\email{shmin@theory.snu.ac.kr}

\author{Sung Gwan Park}
\affiliation{%
  \institution{Seoul National University}
%   \city{Seoul}
%   \country{Korea}
}
\email{sgpark@theory.snu.ac.kr}

\author{Kunsoo Park$^{*}$}
\affiliation{%
  \institution{Seoul National University}
%   \city{Seoul}
%   \country{Korea}
}
\email{kpark@theory.snu.ac.kr}

\author{Dora Giammarresi$^{\dagger}$}
\affiliation{%
  \institution{Università Roma ``Tor Vergata''}
%   \city{Roma}
%   \country{Italy}
}
\email{giammarr@mat.uniroma2.it}

\author{Giuseppe F. Italiano$^{\dagger}$}
\affiliation{%
  \institution{LUISS University}
%   \city{Roma}
%   \country{Italy}
}
\email{gitaliano@luiss.it}

\author{Wook-Shin Han$^{*}$}
\affiliation{%
  \institution{Pohang University of Science and Technology (POSTECH)}
%   \city{Pohang}
%   \country{Korea}
}
\email{wshan@dblab.postech.ac.kr}

\begin{abstract}
In many real datasets such as social media streams and cyber data sources, graphs change over time through a graph update stream of edge insertions and deletions. Detecting critical patterns in such dynamic graphs plays an important role in various application domains such as fraud detection, cyber security, and recommendation systems for social networks. Given a dynamic data graph and a query graph, the continuous subgraph matching problem is to find all positive matches for each edge insertion and all negative matches for each edge deletion. The state-of-the-art algorithm \textsf{TurboFlux} uses a spanning tree of a query graph for filtering. However, using the spanning tree may have a low pruning power because it does not take into account all edges of the query graph. In this paper, we present a symmetric and much faster algorithm \textsf{SymBi} which maintains an auxiliary data structure based on a directed acyclic graph instead of a spanning tree, which maintains the intermediate results of bidirectional dynamic programming between the query graph and the dynamic graph. Extensive experiments with real and synthetic datasets show that \textsf{SymBi} outperforms the state-of-the-art algorithm by up to three orders of magnitude in terms of the elapsed time.
\end{abstract}

\maketitle

\renewcommand\thefootnote{}\footnote{\noindent
$*$ Contact author\\
${\dagger}$ Work partially done while visiting Seoul National University

}

\pagestyle{\vldbpagestyle}
% \begingroup\small\noindent\raggedright\textbf{PVLDB Reference Format:}\\
% \vldbauthors. \vldbtitle. PVLDB, \vldbvolume(\vldbissue): \vldbpages, \vldbyear.\\
% \href{https://doi.org/\vldbdoi}{doi:\vldbdoi}
% \endgroup
% \begingroup
% \renewcommand\thefootnote{}\footnote{\noindent
% This work is licensed under the Creative Commons BY-NC-ND 4.0 International License. Visit \url{https://creativecommons.org/licenses/by-nc-nd/4.0/} to view a copy of this license. For any use beyond those covered by this license, obtain permission by emailing \href{mailto:info@vldb.org}{info@vldb.org}. Copyright is held by the owner/author(s). Publication rights licensed to the VLDB Endowment. \\
% \raggedright Proceedings of the VLDB Endowment, Vol. \vldbvolume, No. \vldbissue\ %
% ISSN 2150-8097. \\
% \href{https://doi.org/\vldbdoi}{doi:\vldbdoi} \\
% }\addtocounter{footnote}{-1}\endgroup

\ifdefempty{\vldbavailabilityurl}{}{
\vspace{.3cm}
\begingroup\small\noindent\raggedright\textbf{PVLDB Availability Tag:}\\
The source code of this research paper has been made publicly available at \url{\vldbavailabilityurl}.
\endgroup
}

\section{Introduction}\label{sec:introduction}

A dynamic graph is a graph that changes over time through a graph update stream of edge insertions and deletions. In the last decade, the topic of massive dynamic graphs has become popular. Social media streams and cyber data sources, such as computer network traffic and financial transaction networks, are examples of dynamic graphs. A social media stream can be modeled as a graph where vertices represent people, movies, or images, and edges represent relationship such as friendship, like, post, etc. A computer network traffic consists of vertices representing IP addresses and edges representing protocols of network traffic \cite{joslyn2013massive}. 

Extensive research has been done for the efficient analysis of dynamic graphs \cite{mcgregor2014graph, abdelhamid2017incremental, kumar2019graphone, tang2016graph, namaki2017beams}. Among them, detecting critical patterns or events in a dynamic graph is an important issue since it lies at the core of various application domains such as fraud detection \cite{sadowski2014fraud, qiu2018real}, cyber security \cite{choudhury2013streamworks, choudhury2015selectivity}, and recommendation systems for social networks \cite{gupta2014real, kankanamge2017graphflow}. For example, various cyber attacks such as denial-of-service attack and data exfiltration attack can be represented as graphs \cite{choudhury2015selectivity}. Moreover, US communications company Verizon reports that 94\% of the cyber security incidents fell into nine patterns, many of which can be described as graph patterns in their study, ``2020 Data Breach Investigations Report'' \cite{verizon2020}. Cyber security applications should detect in real-time that such graph patterns appear in network traffic, which is one of dynamic graphs \cite{choudhury2013streamworks}.

In this paper, we focus on the problem of detecting and reporting such graph patterns in a dynamic graph, called \textit{continuous subgraph matching}. Many researchers have developed efficient solutions for continuous subgraph matching \cite{chen2010continuous, gao2014continuous, fan2013incremental, choudhury2015selectivity, kankanamge2017graphflow, kim2018turboflux, gao2016toward, pugliese2014efficient} and its variants \cite{song2014event, li2019time, zhang2019continuously, mondal2014eagr, fan2020extending, zervakis2019efficient} over the past decade. Due to the NP-hardness of continuous subgraph matching, Chen et al. \cite{chen2010continuous} and Gao et al. \cite{gao2014continuous} propose algorithms that cannot guarantee the exact solution for continuous subgraph matching. The results of these algorithms may include false positive matches, which is far from being desirable. Since several algorithms such as \textsf{IncIsoMat} \cite{fan2013incremental} and \textsf{Graphflow} \cite{kankanamge2017graphflow} do not maintain any intermediate results, these algorithms need to perform subgraph matching for each graph update even if the update does not incur any match of the pattern, which leads to significant overhead. Unlike \textsf{IncIsoMat} and \textsf{Graphflow}, \textsf{SJ-Tree} \cite{choudhury2015selectivity} stores all partial matches for each subgraph of the pattern to get better performance, but this method requires expensive storage space. The state-of-the-art algorithm \textsf{TurboFlux} \cite{kim2018turboflux} uses the idea of \textsf{Turbo\textsubscript{iso}} \cite{han2013turboiso} which is one of state-of-the-art algorithms for the subgraph matching problem. It proposes an auxiliary data structure called \textit{data-centric graph} (\textsf{DCG}), which is an updatable graph to store the partial matches for a spanning tree of the pattern graph. \textsf{TurboFlux} uses less storage space for the auxiliary data structure than \textsf{SJ-Tree} and outperforms the other algorithms. According to experimental results, however, \textsf{TurboFlux} has the disadvantage that processing edge deletions is much slower than edge insertions due to the asymmetric update process of \textsf{DCG}.

%Previous studies show that which intermediate results to be stored (i.e., which auxiliary data structure is used) is important for solving continuous subgraph matching. 
Previous studies show that what information is stored as intermediate results in an auxiliary data structure is important for solving continuous subgraph matching.
An auxiliary data structure should be designed such that it doesn't take long time to update while containing enough information to help detect the pattern quickly (i.e., balancing update time vs.~amount of information to keep). It was shown in \cite{han2019efficient} that the \textit{weak embedding} of a directed acyclic graph is more effective in filtering candidates than the embedding of a spanning tree. In this paper we embed the weak embedding into our data structure so that the intermediate results (i.e., weak embeddings of directed acyclic graphs) contain information that helps detect the pattern quickly and can be updated efficiently. We propose an algorithm \textsf{SymBi} for continuous subgraph matching which uses the proposed data structure. Compared to the state-of-the-art algorithm \textsf{TurboFlux}, this is a substantial benefit since directed acyclic graphs have better pruning power than spanning trees due to non-tree edges, while the update of intermediate results is fast. The contributions of this paper are as follows:  

% In this paper, we propose an algorithm \textsf{SymBi} for continuous subgraph matching which uses a directed acyclic graph of the pattern graph instead of a spanning tree. This is a substantial benefit since directed acyclic graphs have better pruning power than spanning trees due to non-tree edges. The contributions of this paper are as follows: 
\begin{itemize}
    \item We propose an auxiliary data structure called \textit{dynamic candidate space} (\textsf{DCS}), which maintains the intermediate results of bidirectional (i.e., top-down and bottom-up) dynamic programming between a directed acyclic graph of the pattern graph and the dynamic graph. \textsf{DCS} serves as a complete search space to find all matches of the pattern graph in the dynamic graph, and it enables us to symmetrically handle edge insertions and edge deletions. Also, we propose an efficient algorithm to maintain \textsf{DCS} for each graph update. Rather than recomputing the entire structure, this algorithm updates only a small portion of \textsf{DCS} that changes.
    \item We introduce a new matching algorithm using \textsf{DCS} that works for both edge insertions and edge deletions. Unlike the subgraph matching problem, in continuous subgraph matching we need to find matches that contain the updated data graph edge. Thus, we propose a new matching order which is different from the matching orders used in existing subgraph matching algorithms. This matching order starts from an edge of the query graph corresponding to the updated data graph edge, and then selects a next query vertex to match from the neighbors of the matched vertices. When selecting the next vertex, we use an \textit{estimate of the candidate size} of the vertex instead of the exact candidate size \cite{han2019efficient} for efficiency. In addition, we introduce the concept of \textit{isolated vertices} which is an extension of the leaf decomposition technique from \cite{bi2016efficient}.
\end{itemize}

\begin{figure}[t]
\centering
    \subcaptionbox{Query graph $q$\label{fig:query_graph_q}}{
        \includegraphics[width=0.34\linewidth,scale=0.2]{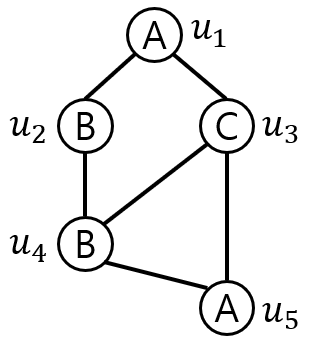}
    }
    \subcaptionbox{Dynamic data graph $g$ with an initial data graph $g_0$ and two edge insertions\label{fig:data_graph_g0}}{
        \includegraphics[width=0.60\linewidth,scale=0.2]{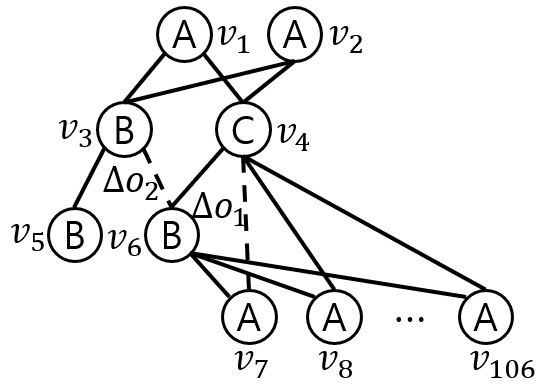}
    }
  \vspace*{-3mm}
\caption{\textbf{A running example of query graph and data graph for continuous subgraph matching}}
\label{fig:query_and_data_graph}
\vspace{-3mm}
\end{figure}

Experiments show that \textsf{SymBi} outperforms \textsf{TurboFlux} by up to three orders of magnitude. In particular, when edge deletions are included in the graph update stream, the performance gap between the two algorithms becomes larger. In an experiment where all query graphs are solved within the time limit by both algorithms, for example, when the ratio of the number of edge deletions to the number of edge insertions increases from 0\% to 10\%, the performance improvement of \textsf{SymBi} over \textsf{TurboFlux} increases from 224.61 times to 309.45 times. While the deletion ratio changes from 0\% to 10\%, the average elapsed time of \textsf{SymBi} increases only 1.54 times, but \textsf{TurboFlux} increases 2.13 times. This supports the fact that \textsf{SymBi} handles edge deletions better than \textsf{TurboFlux}.

The remainder of the paper is organized as follows. Section \ref{sec:preliminaries} formally defines the problem of continuous subgraph matching and describes some related work. Section \ref{sec:overview} describes a brief overview of our algorithm. Section \ref{sec:dcs_structure} introduces \textsf{DCS} and proposes an algorithm to maintain \textsf{DCS} efficiently. Section \ref{sec:backtracking} presents our matching algorithm. Section \ref{sec:performance_evaluation} presents the results of our performance evaluation. Finally, we conclude in Section \ref{sec:conclusion}.

\section{Preliminaries}\label{sec:preliminaries}
% \textbf{(Definition of graph, DAF, embedding, isomorphism, graph update stream, and problem)}

% Graph(labeled, connected, undirected)
For simplicity of presentation, we focus on undirected, connected, and vertex-labeled graphs. Our algorithm can be easily extended to directed or disconnected graphs with multiple labels on vertices or edges. A \textit{graph} $g$ is defined as $(V(g), E(g), l_g)$, where $V(g)$ is a set of vertices, $E(g)$ is a set of edges, and $l_g:V(g)\rightarrow \Sigma$ is a labeling function, where $\Sigma$ is a set of labels. Given $S \subseteq V(g)$, an \textit{induced subgraph} $g[S]$ of $g$ is a graph whose vertex set is $S$ and whose edge set consists of all the edges in $E(g)$ that have both endpoints in $S$.

\begin{figure}[t]
\centering
    \subcaptionbox{Spanning tree $q_T$\label{fig:spanning_tree}}{
        \includegraphics[width=0.28\linewidth,scale=0.2]{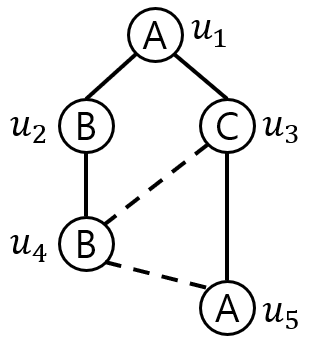}
    }
    \subcaptionbox{Query DAG $\hat{q}$\label{fig:DAG_q}}{
        \includegraphics[width=0.28\linewidth,scale=0.2]{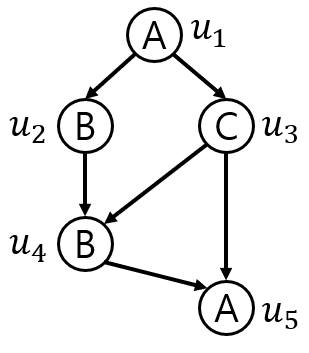}
    }
    \subcaptionbox{Path tree of $\hat{q}$\label{fig:path_tree}}{
        \includegraphics[width=0.33\linewidth,scale=0.2]{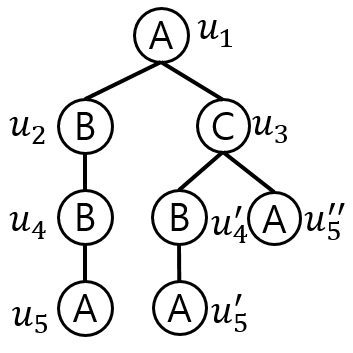}
    }
    \vspace*{-2mm}
\caption{\textbf{Spanning tree, DAG, and path tree for the running example}} 
\label{fig:tree_DAG}
\vspace{-5mm}
\end{figure}

% DAG
A \textit{directed acyclic graph} (DAG) $\hat{q}$ is a directed graph that contains no cycles. A \textit{root} (resp., \textit{leaf}) of a DAG is a vertex with no incoming (resp., outgoing) edges. A DAG $\hat{q}$ is a \textit{rooted DAG} if there is only one root (e.g., the rooted DAG in Figure \ref{fig:DAG_q} can be obtained by directing edges of $q$ in Figure \ref{fig:query_graph_q} in such a way that $u_1$ is the root).
% A DAG $\hat{q}$ is a \textit{rooted DAG} if there is only one vertex $r\in V(\hat{q})$ (i.e., root) that has no incoming edges (e.g., the rooted DAG in Figure \ref{fig:DAG_q} can be obtained by directing edges of $q$ in Figure \ref{fig:query_graph_q} in such a way that $u_1$ is the root). A \textit{leaf} of a DAG is a vertex with no outgoing edges. 
Its reverse $\hat{q}^{-1}$ is the same as $\hat{q}$ with all of the edges reversed. We say that $u$ is a \textit{parent} of $v$ ($v$ is a \textit{child} of $u$) if there exists a directed edge from $u$ to $v$. An \textit{ancestor} of a vertex $v$ is a vertex which is either a parent of $v$ or an ancestor of a parent of $v$. A \textit{descendant} of a vertex $v$ is a vertex which is either a child of $v$ or a descendant of a child of $v$. A \textit{sub-DAG of $\hat{q}$ rooted at $u$}, denoted by $\hat{q}_u$, is the induced subgraph of $\hat{q}$ whose vertices set consists of $u$ and all the descendants of $u$. The \textit{height} of a rooted DAG $\hat{q}$ is the maximum distance between the root and any other vertex in $\hat{q}$, where the distance between two vertices is the number of edges in a shortest path connecting them. Let $\textrm{Child}(u)$, $\textrm{Parent}(u)$, and $\textrm{Nbr}(u)$ denote the children, parents, and neighbors of $u$ in $\hat{q}$, respectively.

% embedding (+ Partial/Full embedding?)
% isomorphism (+ homomorphism)
\begin{MyDefinition}\label{def:embedding}
Given a query graph $q=(V(q), E(q), l_q)$ and a data graph $g=(V(g), E(g), l_g)$, a \textit{homomorphism} of $q$ in $g$ is a mapping $M:V(q)\rightarrow V(g)$ such that (1) $l_q(u) = l_g(M(u))$ for every $u\in V(q)$, and (2) $(M(u), M(u'))\in E(g)$ for every $(u, u')\in E(q)$. An \textit{embedding} of $q$ in $g$ is an injective (i.e., $\forall u, u' \in V(q) \text{ such that } u \neq u' \Rightarrow M(u) \neq M(u')$) homomorphism.
\end{MyDefinition}

An embedding of an induced subgraph of $q$ in $g$ is called a \textit{partial embedding} of $q$ in $g$. We say that $q$ is \textit{subgraph-isomorphic} (resp., \textit{subgraph-homomorphic}) to $g$, if there is an embedding (resp., homomorphism) of $q$ in $g$. We use subgraph isomorphism as our default matching semantics. Subgraph homomorphism can be easily obtained by omitting the injective constraint. 

% path tree
\begin{MyDefinition}\label{def:path_tree}
\cite{han2019efficient} The \textit{path tree} of a rooted DAG $\hat{q}$ is defined as the tree $\hat{q}_T$ such that each root-to-leaf path in $\hat{q}_T$ corresponds to a distinct root-to-leaf path in $\hat{q}$, and $\hat{q}_T$ shares common prefixes of its root-to-leaf paths. Figure \ref{fig:path_tree} shows the path tree of $\hat{q}$ in Figure \ref{fig:DAG_q}.
\end{MyDefinition}

% weak embedding
\begin{MyDefinition}\label{def:weak_embedding}
\cite{han2019efficient} For a rooted DAG $\hat{q}$ with root $u$, a \textit{weak embedding} $M'$ of $\hat{q}$ at $v\in V(g)$ is defined as a homomorphism of the path tree of $\hat{q}$ in $g$ such that $M'(u)=v$.
\end{MyDefinition}

\begin{MyExample}
We will use the query graph and the dynamic data graph in Figure \ref{fig:query_and_data_graph} and the DAG of the query graph in Figure \ref{fig:DAG_q} as a running example throughout this paper. For example, $\{(u_3,v_4),(u_4',\allowbreak v_6),\allowbreak (u_5',v_7),(u_5'',v_8)\}$ is a weak embedding of $\hat{q}_{u_3}$ (Figure \ref{fig:DAG_q}) at $v_4$ in $g_0$ (Figure \ref{fig:data_graph_g0}), where $\hat{q}_{u_3}$ is a sub-DAG of $\hat{q}$ rooted at $u_3$. Note that $u_5$ in $\hat{q}$ is mapped to two different vertices $v_7$ and $v_8$ of $g_0$ via the path tree. If $\Delta o_1$ is inserted to $g_0$, $\{(u_3,v_4),(u_4',v_6),(u_5',v_7),(u_5'',v_7)\}$ is a weak embedding (also an embedding) of $\hat{q}_{u_3}$ at $v_4$.
\end{MyExample}
Every embedding of $q$ in $g$ is a weak embedding of $\hat{q}$ in $g$, but the converse is not true. Hence a weak embedding is a necessary condition for an embedding. The weak embedding is a key notion in our filtering.

% graph update stream
\begin{MyDefinition}\label{def:graph_update_stream}
A \textit{graph update stream} $\Delta g$ is a sequence of update operations $(\Delta o_1, \Delta o_2, \cdots)$, where $\Delta o_i$ is a triple $(\textit{op},v,v')$ such that $v,v'\in V(g)$ and $\myop$ is the type of the update operation which is one of edge insertion (denoted by $+$) or edge deletion (denoted by $-$) of an edge $(v,v')$.
%consists of the type of the update operation such as edge insertion/deletion and an edge $(v, v')$.
\end{MyDefinition}

% how to handle vertex insertion / deletion
Update operations are defined only as inserting and deleting edges between existing vertices, but inserting new vertices or deleting existing vertices is also easy to handle. We can insert a new vertex $v$ by putting $v$ in $V(g)$ and defining a labeling function for $v$. To delete an existing vertex $v$, we first delete all edges connected to $v$ and then remove $v$ from $V(g)$ and the labeling function.

% problem statement
\noindent\textbf{Problem Statement.}
Given an initial data graph $g_0$, a graph update stream $\Delta g$, and a query graph $q$, the \textit{continuous subgraph matching problem} is to find all positive/negative matches for each update operation in $\Delta g$. For example, given a query graph $q$ and an initial data graph $g_0$ with two edge insertion operations $\Delta o_1=(+,v_4,v_7)$ and $\Delta o_2=(+, v_3,v_6)$ in Figure \ref{fig:query_and_data_graph}, continuous subgraph matching finds 200 positive matches when $\Delta o_2$ occurs.

\begin{table}[h]
    \centering
    \caption{Frequently used notations}
    \vspace*{-1mm}
    \renewcommand{\arraystretch}{1.05}
    \begin{tabular}{p{1.1cm} l}
        \toprule
        Symbol & Description \\
        \midrule
        $g$ & Data graph \\
        $\Delta g$ & Graph update stream \\
        $q$ & Query graph \\
        $\hat{q}$ & Query DAG \\
        $C(u)$ & Candidate set for query vertex $u$ \\
        $M(u)$ & Mapping of $u$ in (partial) embedding $M$ \\
        $C_M(u)$ & Set of extendable candidates of $u$ regarding partial \\ & embedding $M$ \\
        $\textrm{Nbr}_M(u)$ & Set of matched neighbors of $u$ in $q$ regarding partial \\ & embedding $M$ \\
        \bottomrule
    \end{tabular}
    \label{tab:notations}
    \vspace{-4mm}
\end{table}

\subsection{Related Work}\label{sec:related_work}
\noindent\textbf{Labeled Subgraph Matching.} There are many studies for practical subgraph matching algorithms for labeled graphs \cite{cordella2004sub, han2013turboiso, ren2015exploiting, bi2016efficient, han2019efficient, bhattarai2019ceci, sun2020memory, he2008graphs, shang2008taming, zhao2010graph, rivero2017efficient}, which are initiated by Ullmann's backtracking algorithm \cite{ullmann1976algorithm}. Given a query graph $q$ and a data graph $g$, this algorithm finds all embeddings by mapping a query vertex to a data vertex one by one. Extensive research has been done to improve the backtracking algorithm. Recently, there are many efficient algorithms solving the subgraph matching problem, such as \textsf{Turbo\textsubscript{iso}} \cite{han2013turboiso}, \textsf{CFL-Match} \cite{bi2016efficient}, and \textsf{DAF} \cite{han2019efficient}.

\textsf{Turbo\textsubscript{iso}} finds all the embeddings of a spanning tree $q_T$ of $q$ (e.g., solid edges in Figure \ref{fig:spanning_tree} form a spanning tree of $q$ in Figure \ref{fig:query_graph_q}) in the data graph. Based on the result, it extracts candidate regions from the data graph that may have embeddings of the query graph, and decides an effective matching order for each candidate region by the \emph{path-ordering} technique. Furthermore, it uses a technique called \textit{neighborhood equivalence class}, which compresses equivalent vertices in the query graph.

\textsf{CFL-Match} also uses a spanning tree for filtering to solve the subgraph matching problem, while it proposes additional techniques to improve \textsf{Turbo\textsubscript{iso}}. It focuses on the fact that \textsf{Turbo\textsubscript{iso}} may check the non-tree edges of $q$ too late, and thus result in a huge search space. To handle this issue, it proposes the \emph{core-forest-leaf decomposition} technique, which decomposes the query graph into a core including the non-tree edges, a forest adjacent to the core, and leaves adjacent to the forest. It is shown in \cite{bi2016efficient} that this technique reduces the search space effectively.

\textsf{DAF} proposes a new approach to solve the subgraph matching problem, by building a \emph{query DAG} instead of a spanning tree. It gives three techniques to solve the subgraph matching problem using query DAG, which are dynamic programming on DAG, adaptive matching order with DAG ordering, and pruning by failing sets. It is shown in \cite{han2019efficient} that the query DAG results in the high pruning power and better matching order. For example, \textsf{DAF} finds that there is no embeddings of $q$ in $g_0$ in Figure \ref{fig:query_and_data_graph} without backtracking process, while \textsf{Turbo\textsubscript{iso}} and \textsf{CFL-Match} need backtracking.

\noindent\textbf{Continuous Subgraph Matching.} Extensive studies have been done to solve continuous subgraph matching, such as \textsf{IncIsoMat} \cite{fan2013incremental}, \textsf{Graphflow} \cite{kankanamge2017graphflow}, \textsf{SJ-Tree} \cite{choudhury2015selectivity}, and \textsf{TurboFlux} \cite{kim2018turboflux}.

\textsf{IncIsoMat} finds a subgraph of a data graph that is affected by a graph update operation, executes subgraph matching to it before and after a graph update operation, and computes the difference between them. The affected range within a data graph is computed based on the diameter of a query graph, where the diameter of a query graph $q$ is defined as the maximum of the length of the shortest paths between arbitrary two vertices in $q$. Since subgraph matching is an NP-hard problem, it costs a lot of time to execute subgraph matching for each graph update operation.

\textsf{Graphflow} uses a worst-case optimal join algorithm \cite{ngo2014skew, mhedhbi2019optimizing}. Starting from each query edge $(u, u')$ that matches a graph edge $(v, v')$, it solves the subgraph matching starting from partial embedding $\{(u, v), \allowbreak(u', v')\}$ and incrementally joins the other edges in the query graph until it gets the set of full embeddings of a query graph. Since it does not maintain any intermediate results, it starts subgraph matching from scratch every time the graph update operation occurs.

\textsf{SJ-Tree} decomposes a query graph $q$ into smaller graphs recursively until each graph consists of only one edge, and build a tree structure called \textsf{SJ-Tree} based on them, where each node in the tree corresponds to a subgraph of $q$. For each node, it stores an intermediate result for subgraph matching between a data graph and a subgraph of $q$ the node represents. When the graph update operation occurs, it updates the intermediate results starting from the leaves of \textsf{SJ-Tree} and recursively perform join operations between the neighbors in \textsf{SJ-Tree}, until it reaches the root of the tree. Since it stores all the intermediate results in an auxiliary data structure, it may cost an exponential space to the size of the query graph.

\textsf{TurboFlux} uses the idea of \textsf{Turbo\textsubscript{iso}}, and modifies it to solve continuous subgraph matching efficiently. It maintains an auxiliary data structure called \emph{data-centric graph}, or \textsf{DCG}, to maintain the intermediate results efficiently. For every pair of an edge in the data graph and an edge in a spanning tree of $q$, it stores a filtering information whether the two edges can be matched or not. For each graph update operation, it updates whether each pair of edges in \textsf{DCG} can be used to compose an embedding of a query graph, based on \emph{edge transition model}. It is shown in \cite{kim2018turboflux} that \textsf{TurboFlux} is more than two orders of magnitude faster in solving continuous subgraph matching than the previous results. Note that both \textsf{Turbo\textsubscript{iso}} and \textsf{TurboFlux} use a spanning tree of the query graph to filter the candidates, while \textsf{DAF} uses a DAG built from the query graph for filtering.

\vspace{-2mm}
\section{Overview of our algorithm}\label{sec:overview}

Algorithm \ref{alg:continous_subgraph_matching} shows the overview of \textsf{SymBi}, which takes a data graph $g$, a graph update stream $\Delta g$, and a query graph $q$ as input, and find all positive/negative matches of $q$ for each update operation in $\Delta g$. \textsf{SymBi} uses three main procedures below.

\begin{enumerate}[1.]
    % build DAG(select root with max depth)
    \item We first build a rooted DAG $\hat{q}$ from $q$. In order to build $\hat{q}$, we traverse $q$ in a BFS order and direct all edges from earlier to later visited vertices. In \textsc{BuildDAG}, we select a vertex as root $r$ such that the DAG has the highest height. Figure \ref{fig:DAG_q} shows a rooted DAG $\hat{q}$ built from query graph $q$ in Figure \ref{fig:query_graph_q} when $u_1$ is the root.
    % build initial CS structure
    \item \textsc{BuildDCS} is called to build an initial \textsf{DCS} structure by using bidirectional dynamic programming between the rooted DAG $\hat{q}$ and the initial data graph $g_0$ (Section \ref{subsec:dcs_structure}). 
    % handling continuous subgraph matching for each update operation(data graph update + CS update + backtracking)
    \item For each update operation, we update the data graph $g$ and the \textsf{DCS} structure, and perform continuous subgraph matching. For insertion of edge $e$, we first invoke \textsc{DCSChangedEdge} to compute a set $E_{DCS}$ which consists of updated edges in \textsf{DCS} due to the inserted edge $e$. We also update the data graph by inserting the edge $e$ into $g$ and update the \textsf{DCS} structure with $E_{DCS}$ (Section \ref{subsec:dcs_update}). Finally, we find positive matches from the updated \textsf{DCS} and $E_{DCS}$ by calling the backtracking procedure (Section \ref{sec:backtracking}). For deletion of edge $e$, we find negative matches first and then update data structures because the information related to $e$ is deleted during the update.
    % For edge insertion, we first insert the edge $e$ into $g$, and update \textsf{DCS} with a set $E_{DCS}$ which consists of updated edges in \textsf{DCS} by $e$ (Section \ref{subsec:dcs_update}). Finally, we find positive matches from \textsf{DCS} and $E_{DCS}$ by calling the backtracking procedure (Section \ref{sec:backtracking}). For edge deletion, we find negative matches first and then update data structures. Note that the order of the computation varies according to the update type. 
\end{enumerate}

%\vspace*{-5mm}
\begin{algorithm}
\KwIn{A data graph $g$, a graph update stream $\Delta g$, and a query graph $q$}
\KwOut{all positive/negative matches}
$\hat{q} \gets$ \textsc{BuildDAG}$(q)$\;
\textsf{DCS} $\gets$ \textsc{BuildDCS}$(g, \hat{q})$\;
\ForEach{$\Delta o \in \Delta g$}{
    $e \gets (\Delta o.v, \Delta o.v')$\;
    \If{$\Delta o.\textit{op}=+$}{
        $E_{DCS} \gets$ \textsc{DCSChangedEdge}$(g, q, e)$\;
        \textsc{InsertEdgeToDataGraph}$(g, e)$\;
        \textsc{DCSInsertionUpdate}$(\textsf{DCS},E_{DCS})$\;
        \textsc{FindMatches}$(\textsf{DCS},E_{DCS},\emptyset )$\;
    }
    \If{$\Delta o.\textit{op}=-$}{
        $E_{DCS} \gets$ \textsc{DCSChangedEdge}$(g, q, e)$\;
        \textsc{FindMatches}$(\textsf{DCS},E_{DCS},\emptyset )$\;
        \textsc{DeleteEdgeFromDataGraph}$(g, e)$\;
        \textsc{DCSDeletionUpdate}$(\textsf{DCS},E_{DCS})$\;
    }
}
\caption{\textsc{Continuous Subgraph Matching}}
\label{alg:continous_subgraph_matching}
\vspace*{-1mm}
\end{algorithm}
%\vspace*{-6mm}

\begin{figure*}[t]
\centering
    \subcaptionbox{Initial \textsf{DCS}\textsubscript{0}\label{fig:DCS0}}{
        \includegraphics[width=0.321\linewidth,scale=0.5]{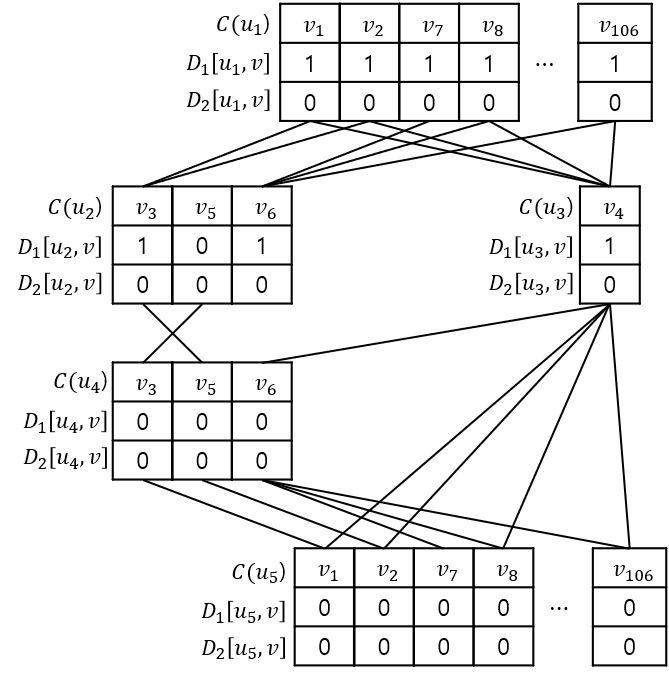}
    }
    \subcaptionbox{\textsf{DCS}\textsubscript{1} after $\Delta o_1=(+,v_4,v_7)$ occurs\label{fig:DCS1}}{
        \includegraphics[width=0.321\linewidth,scale=0.5]{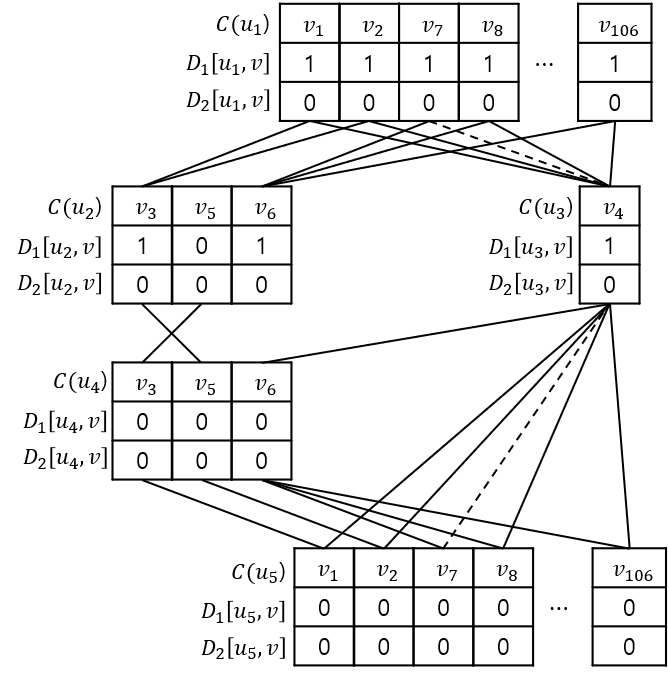}
    }
    \subcaptionbox{\textsf{DCS}\textsubscript{2} after $\Delta o_2=(+, v_3,v_6)$ occurs\label{fig:DCS2}}{
        \includegraphics[width=0.321\linewidth,scale=0.5]{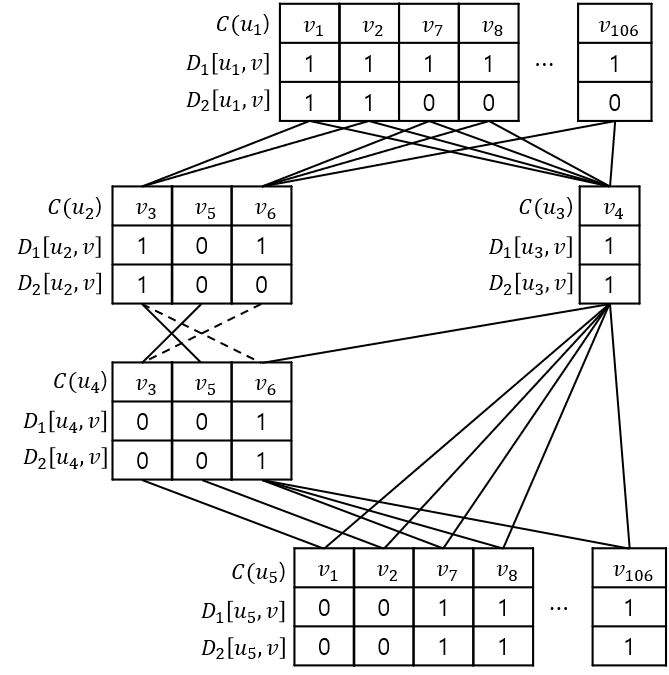}
    }
    \vspace*{-2mm}
\caption{\textbf{A running example of \textsf{DCS} structure} on DAG $\hat{q}$ in Figure \ref{fig:DAG_q} and the dynamic data graph $g$ in Figure \ref{fig:data_graph_g0}} 
\vspace*{-3mm}
\label{fig:DCS_example}
\end{figure*}

\section{DCS Structure}\label{sec:dcs_structure}
\subsection{DCS Structure}\label{subsec:dcs_structure}
% DAF \cite{han2019efficient} constructs an auxiliary data structure called \textsf{CS} (Candidate Space) consisting of candidate vertices and corresponding edges to handle the subgraph matching problem. To deal with continuous subgraph matching, we introduce an auxiliary data structure called the \textit{dynamic candidate space} (\textsf{DCS}) which stores the intermediate results of top-down and bottom-up dynamic programming between a DAG of a query graph and a dynamic data graph. Using \textsf{DCS}, we can symmetrically handle edge deletions and edge insertions.
To deal with continuous subgraph matching, we introduce an auxiliary data structure called the \textit{dynamic candidate space} (\textsf{DCS}) which stores weak embeddings of DAGs as intermediate results that help reduce the search space of backtracking based on the fact that a weak embedding is a necessary condition for an embedding. These intermediate results are obtained through top-down and bottom-up dynamic programming between a DAG of a query graph and a dynamic data graph. Compared to the auxiliary data structure \textsf{DCG} used in \textsf{TurboFlux}, \textsf{DCS} has non-tree edge information which \textsf{DCG} does not have, so it is advantageous in the backtracking process.
The auxiliary data structure \textsf{CS} (Candidate Space) which DAF \cite{han2019efficient} uses to solve the subgraph matching problem does not store intermediate results, and thus it cannot respond efficiently to the update operations.

% \textbf{(Definition of DCS structure)}
\noindent\textbf{\textsf{DCS} Structure.} Given a rooted DAG $\hat{q}$ from $q$ and a data graph $g$, a \textit{\textsf{DCS}} on $\hat{q}$ and $g$ consists of the following.
\begin{itemize}
    \item For each $u\in V(q)$, a candidate set $C(u)$ is a set of vertices $v\in V(g)$ such that $l_q(u)=l_g(v)$. Let $\langle u,v\rangle$ denote $v$ in $C(u)$.
    \item For each $u\in V(q)$ and $v\in C(u)$, $D_1[u,v]=1$ if there exists a weak embedding of sub-DAG $\hat{q}_{u}^{-1}$ at $v$; $D_1[u,v]=0$ otherwise.
    \item For each $u\in V(q)$ and $v\in C(u)$, $D_2[u,v]=1$ if there exists a weak embedding $M'$ of sub-DAG $\hat{q}_{u}$ at $v$ such that $D_1[u',v']=1$ for every mapping $(u',v')\in M'$; $D_2[u,v]=0$ otherwise.
    % \item For each $u\in V(q)$ and $v\in C(u)$, $D_i(u,v)$ is a boolean data which represents the survival of the candidate after the $i$th refinement, where $1 \leq i \leq n$. $D_i(u,v)=1$ if $D_{i-1}(u,v)=1$ and there exists a weak embedding $M'$ of $q^{(i)}_u$ at $v$ such that $D_{i-1}(u',v')=1$ for all $(u',v')\in M'$; $D_i(u,v)=0$ otherwise. Initially, $D_0(u,v)=1$ for every $u\in V(q)$ and $v\in C(u)$.
    % \item For each $(u,u')\in E(q)$ and $(v,v')\in E(g)$, if $v \in C(u)$ and $v' \in C(u')$, there is an edge $(\langle u,v\rangle ,\langle u',v'\rangle )$ between $\langle u,v\rangle$ and $\langle u',v'\rangle$. We say that $\langle u,v\rangle$ is a parent (or child) of $\langle u',v'\rangle$ if $u$ is a parent (or child) of $u'$ in $\hat{q}$.
    \item There is an edge $(\langle u,v\rangle ,\langle u',v'\rangle )$ between $\langle u,v\rangle$ and $\langle u',v'\rangle$ if and only if $(u,u')\in E(q)$ and $(v,v')\in E(g)$. We say that $\langle u,v\rangle$ is a parent (or child) of $\langle u',v'\rangle$ if $u$ is a parent (or child) of $u'$ in $\hat{q}$.
\end{itemize}

The \textsf{DCS} structure can be viewed as a labeled graph (labeled with $D_1$ and $D_2$) whose vertices are $\langle u,v\rangle$'s and edges are $(\langle u,v\rangle ,\allowbreak \langle u',v'\rangle )$'s.
Note that the intermediate results $D_1$ and $D_2$ which \textsf{DCS} stores are weak embeddings of sub-DAGs.
$D_1$ and $D_2$ store the results of top-down and bottom-up dynamic programming, respectively, which are used to filter candidates. For any embedding $M$ of $q$ in $g$, $D_2[u,v]=1$ must hold for every $(u,v)\in M$, since a weak embedding of a sub-DAG of $q$ is a necessary condition for an embedding of $q$. From this observation, we need only consider $(u,v)$ pairs such that $D_2[u,v]=1$ when computing an embedding of $q$ in $g$. 

% example of DCS + neighbor <u,v>
\begin{MyExample}
Figure \ref{fig:DCS_example} shows the \textsf{DCS} structure on the DAG $\hat{q}$ in Figure \ref{fig:DAG_q} and the dynamic data graph $g$ in Figure \ref{fig:data_graph_g0}. Figure \ref{fig:DCS0} shows the initial \textsf{DCS}\textsubscript{0} on $\hat{q}$ in Figure \ref{fig:DAG_q} and $g_0$ in Figure \ref{fig:data_graph_g0}. Figure \ref{fig:DCS1} and \ref{fig:DCS2} show \textsf{DCS} after $\Delta o_1$ and $\Delta o_2$ occur, respectively. Dashed lines $(\langle u_1,v_7\rangle , \langle u_3,v_4 \rangle )$ and $(\langle u_3,v_4\rangle , \langle u_4,v_7 \rangle )$ in Figure \ref{fig:DCS1} represent inserted edges due to $\Delta o_1$. Note that multiple edges can be inserted to \textsf{DCS} by one edge insertion to the data graph. In the initial \textsf{DCS}\textsubscript{0} (Figure \ref{fig:DCS0}), $C(u_2)=\{v_3,v_5,v_6\}$ because $v_3$, $v_5$, and $v_6$ have the same label as $u_2$, $D_1[u_2,v_3]=1$ since there exists a weak embedding $M'=\{(u_2,v_3),(u_1,v_1)\}$ of sub-DAG $\hat{q}^{-1}_{u_2}$ at $v_3$, and $D_2[u_2,v_3]=0$ because there is no weak embedding of sub-DAG $\hat{q}_{u_2}$ at $v_3$. Since there are no $(u,v)$ pairs such that $D_2[u,v]=1$ in Figure \ref{fig:DCS1}, \textsf{SymBi} reports that there are no positive matches for $\Delta o_1$ without backtracking. In contrast, \textsf{TurboFlux}, which uses the spanning tree in Figure \ref{fig:spanning_tree}, needs to perform backtracking only to find that there are no positive matches for $\Delta o_1$, because there exists a spanning tree that includes the inserted edge $(v_4,v_7)$ in the data graph.

\end{MyExample}

% edge conceptual? for reducing the storage cost

% D_2[u,v] = 0 => filtering

% \noindent\textbf{DAG-Graph DP} 
To compute $D_1$ and $D_2$, we use following recurrences which can be obtained from the definition:
\begin{align*}
    D_1[u,v]=1 & \text{ iff } \exists v_p \in C(u_p) \text{ adjacent to } v \text{ such that } D_1[u_p,v_p]=1 \\
    &  \text{ for every parent } u_p \text{ of } u \text{ in } \hat{q} \hspace{-3mm}\eqnlabel \label{eqn} \\
%\end{align*}
%\begin{align*}
    D_2[u,v]=1 & \text{ iff } D_1[u,v]=1 \text{ and } \exists v_c \in C(u_c) \text{ adjacent to } v \text{ such}\\ 
    &  \text{that } D_2[u_c,v_c]=1 \text{ for every child } u_c \text{ of } u \text{ in } \hat{q} \hspace{-3mm}\eqnlabel \label{eqn2}\\
\end{align*}

Based on the above recurrences, we can compute $D_1$ and $D_2$ by dynamic programming in a top-down and bottom-up fashion in DAG $\hat{q}$, respectively. Note that we reverse the parent-child relationship in the first recurrence in order to take only one DAG $\hat{q}$ into account.

\begin{MyLemma}\label{lma:construction_correct}
Recurrences (\ref{eqn}) and (\ref{eqn2}) correctly compute $D_1$ and $D_2$ according to the definition.
\end{MyLemma}

\color{black}

%$Parent(u)$ denotes the parents of $u$ in $\hat{q}$ and $Child(u)$ denotes the children of $u$ in $\hat{q}$ from the remaining part. 

\begin{MyLemma}\label{lma:construction_complexity}
Given a query graph $q$ and a data graph $g$, the space complexity of the \textsf{DCS} structure and the time complexity of \textsf{DCS} construction are $O(|E(q)|\times |E(g)|)$.
\end{MyLemma}

\subsection{DCS Update}\label{subsec:dcs_update}
In this subsection, we describe how to update the \textsf{DCS} structure for each update operation. An edge update in a data graph causes insertion or deletion of a set of edges in \textsf{DCS}, and makes changes on $D_1$ and $D_2$. Since the update algorithm works symmetrically for edge insertions and edge deletions, we describe how to update $D_1$ and $D_2$ when an edge is inserted, and then describe what changes when an edge is deleted.

We first explain \textsc{DCSChangedEdge} (Lines 6 and 11 in Algorithm \ref{alg:continous_subgraph_matching}) which returns a set of inserted/deleted edges in \textsf{DCS} due to the updated edge $e=(v,v')$. We traverse the query graph and find an edge $(u,u')$ such that $l_q(u)=l_g(v)$ and $l_q(u')=l_g(v')$. We then insert the edge $(\langle u,v\rangle ,\langle u',v'\rangle )$ into the set $E_{DCS}$. In Figure \ref{fig:query_and_data_graph}, \textsc{DCSChangedEdge} returns $\{(\langle u_2,v_3 \rangle ,\langle u_4,v_6\rangle )  ,(\langle u_2,v_6 \rangle ,\langle u_4,v_3\rangle )\}$ when $\Delta o_2=(+,v_3,v_6)$ occurs.

\noindent\textbf{Edge Insertion.} Now, we focus on updating $D_1$ for the case of edge insertion. Obviously, it is inefficient to recompute the entire process of top-down dynamic programming to update $D_1$ for each update. Instead of computing the whole $D_1$, we want to compute only the elements of $D_1$ whose values may change. To update $D_1$, we start with an edge $(\langle u_p,v_p\rangle ,\langle u,v\rangle )$ of $E_{DCS}$ where $\langle u_p,v_p\rangle$ is a parent of $\langle u,v\rangle$. If $D_1[u_p,v_p]$ is 0, then the edge $(\langle u_p,v_p\rangle ,\langle u,v\rangle )$ does not affect $D_1[u,v]$ nor its descendants, so we stop the update and move to the next edge of $E_{DCS}$. On the other hand, if $D_1[u_p,v_p]$ is 1, then $D_1$ of $\langle u,v\rangle$ and its descendants may be changed due to this edge. First, we compute $D_1[u,v]$. If $D_1[u,v]$ changes from 0 to 1, then we repeat this process for the children of $\langle u,v\rangle$ until $D_1$ has no changes. Next, we try to update $D_1$ with the next edge in $E_{DCS}$.

% Instead, we want to recompute only $D_1[u,v]$ whose value may change. There are two cases where $\langle u_p,v_p\rangle$ changes $D_1[u,v]$ where $u_p\in \textrm{Parent}(u)$: (1) if $D_1[u_p,v_p]=1$, and an edge between $\langle u,v\rangle$ and $\langle u_p,v_p\rangle$ is inserted or deleted, or (2) if $D_1[u_p,v_p]$ is changed, and there is an edge between $\langle u,v\rangle$ and $\langle u_p,v_p\rangle$. In both of these cases, we call $\langle u_p, v_p\rangle$ an \emph{updated parent} of $\langle u, v\rangle$. Since there is no changed value in $D_1$ at the beginning of update, we start with the first case. If some change occurs on $D_1[u,v]$, then we compute $D_1[u_c,v_c]$ where $\langle u_c,v_c\rangle$ is child of $\langle u,v\rangle$, and repeat this process on $\langle u_c,v_c\rangle$, so we update $D_1$ in a top-down fashion. 

% example - D_1 update & skip computing some D_1
\begin{MyExample}\label{ex:dcs_update}
When $\Delta o_2$ occurs, we update $D_1$ in Figure \ref{fig:DCS1} with a set $E_{DCS}=\{(\langle u_2,v_3 \rangle ,\langle u_4,v_6\rangle )  ,(\langle u_2,v_6 \rangle ,\allowbreak\langle u_4,v_3\rangle )\}$. As mentioned above, we start with the edge $(\langle u_2,v_3\rangle ,\allowbreak\langle u_4,v_6\rangle )$. Since $D_1[u_2,v_3]$ is 1, we recompute $D_1[u_4,v_6]$ and it changes from 0 to 1 because now every parent of $u_4$ has a candidate adjacent to $\langle u_4,v_6\rangle$ whose $D_1$ value is 1. Since $D_1[u_4,v_6]$ becomes 1, we iterate for the children of $\langle u_4,v_6\rangle$ (e.g., $\langle u_5, v_7\rangle ,\allowbreak\ldots , \langle u_5, v_{106} \rangle$), and then we stop updating $D_1$ because $u_5$ has no children. Now, we update $D_1$ with the next edge $(\langle u_2,v_6\rangle ,\langle u_4,v_3\rangle )$. Similarly to the previous case, we recompute $D_1[u_4,v_3]$, but it remains 0 because there are no edges between $\langle u_4,v_3 \rangle$ and $\langle u_3,v_4\rangle$. So, we stop the update with $(\langle u_2,v_6\rangle ,\langle u_4,v_3\rangle )$. Since $E_{DCS}$ has no more edges, we finish the update and obtain $D_1$ in Figure \ref{fig:DCS2}.

% We start updating $D_1$ with given a set $E_{DCS}=\{(\langle u_2,v_3 \rangle ,\langle u_4,v_6\rangle )  ,(\langle u_2,v_6 \rangle ,\langle u_4,v_3\rangle )\}$ in Figure \ref{fig:DCS2}. Since $D_1[u_2,v_3]$ and $D_1[u_2,v_6]$ are 1, it satisfy the condition of the first case, so we compute $D_1[u_4,v_6]$ and $D_1[u_4,v_3]$. $D_1[u_4,v_6]$ changes from 0 to 1, however $D_1[u_4,v_3]$ remains 0 because there are no edges between $\langle u_4,v_3 \rangle$ and $\langle u_3,v_4\rangle$. Since $\langle u_4, v_6\rangle$ became an updated parent of its children, we iterate for children of $\langle u_4,v_6\rangle$ (e.g., $\langle u_5, v_7\rangle ,\ldots , \langle u_5, v_{106} \rangle$), and then we stop updating $D_1$ because $u_5$ has no children. Note that we just recompute $D_1$ for 102 $(u,v)$ pairs rather than the whole $D_1$.
\end{MyExample}

We can see that there are two cases that $\langle u_p,v_p\rangle$ affects $D_1[u,v]$: 
\begin{enumerate}[(i)]
    \item If $D_1[u_p,v_p]=1$ and an edge between $\langle u_p,v_p\rangle$ and $\langle u,v\rangle$ is inserted.
    \item If $D_1[u_p,v_p]$ changes from 0 to 1 and there is an edge between $\langle u_p,v_p\rangle$ and $\langle u,v\rangle$.
\end{enumerate}
In both of these cases, we say that $\langle u_p,v_p\rangle$ is an \emph{updated parent} of $\langle u,v\rangle$. In Example \ref{ex:dcs_update}, $\langle u_2,v_3\rangle$ is an updated parent of $\langle u_4,v_6\rangle$ from case {(\romannumeral 1)} and $\langle u_4,v_6\rangle$ is an updated parent of its children from case {(\romannumeral 2)}.

% start from the previous example, skip some computation
% 2 problem - 1) compute D[u,v] many times, 2) we can skip some computation
% However, the above method has some problems. Suppose that $\langle u,v\rangle$ has $n$ updated parents. The above method computes $D_1[u,v]$ $n$ times in the worst case. Also, to compute ~~~~
However, the above method has redundant computations in two aspects. First, if $\langle u,v\rangle$ has $n$ updated parents then the above method computes $D_1[u,v]$ $n$ times in the worst case. Second, to compute $D_1[u,v]$, we need to reference the non-updated parents of $\langle u,v\rangle$ even if they do not change during the update. To handle these issues, we store additional information between $\langle u,v\rangle$ and its parents. When $\langle u_p,v_p\rangle$ becomes an updated parent of $\langle u,v\rangle$, instead of computing $D_1[u,v]$ from scratch, we update the information of $\langle u,v\rangle$ related to $\langle u_p,v_p\rangle$, and then update $D_1[u,v]$ using the stored information.

We store the aforementioned information using two additional arrays, $N^1_{u,v}[u_p]$ and $N^1_P[u,v]$:
\begin{itemize}
    \item $N^1_{u,v}[u_p]$ stores the number of candidates $v_p$ of $u_p$ such that there exists an edge $(\langle u_p,v_p\rangle, \langle u,v\rangle )$ and $D_1[u_p,v_p]=1$, where $u_p$ is a parent of $u$. For example, $N^1_{u_2,v_3}[u_1]=2$ in Figure \ref{fig:DCS1} because $v_1$ and $v_2$ in $C(u_1)$ satisfy the condition. By definition of updated parents and $N^1_{u,v}[u_p]$, we can easily update $N^1_{u,v}[u_p]$ while updating \textsf{DCS}: when $\langle u_p,v_p\rangle$ becomes an updated parent of $\langle u,v\rangle$, we increase $N^1_{u,v}[u_p]$ by 1.
    \item $N^1_P[u,v]$ stores the number of parents $u_p$ of $u$ such that $N^1_{u,v}[u_p]\neq 0$. When $N^1_{u,v}[u_p]$ changes from 0 to 1 during the update, we increase $N^1_P[u,v]$ by 1. We can update $D_1[u,v]$ using $N^1_P[u,v]$ from the following equation obtained from Recurrence \eqref{eqn} in Section \ref{subsec:dcs_structure}:
    \begin{align*}
        D_1[u,v]=1 \text{ if and only if } N^1_P[u,v]=|\textrm{Parent}(u)|.
    \end{align*}
    \vspace{-3mm}
\end{itemize}
 
% the number of candidates v_p such that <u_p,v_p> is a parent of <u,v> and D[u_p,v_p]=1

Back to the situation in Example \ref{ex:dcs_update}, we increase $N^1_{u_4,v_6}[u_2]$ by 1 instead of recomputing $D_1[u_4,v_6]$. Since $N^1_{u_4,v_6}[u_2]$ becomes 1 from 0,  we increase $N^1_P[u_4,v_6]$ from 1 to 2. Because $N^1_P[u_4,v_6]=|\textrm{Parent}(u_4)|=2$ now holds, $D_1[u_4,v_6]$ becomes 1. Thus, we can update $D_1$ correctly without redundant computations.

The revised method solves the two problems described earlier. The first problem is solved in two aspects. First, the revised method still performs the update for $\langle u,v\rangle$ as many times as the number of updated parents of $\langle u,v\rangle$. However, when there is an updated parent of $\langle u,v\rangle$, we update the corresponding arrays and $D_1[u,v]$ in constant time. So, during the $D_1$ update, the total computational cost to update $D_1[u,v]$ is proportional to the number of updated parents of $\langle u,v\rangle$. Second, even if we update $D_1[u,v]$ more than once, $\langle u,v\rangle$ affects its children at most once because $\langle u,v\rangle$ affects its children only when $D_1[u,v]$ changes from 0 to 1.  The second problem is solved because now we update only the information of $\langle u,v\rangle$ related to its updated parents (i.e., $N^1_{u,v}$ and $N^1_P[u, v]$) during the update process.

Similarly with the $D_1$ update, we can define \emph{updated child}, $N^2_{u,v}$, and $N^2_C$ to update $D_2$ efficiently in a bottom-up fashion. 
\begin{itemize}
    \item $N^2_{u,v}[u']$ stores the number of candidates $v'$ of $u'$ such that there exists an edge $(\langle u',v'\rangle ,\langle u,v\rangle )$ and $D_2[u',v']=1$, where $u'$ is a neighbor of $u$.
    \item $N^2_C[u,v]$ stores the number of children $u_c$ of $u$ such that $N^2_{u,v}[u_c]\neq 0$.
\end{itemize}  
While we need to define $N^2_{u, v}$ only for the children of $u$ in the $D_2$ update, we define $N^2_{u,v}$ for every neighbor of $u$ in order to use it in the backtracking process (Section \ref{subsec:computing_extendable_candidates}).
The difference between the $D_1$ update and the $D_2$ update arises from the condition that $D_1[u,v]$ should be 1 in order for $D_2[u,v]$ to be 1. There is one more case where $D_2[u,v]$ changes from 0 to 1, except when it changes due to its updated children. If $D_1[u,v]$ becomes 1 and $N^2_C[u,v]=|\textrm{Child}(u)|$ already holds, $D_2[u,v]$ changes from 0 to 1. For example, $D_2[u_5,v_7]$ in Figure \ref{fig:DCS2} changes to 1 after $D_1[u_5,v_7]$ changes to 1, and then $\langle u_5,v_7\rangle$ becomes an updated child of its parents.

% We can also define \emph{updated child} similarly with updated parent, and update $D_2$ in a bottom-up fashion. We can define $N^2_{u,v}$ and $N^2_C$ similarly with $N^1_{u,v}$ and $N^1_P$, defined as follows.
% \begin{align*}
%     & N^2_{u,v}[u_c]=|\{v_c\in C(u_c)\mid D_2[u_c,v_c]=1 \text{ and } \exists (v,v_c)\in E(g)\}| \\
%     & N^2_C[u,v]= |\{u_c\mid u_c\in \textrm{Child}(u) \text{ and } N^2_{u,v}[u_c]\neq 0\}|
% \end{align*}
% Now, we present the detailed algorithm for updating the \textsf{DCS} for each case. 

Algorithm \ref{alg:DCS_insertion_update} shows the process of updating $D_1$, $D_2$ and the additional arrays for edge insertion. There are two queues $Q_1$ and $Q_2$ which store $\langle u,v\rangle$ such that $D_1[u,v]$ and $D_2[u,v]$ changed from $0$ to $1$, respectively. \textsc{DCSInsertionUpdate} performs the update process described above for each inserted edge $(\langle u_1,v_1\rangle ,\langle u_2,v_2\rangle )$ in $E_{DCS}$. Suppose that $\langle u_1, v_1\rangle$ is a parent of $\langle u_2,v_2\rangle$. Lines 5-8 describe the update by case {(\romannumeral 1)} of updated parents and updated children. It invokes the following two algorithms. \textsc{InsertionTopDown} (Algorithm \ref{alg:insertion_topdown}) updates $N^1_{u_c,v_c}[u]$, $N^1_P[u_c,v_c]$, and $D_1[u_c,v_c]$ when $\langle u, v\rangle$ is an updated parent of $\langle u_c, v_c\rangle$. Also, when $D_1[u_c,v_c]$ becomes 1, it pushes $\langle u_c,v_c\rangle$ into $Q_1$ and check the condition ($N^2_c[u,v]=|\textrm{Child}(u)|$) to see if $D_2[u_c,v_c]$ can change to 1. \textsc{InsertionBottomUp} (Algorithm \ref{alg:insertion_bottomup}) works similarly.
%updates the corresponding arrays and pushes $\langle u_p,v_p\rangle$ into $Q_2$ if the condition holds when $\langle u, v\rangle$ is an updated child of $\langle u_p, v_p\rangle$.
Lines 11-14 (or Lines 15-18) perform the update process of case {(\romannumeral 2)} of updated parents (or updated children) until $Q_1$ (or $Q_2$) is not empty. 
%Since we define $N^2_{u,v}$ for every neighbor of $u$, we update $N^2_{u,v}$ for the parent of $u$ in Lines 9-10 and 19-20.

%\textsc{DCSInsertionUpdate} invokes two algorithms \textsc{InsertionTopDown}$(DCS,u,u_c,v,v_c)$ and \textsc{InsertionBottomUp}$(DCS,u_p,u,v_p,v)$. \textsc{InsertionTopDown}$(DCS,u,u_c,v,v_c)$ is invoked when $N^1_{u_c,v_c}[u]$ is increased by $v$. \textsc{InsertionTopDown}$(DCS,u,u_c,v,v_c)$ increases $N^1_{u_c,v_c}[u]$ by 1 and increases $N^1_P[u_c,v_c]$ by 1 if $N^1_{u_c,v_c}[u]$ is 0 before updating. Finally, if $N^1_P[u_c,v_c]$ becomes $|Parent(u_c)|$ then push $\langle u_c,v_c\rangle$ in $Q_1$. Similarly, \textsc{InsertionBottomUp}$(DCS,u_p,u,v_p,v)$ is invoked to update $\langle u_p,v_p\rangle$ affected by $\langle u,v\rangle$.

 %We start from the first case with a set $E_{DCS}$ which contains inserted edges. For each inserted edge $(\langle u,v\rangle , \langle u',v'\rangle )$ in $E_{DCS}$, we suppose that $\langle u,v\rangle$ is parent of $\langle u',v'\rangle$ (Lines 3-4). If $D_1[u,v]$ is $1$ or $D_2[u',v']$ is $1$, then we invoke \textsc{InsertionTopDown}$(DCS,u,u',v,v')$ and \textsc{InsertionBottomup}$(DCS,u',u,v',v)$ to update $D_1[u',v']$ and $D_2[u,v]$ (Lines 5-8).
 
% Next, we update $D_1[u,v]$ to 1 and make changes on child of $\langle u,v\rangle$ until $Q_1$ is not empty. If $N^2_C[u,v]=|Child(u)|$ holds and $D_1[u,v]$ becomes 1, then push $\langle u,v\rangle$ to $Q_2$ (Lines 9-16). Finally, we update $D_2[u,v]$ to 1 and makes changes on parent of $\langle u,v\rangle$ until $Q_2$ is not empty (Lines 17-22).

\begin{algorithm}
% \KwIn{A data graph $g$, a query graph $q$, and edge $e=(v,v')$}
% \KwOut{all positive/negative matches}
$Q_1,Q_2\gets$ empty queue\;
\ForEach{$(\langle u_1,v_1\rangle ,\langle u_2,v_2\rangle )\in E_{DCS}$}{
    \If{$\langle u_2, v_2\rangle$ \emph{is a parent of} $\langle u_1, v_1\rangle$}{
        swap($\langle u_1,v_1\rangle,\langle u_2,v_2\rangle$)\;
    }
    \If{$D_1[u_1,v_1]=1$}{
        \textsc{InsertionTopDown}$(\langle u_1,v_1\rangle,\langle u_2,v_2\rangle)$\;
    }
    \If{$D_2[u_2,v_2]=1$}{
        \textsc{InsertionBottomUp}$(\langle u_2,v_2\rangle,\langle u_1,v_1\rangle)$\;
    }
    \If{$D_2[u_1,v_1]=1$}{
        $N^2_{u_2,v_2}[u_1]\gets N^2_{u_2,v_2}[u_1]+1$
    }
    \While{$Q_1\neq \emptyset$}{
        $\langle u,v\rangle\gets Q_1.pop$\;
        \ForEach{$\langle u_c, v_c\rangle$ \emph{which is a child of} $\langle u,v\rangle$}{
            \textsc{InsertionTopDown}$(\langle u,v\rangle,\langle u_c,v_c\rangle)$\;
        }
    }
    \While{$Q_2\neq \emptyset$}{
        $\langle u,v\rangle\gets Q_2.pop$\;
        \ForEach{$\langle u_p, v_p\rangle$ \emph{which is a parent of} $\langle u, v\rangle$}{
            \textsc{InsertionBottomUp}$(\langle u,v\rangle,\langle u_p,v_p\rangle)$\;
        }
        \ForEach{$\langle u_c, v_c\rangle$ \emph{which is a child of} $\langle u, v\rangle$}{
            $N^2_{u_c,v_c}[u]\gets N^2_{u_c,v_c}[u]+1$
        }
    }
}

\caption{\textsc{DCSInsertionUpdate($DCS,E_{DCS}$)}}
\label{alg:DCS_insertion_update}
\vspace*{-1mm}
\end{algorithm}

\begin{algorithm}
% \KwIn{A data graph $g$, a query graph $q$, and edge $e=(v,v')$}
% \KwOut{all positive/negative matches}
\If{$N^1_{u_c,v_c}[u]=0$}{
    $N^1_P[u_c,v_c]\gets N^1_P[u_c,v_c]+1$ \;
    \If{$N^1_P[u_c,v_c]=|\textrm{\emph{Parent}}(u_c)|$}{
        $D_1[u_c,v_c]\gets 1$\;
        $Q_1.push(\langle u_c,v_c\rangle )$\;
        \If{$N^2_C[u_c,v_c]=|\textrm{\emph{Child}}(u_c)|$}{
            $D_2[u_c,v_c]\gets 1$\;
            $Q_2.push(\langle u_c,v_c\rangle)$\;
        }
    }
        
}
$N^1_{u_c,v_c}[u]\gets N^1_{u_c,v_c}[u]+1$
\vspace{1mm}
\caption{\textsc{InsertionTopDown($\langle u,v\rangle ,\langle u_c,v_c\rangle $)}}
\label{alg:insertion_topdown}
\vspace*{-1mm}
\end{algorithm}

\begin{algorithm}
% \KwIn{A data graph $g$, a query graph $q$, and edge $e=(v,v')$}
% \KwOut{all positive/negative matches}
\If{$N^2_{u_p,v_p}[u]=0$}{
    $N^2_C[u_p,v_p]\gets N^2_C[u_p,v_p]+1$ \;
    \If{$D_1[u_p,v_p]=1$ and $N^2_C[u_p,v_p]=|\textrm{\emph{Child}}(u_p)|$}{
        $D_2[u_p,v_p]\gets 1$\;
        $Q_2.push(\langle u_p,v_p\rangle )$\;
    }
}
$N^2_{u_p,v_p}[u]\gets N^2_{u_p,v_p}[u]+1$
\vspace{1mm}
\caption{\textsc{InsertionBottomUp($\langle u,v\rangle ,\langle u_p,v_p\rangle $)}}
\label{alg:insertion_bottomup}
\vspace*{-1mm}
\end{algorithm}

Now we show that Algorithm \ref{alg:DCS_insertion_update} correctly updates $D_1$ and $D_2$ for the edge insertion.

\begin{MyLemma}\label{lma:update_correct}
If we have a correct \textsf{DCS}, and edges in $E_{DCS}$ are inserted into \textsf{DCS} by running Algorithm \ref{alg:DCS_insertion_update}, the \textsf{DCS} structure is still correct after the insertion.
\end{MyLemma}
\color{black}

%\vspace{-2mm}
\noindent\textbf{Edge Deletion.} We can update \textsf{DCS} for edge deletion with a small modification of the previous method. The first case of the updated parent (or updated child) is changed to when an edge is deleted and the second case is changed to when $D_1[u_p,v_p]$ (or $D_2[u_p,v_p]$) changes from 1 to 0. Next, if $D_2[u,v]\allowbreak=1$ and $D_1[u,v]$ becomes 0 during the $D_1$ update, then $D_2[u,v]$ also changes to 0.

%Similarly to edge insertion case, we can update \textsf{DCS} for edge deletion using two queues $Q_1$ and $Q_2$ (Algorithm \ref{alg:DCS_deletion_update}). The main difference is that $Q_1$ stores not only $\langle u,v\rangle$ where $D_1[u,v]$ becomes 0 from 1, but also $\langle u,v\rangle$ where $D_1[u,v]$ becomes 0 from 2.

\begin{MyLemma}\label{lma:update_complexity}

Let $P$ be the set of \textsf{DCS} vertices $\langle u,v\rangle$ such that $D_1[u,v]$ or $D_2[u,v]$ is changed during the update. Then the time complexity of the \textsf{DCS} update is $O(\sum_{p\in P}\deg(p)+|E_{DCS}|)$, where $\deg(p)$ is the number of edges connected to $p$. Also, the space complexity of the \textsf{DCS} update excluding \textsf{DCS} itself is $O(|E(q)|\times |V(g)|)$.

\end{MyLemma}

In the worst case, almost all $D_1[u,v]$ and $D_2[u,v]$ in \textsf{DCS} may be changed and the time complexity becomes $O(|E(q)|\times |E(g)|)$, so there is no difference from recomputing \textsf{DCS} from scratch. In Section \ref{sec:performance_evaluation}, however, we will show that there are very few changes in $D_1[u,v]$ or $D_2[u,v]$ in practice, so our proposed update method is efficient.

%\vspace*{-3mm}
\section{Backtracking}\label{sec:backtracking}

In this section, we present our matching algorithm to find all positive/negative matches in the \textsf{DCS} structure. Our matching algorithm works regardless of the cases of edge insertion and edge deletion.

%\vspace*{-3mm}
\subsection{Backtracking Framework}

We find matches by gradually extending a partial embedding until we get an (full) embedding of $q$ in $g$. We extend a partial embedding by matching an \emph{extendable} vertex of $q$, which is defined as below.

\begin{MyDefinition}
Given a partial embedding $M$, a vertex $u$ of query graph $q$ is called \emph{extendable} if $u$ is not mapped to a vertex of $g$ in $M$ and at least one neighbor of $u$ is mapped to a vertex of $g$ in $M$. 
\end{MyDefinition}

Note that the definition of extendable vertices is different from that of \textsf{DAF} \cite{han2019efficient}, which requires \emph{all} parents of $u$ to be mapped to a vertex of $g$ while we require only \emph{one} neighbor of $u$. The difference occurs because \textsf{DAF} has a fixed query DAG and a root vertex for backtracking, while our algorithm has to start backtracking from an arbitrary edge.

We start by mapping one edge from $E_{DCS}$, since we need only find matches including at least one updated edge in $E_{DCS}$. Until we find a full embedding, we recursively perform the following steps. First, we find all extendable vertices, and choose one vertex among them according to the \emph{matching order}. Once we decide an extendable vertex $u$ to match, we compute its \emph{extendable candidates}, which are the vertices in the data graph that can be matched to $u$. Formally, we define an extendable candidate as follows.

\begin{MyDefinition}
    Given a query vertex $u$, a data vertex $v$ is its \textit{extendable candidate} if $v$ satisfies the following conditions:
\begin{enumerate}[1.]
    \item $D_2[u, v] = 1$ (i.e., it is not filtered in the \textsf{DCS} structure)
    \item For all matched neighbors $u'$ of $u$, $(M(u'), v) \in E(g)$
\end{enumerate}
\label{def:extendablecandidates}
\end{MyDefinition}

The set of extendable candidates of $u$ is denoted by $C_M(u)$. Finally, we extend the partial embedding by matching $u$ to one of its extendable candidates and continue the process.

\setlength{\textfloatsep}{0pt}
% \vspace*{-7mm}

Algorithm \ref{alg:backtracking_framework} shows the overall backtracking process. This algorithm is invoked with $M=\emptyset$ in Algorithm \ref{alg:continous_subgraph_matching}. For each edge $(\langle u, v\rangle, \langle u', v'\rangle)$ in $E_{DCS}$, we start backtracking in Lines 6-11 only when $D_2[u, v] = D_2[u', v'] = 1$ (i.e., none of the pairs are filtered). We recursively extend a partial embedding in Lines 13-19. If we get a full embedding, we output it as a match in Line 2. \textsc{UpdateEmbedding} and \textsc{RestoreEmbedding} maintain additional values related to the matching order, every time a new match is augmented to $M$ (i.e., $M$ is updated) or an existing match is removed from $M$ (i.e., $M$ is restored). In Section 5.3, we explain what these functions do in more detail.

\begin{algorithm}
\KwIn{\textsf{DCS}, $E_{DCS}$, and a partial embedding $M$}
\KwOut{all positive/negative matches including an edge in $E_{DCS}$}

\If{$|M| = |V(q)|$}{
    Report $M$ as a match\;
}
\ElseIf{$|M| = 0$}{
    \ForEach{$(\langle u, v\rangle, \langle u', v'\rangle) \in E_{DCS}$}{
        \If{$D_2[u, v] = 1$ and $D_2[u', v'] = 1$}{
            $M \gets \{(u, v), (u', v')\}$\;
            \textsc{UpdateEmbedding}$(M, u)$\; 
            \textsc{UpdateEmbedding}$(M, u')$\; 
            \textsc{FindMatches}$(\textsf{DCS}, E_{DCS}, M)$\;
            \textsc{RestoreEmbedding}$(M, u')$\; 
            \textsc{RestoreEmbedding}$(M, u)$\;
        }
    }
}
\Else{
    $u \gets$ next vertex according to the matching order\;
    
    Compute $C_M(u)$\;
    
    \ForEach{$v \in C_M(u)$}{
        $M' \gets M \cup \{(u, v)\}$\;
        
        \textsc{UpdateEmbedding}$(M, u)$\;
        \textsc{FindMatches}$(\textsf{DCS}, E_{DCS}, M')$\;
        \textsc{RestoreEmbedding}$(M, u)$\;
    }
}
\caption{\textsc{FindMatches}$(\textsf{DCS}, E_{DCS}, M)$}
\label{alg:backtracking_framework}
\vspace*{-1mm}
\end{algorithm}

\vspace{-5mm}
\subsection{Computing Extendable Candidates}\label{subsec:computing_extendable_candidates}

According to Definition \ref{def:extendablecandidates}, the set of extendable candidates $C_M(u)$ of an extendable vertex $u$ is defined as follows:
\begin{equation*}
    C_M(u) = \{ v : D_2[u, v] = 1, \forall u' \in \textrm{Nbr}_M(u), (v, M(u')) \in E(g) \},
\end{equation*}
where $\textrm{Nbr}_M(u)$ represents the set of matched neighbors of $u$. 

We can compute the extendable candidates of $u$ based on the above equation. Even though we can compute $C_M(u)$ by naively iterating through all data vertices $v$ with $D_2[u, v] = 1$ and checking whether $(M(u'), v) \in E(g)$ for all matched neighbors $u'$ of $u$, there can be a large number of $v$'s with $D_2[u, v] = 1$, and thus it costs a lot of time to iterate through them. 

Here we check the conditions in an alternative order to reduce the number of iterations. Given a vertex $u' \in \textrm{Nbr}_M(u)$, we define a set $S_{u'} = \{ v \in C(u) : D_2[u, v] = 1, (v, M(u')) \in E(g) \}$. Among the vertices $u' \in \textrm{Nbr}_M(u)$, we find a vertex with the smallest $|S_{u'}|$ and call it $u_{min}$. We can see that the definition of $|S_{u'}|$ matches the definition of $N^2_{u', M(u')}[u]$ in Section 4.2, since the existence of an edge $(\langle u,v\rangle ,\langle u',M(u')\rangle )$ is equivalent to the existence of an edge $(v, M(u'))$ if $(u, u') \in E(q)$ and $v \in C(u)$. Therefore, we have $|S_{u'}| = N^2_{u', M(u')}[u]$ and thus $u_{min}$ is the vertex $u' \in \textrm{Nbr}_M(u)$ with smallest $N^2_{u', M(u')}[u]$.

Once we compute $u_{min}$, we compute $S_{u_{min}}$ by iterating through the neighbors $v$ of $M(u_{min})$ and checking whether $v \in C(u)$ and $D_2[u, v] = 1$. By using $S_{u_{min}}$, we rewrite $C_M(u)$ as follows: 
\begin{equation*}
\begin{aligned}
C_M(u) = \{ v \in S_{u_{min}} : & \: \forall u' \in \textrm{Nbr}_M(u) \backslash \{u_{min}\}, (v, M(u')) \in E(g) \}.
\end{aligned}
\end{equation*}
Based on the equation, we compute $C_M(u)$ by iterating through the vertices $v$ in $S_{u_{min}}$ and checking whether the edge $(M(u'), v)$ exists for every $u' \in \textrm{Nbr}_M(u) \backslash \{u_{min}\}$. Note that we need only iterate through $S_{u_{min}}$,  which has a considerably smaller size than the number of vertices $v$ with $D_2[u, v] = 1$ in usual. Algorithm \ref{alg:computing_Cmu} shows an algorithm to compute $C_M(u)$.

%We compute the extendable candidates of $u$ by using the \textsf{DCS} structure and considering the neighbors of $u$. We first choose one neighbor $u'$ of $u$, and build a set of vertices $S$ which satisfy the first condition and second condition for $u'$ from Definition \ref{def:extendablecandidates}. Given a set $S$, We check the second conditions for other neighbors of $u$ to compute $C_M(u)$ from $S$. Since we need to iterate through $S$ to compute $C_M(u)$, we want to keep the size of $S$ small as possible.

%Given an extendable vertex $u$ and its matched neighbor $u'$, the number of data vertices $v$ such that $D_2[u, v] = 1$ and $(M(u'), v) \in E(g)$ (which is the size of $S$) is stored in $N^2_{u', M(u')}(u)$. We compare $N^2_{u', M(u')}(u)$ for every $u'$, and choose $u'$ that has the smallest value among them. Let's call it $u_{min}$. Now we compute the set $S$ of data vertices $v$ such that $D_2[u, v] = 1$ and $(M(u_{min}), v) \in E(g)$. Finally, we compute $C_M(u)$ by iterating through $v \in S$ and leave the vertices such that $(M(u'), v) \in E(g)$ for all matched neighbors $u'$ of $u$. Algorithm \ref{alg:computing_Cmu} shows an algorithm to compute $C_M(u)$.

\begin{algorithm}
\KwIn{\textsf{DCS}, a data graph $g$, a query graph $q$, and an extendable query vertex $u$}
\KwOut{A set of extendable candidates $C_M(u)$}

$\textrm{Nbr}_M(u) \gets$ a set of matched neighbors of $u$ in $q$\;

$u_{min} \gets u' \in \textrm{Nbr}_M(u)$ with smallest $N^2_{u', M(u')}[u]$\;

$S_{u_{min}} \gets \{ v \in V(g) : D_2[u, v] = 1, (M(u_{min}), v) \in E(g)\}$\;

$C_M(u) \gets \{ v \in S_{u_{min}} : \textbf{forall } u' \in \textrm{Nbr}_M(u), (M(u'), v) \in E(g)\}$\;

\caption{\textsc{Computing $C_M(u)$}}
\label{alg:computing_Cmu}

\end{algorithm}

% ** get $u_{min}$ with minimum $\textit{validParentCand}(u_{min}, M(u_{min}), u)$

\vspace*{-5mm}
\subsection{Matching Order}
We select a matching order that can reduce the search space (and thus the backtracking time). We choose a matching order based on the size of extendable candidates, and thus it can be adaptively changed during the backtracking process.

In the case of the subgraph matching problem, it is known that the \emph{candidate-size order} is an efficient way \cite{han2019efficient}, which chooses the extendable vertex with smallest $|C_M(u)|$. We basically follow the candidate-size order with an approximation for speed-up. Even though we can compute the exact size of extendable candidates every time we need to decide which extendable vertex to match as in \textsf{DAF} \cite{han2019efficient}, it costs a high overhead in our case because the definition of extendable vertices is different. In the case of \textsf{DAF}, a vertex $u$ is extendable only when all parents of $u$ are matched. Therefore, the extendable candidates of an extendable vertex $u$ do not change. In contrast, in our algorithm, a vertex $u$ is extendable if at least one of neighbors of $u$ is matched. Therefore, $C_M(u)$ may change even if $u$ remains extendable, when an unmatched neighbor of $u$ becomes matched. As a result, our algorithm has more frequent changes in $C_M(u)$, and a higher overhead of maintaining it when compared to \textsf{DAF}.

To handle this issue, we use an estimated size of extendable candidates which can be maintained much faster, and we compute $C_M(u)$ only when all neighbors of $u$ are matched (i.e., when $C_M(u)$ no longer changes). Here we use the fact that in Algorithm \ref{alg:computing_Cmu}, the size of $C_M(u)$ is bounded by $|S_{u_{min}}| = N^2_{u_{min}, M(u_{min})} [u]$. We use this value as an estimated size of extendable candidates, since it provides an upper bound and approximation of $|C_M(u)|$, and it can be easily maintained when we update or restore partial embeddings. In a more formal way, we define $E(u)$, an \textit{estimated size of extendable candidates} of $u$, as follows.

\vspace{-1.5mm}
\begin{equation*}
    E(u) = \min_{u' \in \textrm{Nbr}_M(u)} \{N^2_{u', M(u')}[u]\}
\end{equation*}
% \vspace{-1mm}
Note that for every extendable candidate $u$, $\textrm{Nbr}_M(u)$ is not empty by definition, and thus $E(u)$ is well-defined.

Every time match $(u, v)$ occurs, we iterate through the neighbors $u'$ of $u$ in $q$, and update $E(u')$ for them. Since we have to restore a partial embedding later, we store the old $E(u')$ every time the update occurs, and we revert to the old $E(u')$ when the partial embedding is restored. These are done in \textsc{UpdateEmbedding} and \textsc{RestoreEmbedding} in Algorithm \ref{alg:backtracking_framework}.

\begin{MyExample}
\color{black}
Let's consider an edge insertion operation $\Delta o_2$ in Figure \ref{fig:query_and_data_graph}. As shown in Figure \ref{fig:DCS2}, we get $E_{DCS} = \{(\langle u_2, v_3\rangle, \langle u_4, v_6\rangle), \allowbreak(\langle u_2, v_6\rangle, \langle u_4, v_3\rangle)\}$. We begin with an edge $(\langle u_2,\allowbreak v_3\rangle, \langle u_4, v_6\rangle)$, which results in a partial embedding $M = \{(u_2, v_3), \allowbreak (u_4, v_6)\}$. There are three extendable vertices at this point, which are $u_1$, $u_3$ and $u_5$. We compare the estimated sizes of extendable candidates. Since $E(u_3) =  N^2_{u_4, v_6}\allowbreak[u_3] = 1$ is the smallest compared to $E(u_1) = 2$ and $E(u_5) = 100$, we choose $u_3$ as the next query vertex to match. We get $C_M(u_3) = \{v_4\}$ by Algorithm \ref{alg:computing_Cmu} and extend the partial embedding to $M = \{(u_2, v_3), (u_4, v_6), (u_3, v_4)\}$. Now we choose the next extendable vertex between $u_1$ and $u_5$. We again compare the estimated size of extendable candidates, and match $u_1$ first. Finally, we match $u_5$ with its extendable candidates, and output the matches. For \textsf{DCS} edge $(\langle u_2, v_6\rangle, \langle u_4, v_3\rangle)$, we skip finding matches since $D_2[u_2, v_6] = D_2[u_4, v_3] = 0$.
\end{MyExample}

\color{black}
\subsection{Isolated Vertices}
In this subsection, we describe the leaf decomposition technique from \cite{bi2016efficient} and introduce our new idea, called \emph{isolated vertices}.

For a query graph $q$, its \textit{leaf} vertices are defined as the vertices that have only one neighbor. The main idea of leaf decomposition is to postpone matching the leaf vertices of $q$ until all other vertices are matched. If we match a non-leaf vertex first, its unmatched neighbors can be the new extendable vertices (if none of their neighbors were matched before), or have their extendable candidates pruned (if at least one of their neighbors were matched before), both of which may lead to a smaller search space. These advantages do not apply when we match a leaf vertex first, since there are no unmatched neighbors if the leaf vertex is an extendable vertex.

Based on the above properties, we define \emph{isolated vertices} as follows.
\begin{MyDefinition}
For a query graph $q$, a data graph $g$, and a partial embedding $M$, an \emph{isolated vertex} is an extendable vertex in $q$, where all of its neighbors are mapped in $M$.
\end{MyDefinition}

Note that postponing matching the isolated vertices enjoy the advantages of leaf decomposition, since isolated vertices have no unmatched neighbors and thus have the same properties as leaves in the context of leaf decomposition. Also note that every extendable leaf vertex is also an isolated vertex by definition, but the converse is not true. For example, consider a partial embedding $M = \{(u_2, v_3), (u_3, v_4), (u_4, v_6)\}$ in Figure \ref{fig:query_and_data_graph}. Even though there are no leaf vertices in Figure \ref{fig:query_graph_q}, there are two isolated vertices, $u_1$ and $u_5$. Therefore, the notion of isolated vertices fully includes the leaf decomposition technique, and extends it further.

By combining the discussions in Section 5.3. and 5.4., we use the following matching order.

\begin{enumerate}[1.]
    \item Backtrack if there exists an isolated vertex $u$ such that all data vertices in $C_M(u)$ have already matched.
    \item If there exists at least one non-isolated extendable vertex in $q$, we choose the non-isolated extendable vertex $u$ with smallest $E(u)$.
    \item If every extendable vertex is isolated, we choose the extendable vertex $u$ with smallest $E(u)$.
\end{enumerate}

\section{Performance evaluation}\label{sec:performance_evaluation}
In this section, we present experimental results to show the effectiveness of our algorithm \textsf{SymBi}. Since \textsf{TurboFlux} \cite{kim2018turboflux} outperforms the other existing algorithms (e.g., \textsf{IncIsoMat} \cite{fan2013incremental}, \textsf{SJ-Tree} \cite{choudhury2015selectivity}, \textsf{Graphflow} \cite{kankanamge2017graphflow}), we only compare \textsf{TurboFlux} and \textsf{SymBi}. Experiments are conducted on a machine with two Intel Xeon E5-2680 v3 2.50GHz CPUs and 256GB memory running CentOS Linux. The executable file of \textsf{TurboFlux} was obtained from the authors. 

\noindent\textbf{Datasets.} We use two datasets referred to as \textsf{LSBench} and \textsf{Netflow} which are commonly used in previous works \cite{choudhury2015selectivity, kim2018turboflux}. \textsf{LSBench} is synthetic RDF social media stream data generated by the Linked Stream Benchmark data generator \cite{le2012linked}. We generate three different sizes of datasets with 0.1, 0.5, and 2.5 million users with default settings of Linked Stream Benchmark data generator, and use the first dataset as a default. This dataset contains 23,317,563 triples. \textsf{Netflow} is a real dataset (CAIDA Internet Anonymized Traces 2013 Dataset \cite{caida2013}) which contains anonymized passive traffic traces obtained from CAIDA. \textsf{Netflow} consists of 18,520,759 triples. We split 90\% of the triples of a dataset into an initial graph and 10\% into a graph update stream. 

\noindent\textbf{Queries.} We generate query graphs by random walk over schema graphs. To generate various types of queries, we use schema graphs instead of data graphs to randomly select edge labels regardless of edge distribution \cite{kim2018turboflux}. For each dataset, we set four different query sizes: 10, 15, 20, 25 (denoted by ``G10'', ``G15'', ``G20'', and ``G25''). The size of a query is defined as the number of triples. We exclude queries that have no matches for the entire graph update stream. Also, we use the queries used in \cite{kim2018turboflux} (denoted by ``T3'', ``T6'', ``T9'', ``T12'', ``G6'', ``G9'', and ``G12''), where ``T'' stands for tree and ``G'' stands for graph (having cycles). We generate 100 queries for each dataset and query size. One experiment consists of a dataset and a query set of 100 query graphs with a same size.

\noindent\textbf{Performance Measurement.} We measure the elapsed time of continuous subgraph matching for a dataset and a query graph for the entire update stream $\Delta g$. The preprocessing time (e.g., time to build the initial data graph and the initial auxiliary data structure) is excluded from the elapsed time. Since continuous subgraph matching is an NP-hard problem, some query graphs may not finish in a reasonable time. To address this issue, we set a 2-hours time limit for each query. If some query reaches the time limit, we record the query processing time of that query as 2 hours. We say that a query graph is \textit{solved} if it finishes within 2 hours. To evaluate an algorithm regarding a query set, we report the average elapsed time, the number of solved query graphs, and the average peak memory usage of the program using the ``ps'' utility.

\subsection{Experimental Results}\label{subsec:experimental_resutls}

The performance of \textsf{SymBi} was evaluated in several aspects: (1) varying the query size, (2) varying the insertion rate, (3) varying the deletion rate, and (4) varying the dataset size. Table \ref{tab:experiment_settings} shows the parameters of the experiments. Values in boldface in Table \ref{tab:experiment_settings} are used as default parameters. The insertion rate is defined as the ratio of the number of edge insertions to the number of edges in the original dataset before splitting. Thus, 10\% insertion rate means the entire graph update stream we split. Also, the deletion rate is defined as the ratio of the number of edge deletions to the number of edge insertions in the graph update stream. For example, if the deletion rate is 10\%, one edge deletion occurs for every ten edge insertions. For an edge deletion, we randomly choose an arbitrary edge in the data graph at the time the edge deletion occurs, and delete it.

\vspace*{-2mm}
\begin{table}[h]
    \centering
    \caption{Experiment settings}
    \vspace*{-3mm}
    \begin{tabular}{cc}
    \toprule
    \textbf{Parameter} & \textbf{Value Used} \\ 
    \midrule
    Datasets & \textsf{LSBench}, \textsf{Netflow}\\ 
    Query size & \textbf{G10}, G15, G20, G25, \\
    & T3, T6, T9, T12, G6, G9, G12\\ 
    % & G6, G9, G12 \\
    Insertion rate & 2, 4, 6, 8, \textbf{10} \\ 
    Deletion rate & \textbf{0}, 2, 4, 6, 8, 10 \\ 
    Dataset size & \textbf{0.1}, 0.5, 2.5 million users (\textsf{LSBench}) \\ 
    \bottomrule
    \end{tabular}
    \label{tab:experiment_settings}
\end{table}
\vspace*{-3mm}

\noindent\textbf{Efficiency of \textsf{DCS} Update.} Before we compare our results with \textsf{TurboFlux}, we first show the efficiency of our \textsf{DCS} update algorithm as described in Lemma \ref{lma:update_complexity}. Table \ref{tab:update_recompute} shows the number of updated vertices and the number of visited edges in \textsf{DCS} per update operation for \textsf{Netflow} and \textsf{LSBench}. ``DCS $|V|$'' and ``DCS $|E|$'' denote the average number of vertices and the average number of edges of the DCS structure. ``Updated vertices'' denotes the average number of $\langle u,v\rangle$'s such that $D_1[u,v]$ or $D_2[u,v]$ changes per update operation and ``Visited edges'' denotes the average number of DCS edges visited during the update. This result shows that the portion of \textsf{DCS} we need to update is extremely small compared to the size of \textsf{DCS}.

\begin{table}[h]
    \centering
    \caption{The number of updated vertices and visited edges in DCS per update operation (top: \textsf{Netflow}, bottom: \textsf{LSBench})}
    \vspace*{-1mm}
    \begin{tabular}{ccccc}
    \toprule
    \textbf{Query set} & \textbf{DCS $|V|$} & \textbf{DCS $|E|$} & \textbf{Updated} & \textbf{Visited} \\ 
     & & & \textbf{vertices} & \textbf{edges} \\
    \midrule
    G10	&	30744014	&	9725255	&	0.033	&	4.634	\\
    G15	&	45975850	&	16480557	&	0.052	&	9.655	\\
    G20	&	58217388	&	19570587	&	0.052	&	10.177	\\
    G25	&	76564119	&	25310130	&	0.024	&	12.396	\\
    T3	&	12459580	&	3850193	&	11.219	&	36.284	\\
    T6	&	21804265	&	7888574	&	5.382	&	13.534	\\
    T9	&	31148950	&	12044573	&	2.378	&	12.951	\\
    T12	&	40493635	&	16118240	&	2.596	&	15.999	\\
    G6	&	18689370	&	5583268	&	0.048	&	2.168	\\
    G9	&	28034055	&	8868896	&	0.072	&	4.048	\\
    G12	&	37378740	&	12295665	&	0.069	&	6.093	\\
    \midrule
    G10	&	51111071	&	6872525	&	0.203	&	1.997	\\
    G15	&	75233829	&	10584646	&	0.225	&	3.006	\\
    G20	&	83517886	&	10721872	&	0.141	&	2.536	\\
    G25	&	125354981	&	18676312	&	0.184	&	4.573	\\
    T3	&	18339548	&	2104435	&	0.104	&	0.463	\\
    T6	&	32198412	&	4072171	&	0.07	&	0.773	\\
    T9	&	46421983	&	5999443	&	0.065	&	1.085	\\
    T12	&	60489250	&	8021626	&	0.055	&	1.335	\\
    G6	&	29332857	&	2918077	&	0.042	&	0.675	\\
    G9	&	42410206	&	4969639	&	0.03	&	1.001	\\
    G12	&	56008565	&	6764759	&	0.029	&	1.298	\\
    \bottomrule
    \end{tabular}
    \label{tab:update_recompute}
\end{table}

\noindent\textbf{Varying the query size.} First, we vary the number of triples in query graphs. We set the insertion rate to 10\% and the deletion rate to 0\% (i.e., no edge deletion in the graph update stream).

\begin{figure}[h]
    \centering
    \subcaptionbox{Average elapsed time (in milliseconds)\label{exp:vary_query_size_netflow_time}}{
        \includegraphics[width=\linewidth,scale=0.2]{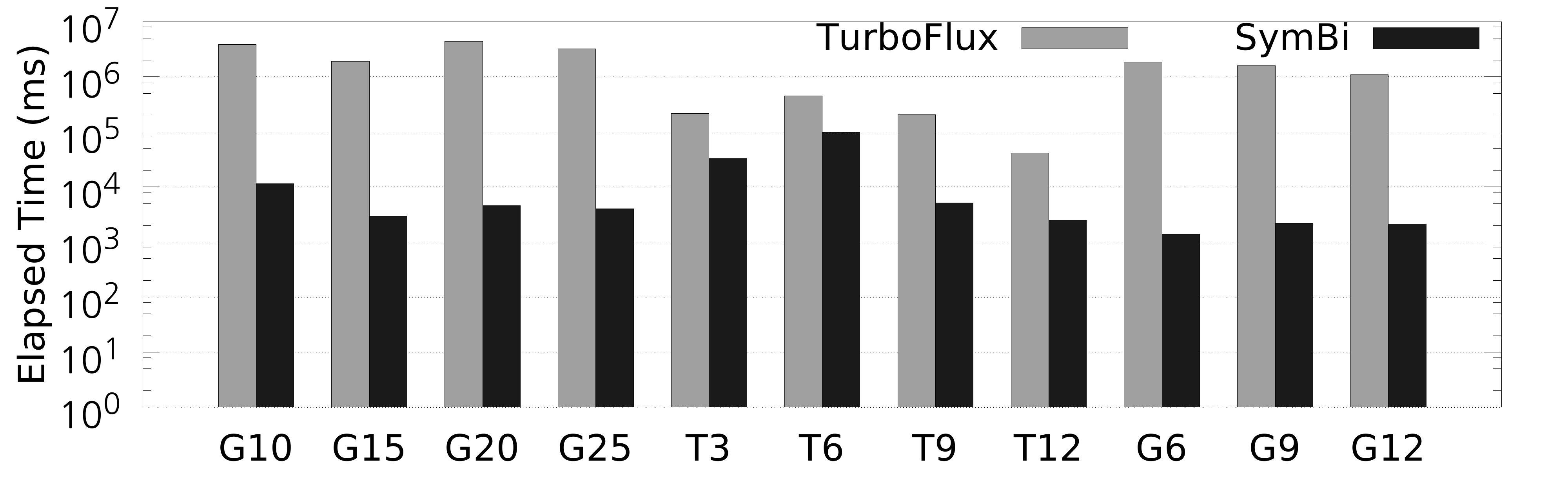}
    }
    \newline
    \vskip 1mm
    \subcaptionbox{Solved queries\label{exp:vary_query_size_netflow_solved}}{
        \includegraphics[width=\linewidth,scale=0.2]{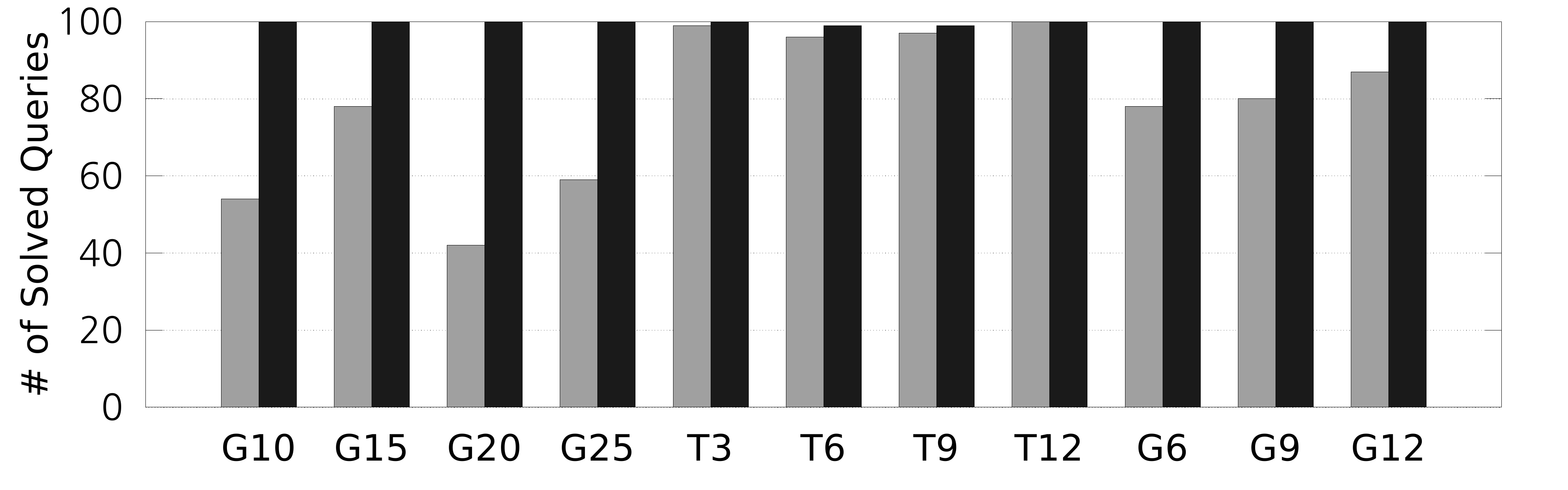}
    }
    \vspace*{-4mm}
    \caption{Varying query size for Netflow}
    \label{exp:vary_query_size_netflow}
\end{figure}

Figure \ref{exp:vary_query_size_netflow_time} shows the average elapsed time for performing continuous subgraph matching for \textsf{Netflow}. When calculating the average elapsed time, we exclude queries that no algorithms can solve within the time limit. \textsf{SymBi} outperforms \textsf{TurboFlux} regardless of query sizes. Specifically, \textsf{SymBi} is 333.13 $\sim$ 947.02 times faster than \textsf{TurboFlux} in our generated queries (G10 $\sim$ G25), 4.54 $\sim$ 16.49 times in tree queries (T3 $\sim$ T12), and 516.48 $\sim$ 1336.24 times in graph queries (G6 $\sim$ G12). The performance gap between \textsf{SymBi} and \textsf{TurboFlux} is larger for graph queries than tree queries. The reason for this is that \textsf{TurboFlux} does not take into account non-tree edges for filtering, whereas \textsf{SymBi} consider all edges for filtering. 

Figure \ref{exp:vary_query_size_netflow_solved} shows the number of queries solved by two algorithms. \textsf{SymBi} solves most queries for every query set (except 1 query in the T6 query set and 1 query in the T9 query set), while \textsf{TurboFlux} has query sets that contains many unsolved queries. Specifically, \textsf{TurboFlux} solves only 42 queries while \textsf{SymBi} solves all queries for G20.

\begin{figure}[h]
    \centering
    \subcaptionbox{Average elapsed time (in milliseconds)\label{exp:vary_query_size_lsbench_time}}{
        \includegraphics[width=\linewidth,scale=0.2]{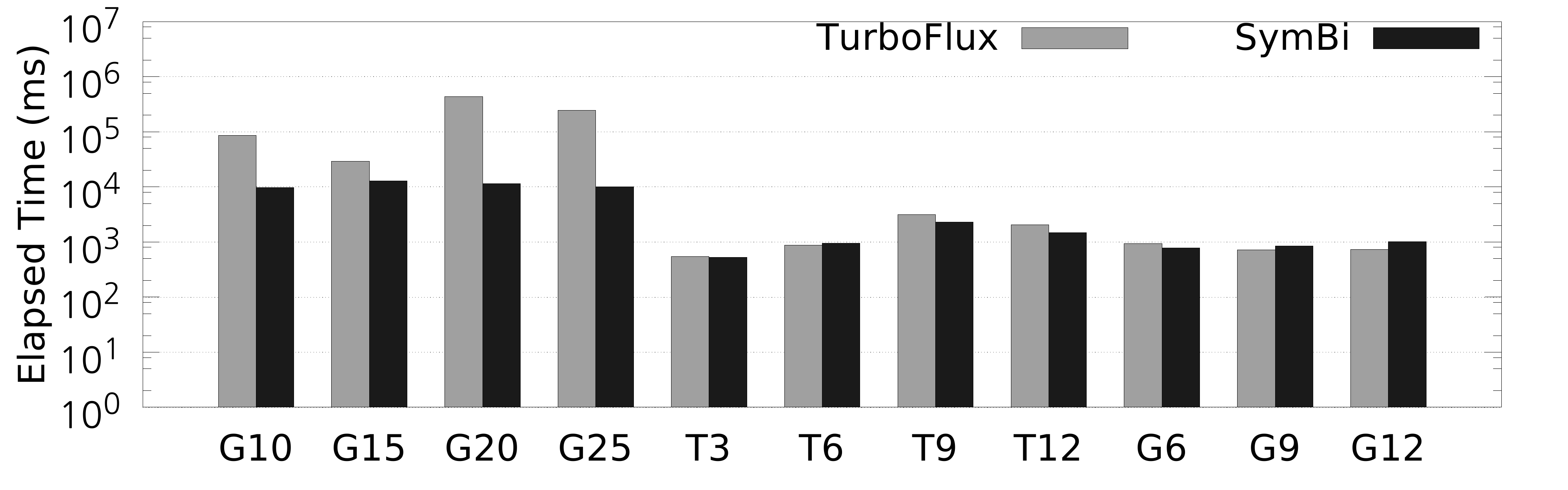}
    }
    \newline
    \vskip 1mm
    \subcaptionbox{Solved queries\label{exp:vary_query_size_lsbench_solved}}{
        \includegraphics[width=\linewidth,scale=0.2]{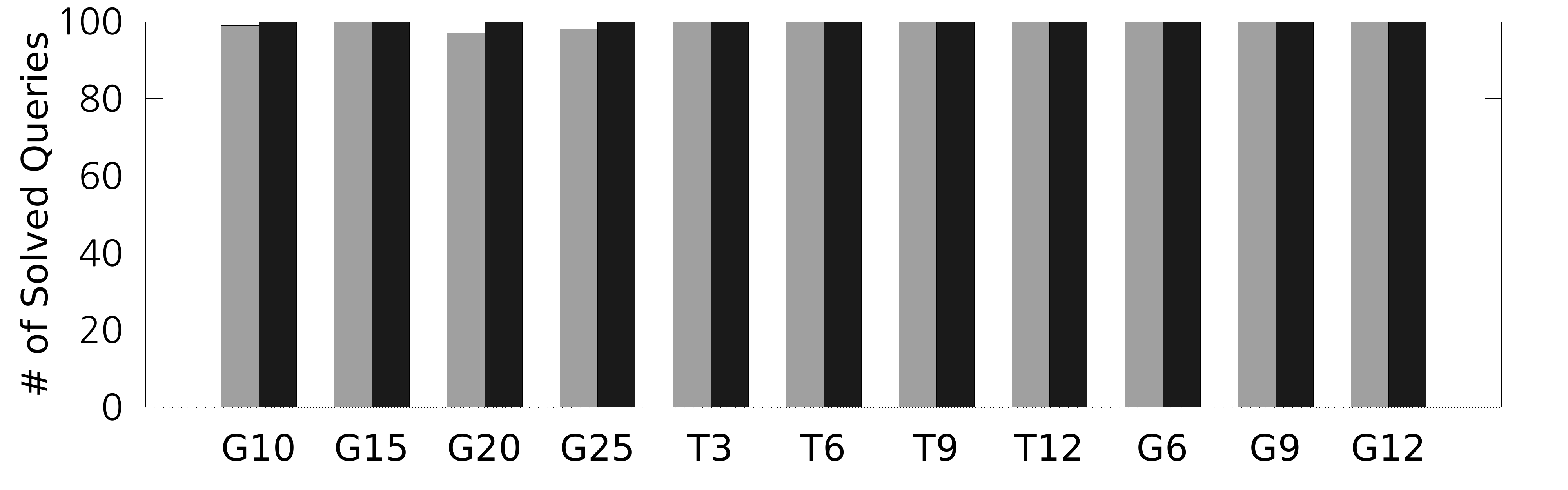}
    }
    \vspace*{-4mm}
    \caption{Varying query size for LSBench}
    \label{exp:vary_query_size_lsbench}
\end{figure}

Figure \ref{exp:vary_query_size_lsbench} shows the performance results for \textsf{LSBench}. In Figure \ref{exp:vary_query_size_lsbench_time}, \textsf{SymBi} outperforms \textsf{TurboFlux} by 2.26$\sim$38.35 times in our generated queries. The reason for the lesser performance gap over \textsf{Netflow} is that \textsf{LSBench} has 45 edge labels, and thus it is an easier dataset to solve than \textsf{Netflow} with 8 edge labels. Also, there is almost no difference in performance for the queries used in \cite{kim2018turboflux}.  Unlike our generated queries, most queries from \cite{kim2018turboflux} are solved in less than one second. For theses cases, \textsf{SymBi} takes most of the elapsed time to update the data graph or auxiliary data structures that takes polynomial time, which is difficult to improve. In Figure \ref{exp:vary_query_size_lsbench_solved}, \textsf{SymBi} solves all queries within the time limit, while \textsf{TurboFlux} reach the time limit for 1, 3 and 2 queries in G10, G20 and G25, respectively. 

\noindent\textbf{Varying the deletion rate.} Next, we vary the deletion rate of the graph update stream. We fix the query set to G10 and the insertion rate to 10\%, and vary the deletion rate from 0\% to 10\% in 2\% increments.

\begin{figure}[h]
    \centering
    \subcaptionbox{Average elapsed time (in milliseconds)\label{exp:vary_deletion_rate_netflow_time}}{
        \includegraphics[width=0.47\linewidth,scale=0.2]{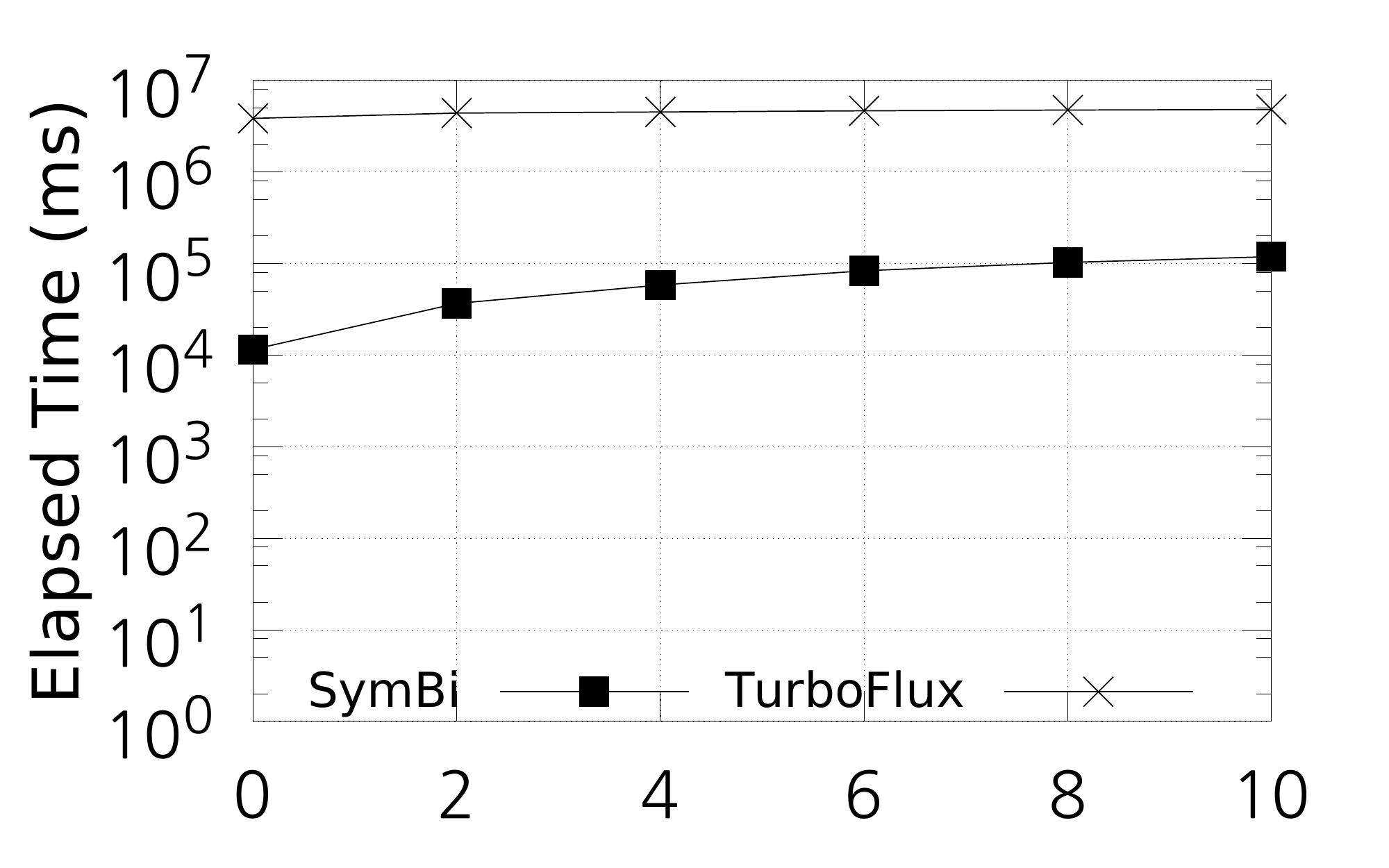}
    }
    \subcaptionbox{Solved queries\label{exp:vary_deletion_rate_netflow_solved}}{
        \includegraphics[width=0.47\linewidth,scale=0.2]{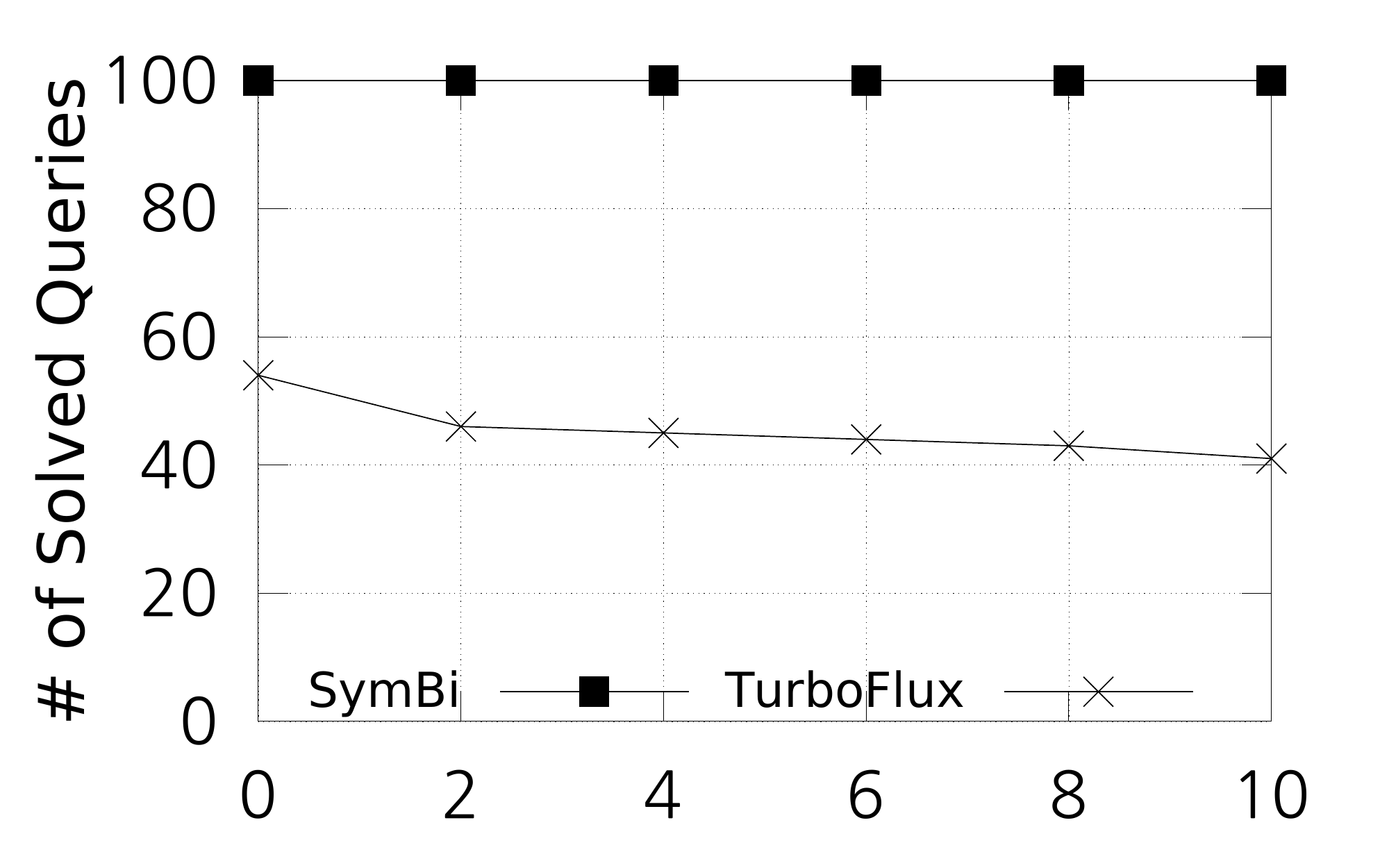}
    }
    \vspace*{-3mm}
    \caption{Varying deletion rate for Netflow}
    \label{exp:vary_deletion_rate_netflow}
    \vspace*{-4mm}
\end{figure}

\begin{figure}[h]
    \centering
    \subcaptionbox{Deletion rate 0\%\label{exp:sorting_deletion_rate0}}{
        \includegraphics[width=0.47\linewidth,scale=0.2]{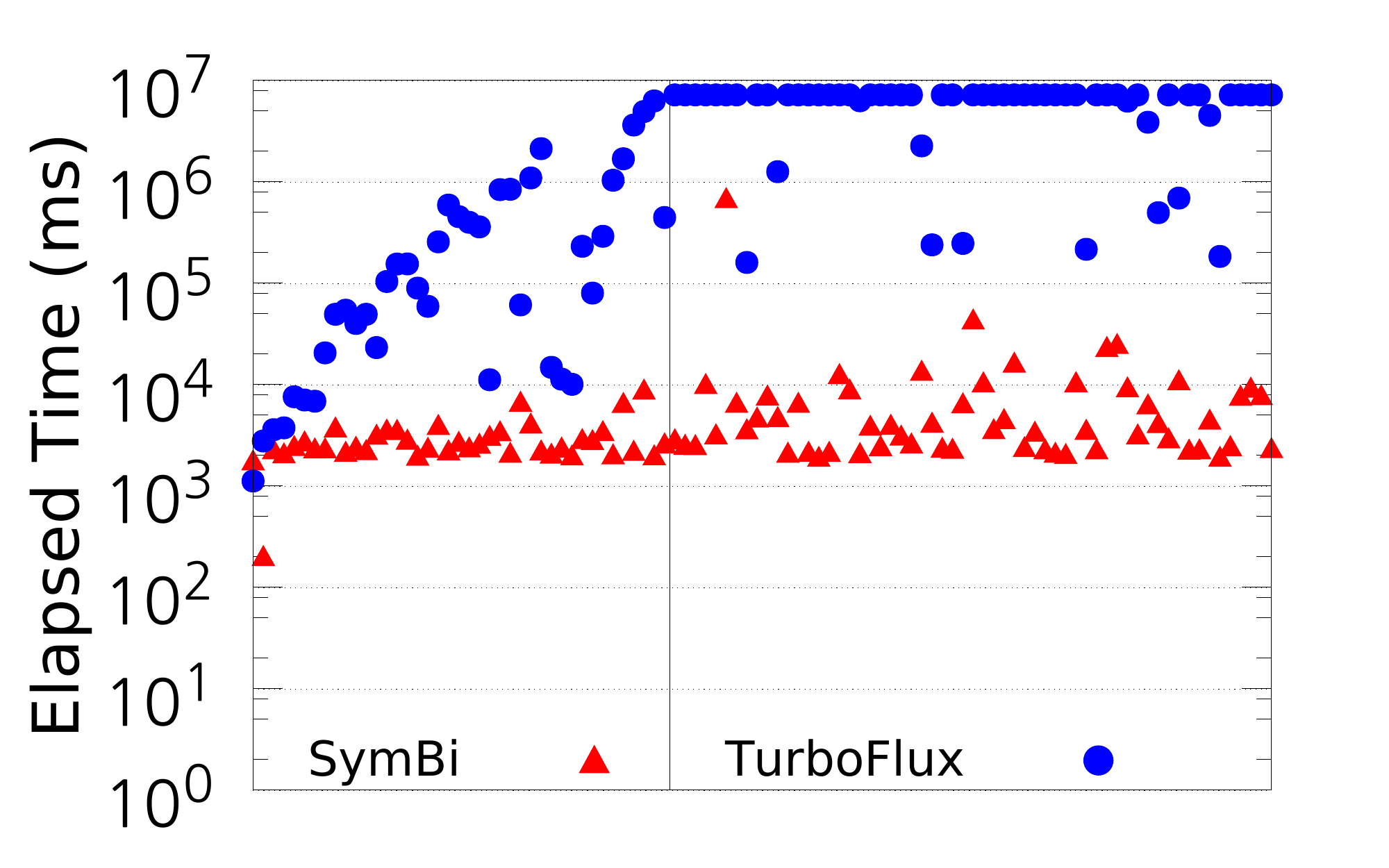}
    }
    \subcaptionbox{Deletion rate 10\%\label{exp:sorting_deletion_rate10}}{
        \includegraphics[width=0.47\linewidth,scale=0.2]{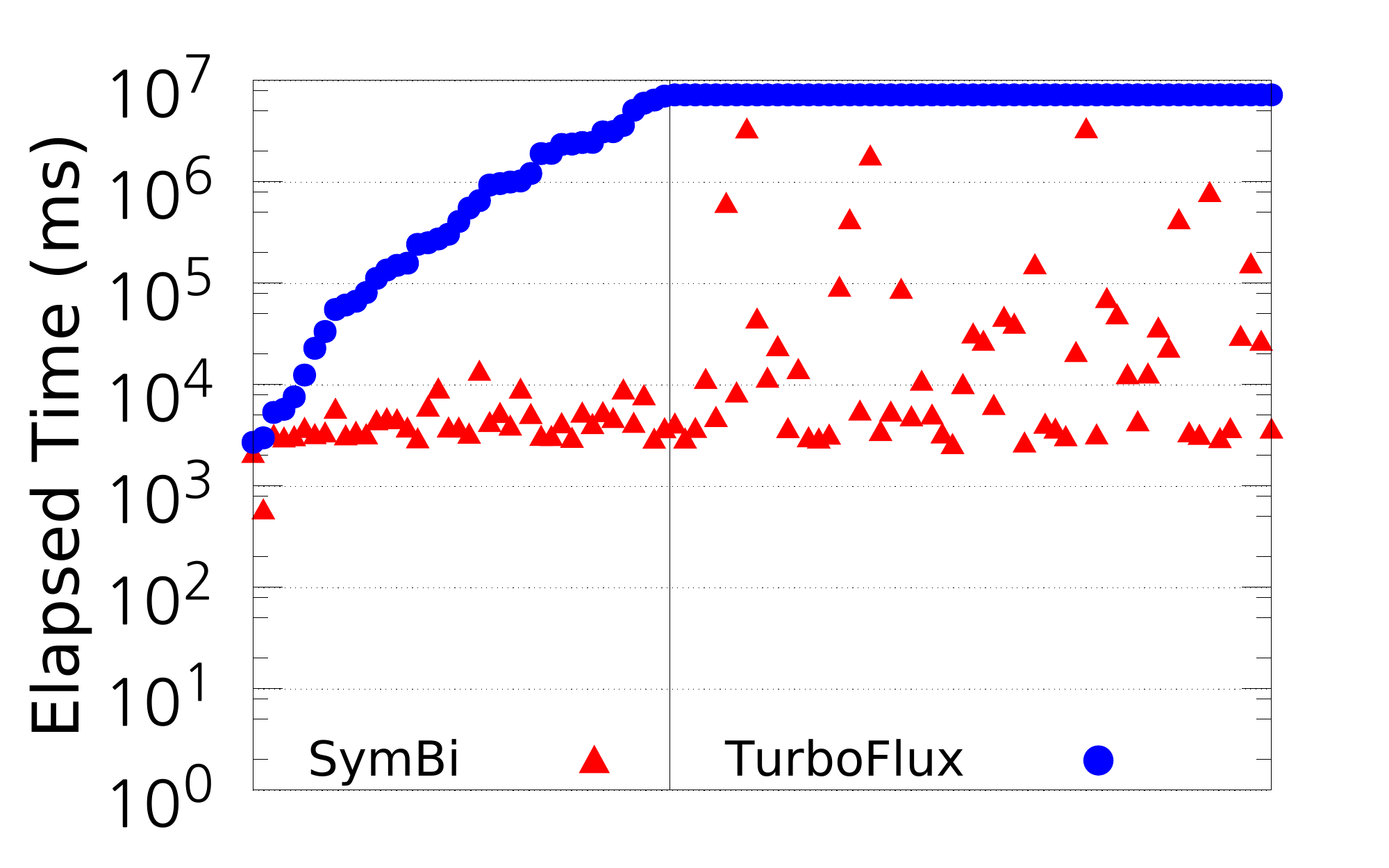}
    }
    \vspace*{-3mm}
    \caption{Elapsed time of all queries for each algorithm with deletion rate 0\% and 10\%}
    \label{exp:sorting_deletion_rate}
    \vspace*{-4mm}
\end{figure}

Figure \ref{exp:vary_deletion_rate_netflow} represents the performance results for \textsf{Netflow}. In Figure \ref{exp:vary_deletion_rate_netflow_solved}, \textsf{SymBi} solves all queries for all deletion rates, but as the deletion rate increases from 0\% to 10\%, the number of solved queries of \textsf{TurboFlux} decreases from 54 to 41. Figure \ref{exp:vary_deletion_rate_netflow_time} shows that the average elapsed time of \textsf{TurboFlux} is almost flat, but the average elapsed time of \textsf{SymBi} increases as the deletion rate increases (i.e., the performance improvement of \textsf{SymBi} over \textsf{TurboFlux} is 333.13 times for 0\% deletion rate and it decreases to 40.44 times as the deletion rate increases to 10\%). The decrease in the performance gap stems from queries that \textsf{TurboFlux} cannot solve, but \textsf{SymBi} solves within the time limit. Figure \ref{exp:sorting_deletion_rate} helps to understand this phenomenon. Figure \ref{exp:sorting_deletion_rate0} and \ref{exp:sorting_deletion_rate10} show the elapsed time of all queries for each algorithm with deletion rate 0\% and 10\%, respectively. Queries on the x-axis of Figure \ref{exp:sorting_deletion_rate0} and \ref{exp:sorting_deletion_rate10} are sorted in ascending order based on the elapsed time of \textsf{TurboFlux} when the deletion rate is 10\%. In Figure \ref{exp:sorting_deletion_rate0} and \ref{exp:sorting_deletion_rate10}, there are many queries for which \textsf{TurboFlux} reaches the time limit. As the deletion rate increases from 0\% (Figure \ref{exp:sorting_deletion_rate0}) to 10\% (Figure \ref{exp:sorting_deletion_rate10}), the elapsed time of \textsf{TurboFlux} for these queries does not increase further beyond the time limit (2 hours), while the elapsed time of \textsf{SymBi} increases. This reduces the performance gap between two algorithms. 

Considering this issue, we focus on 41 queries that \textsf{TurboFlux} solves within the time limit in all deletion rates (queries on the left side of the vertical line in Figure \ref{exp:sorting_deletion_rate}). When we measure the average elapsed time with these 41 queries, the performance ratio between two algorithms increases from 224.61 times to 309.45 times as the deletion rate increases from 0\% to 10\%. While the deletion rate changes from 0\% to 10\%, the average elapsed time of \textsf{SymBi} increases only 1.54 times, but the average elapsed time of \textsf{TurboFlux} increases 2.13 times. When two algorithms are compared, therefore, the number of solved queries as well as the average elapsed time are important measures.

% Especially in some queries, as shown in Figure \ref{exp:sorting_deletion_rate}, the elapsed time of \textsf{SymBi} increases by less than twice, but \textsf{TurboFlux} consumes more than 100 times as the deletion rate increases.

\begin{figure}[h]
    \centering
    \subcaptionbox{Average elapsed time (in milliseconds)\label{exp:vary_deletion_rate_lsbench_time}}{
        \includegraphics[width=0.47\linewidth,scale=0.2]{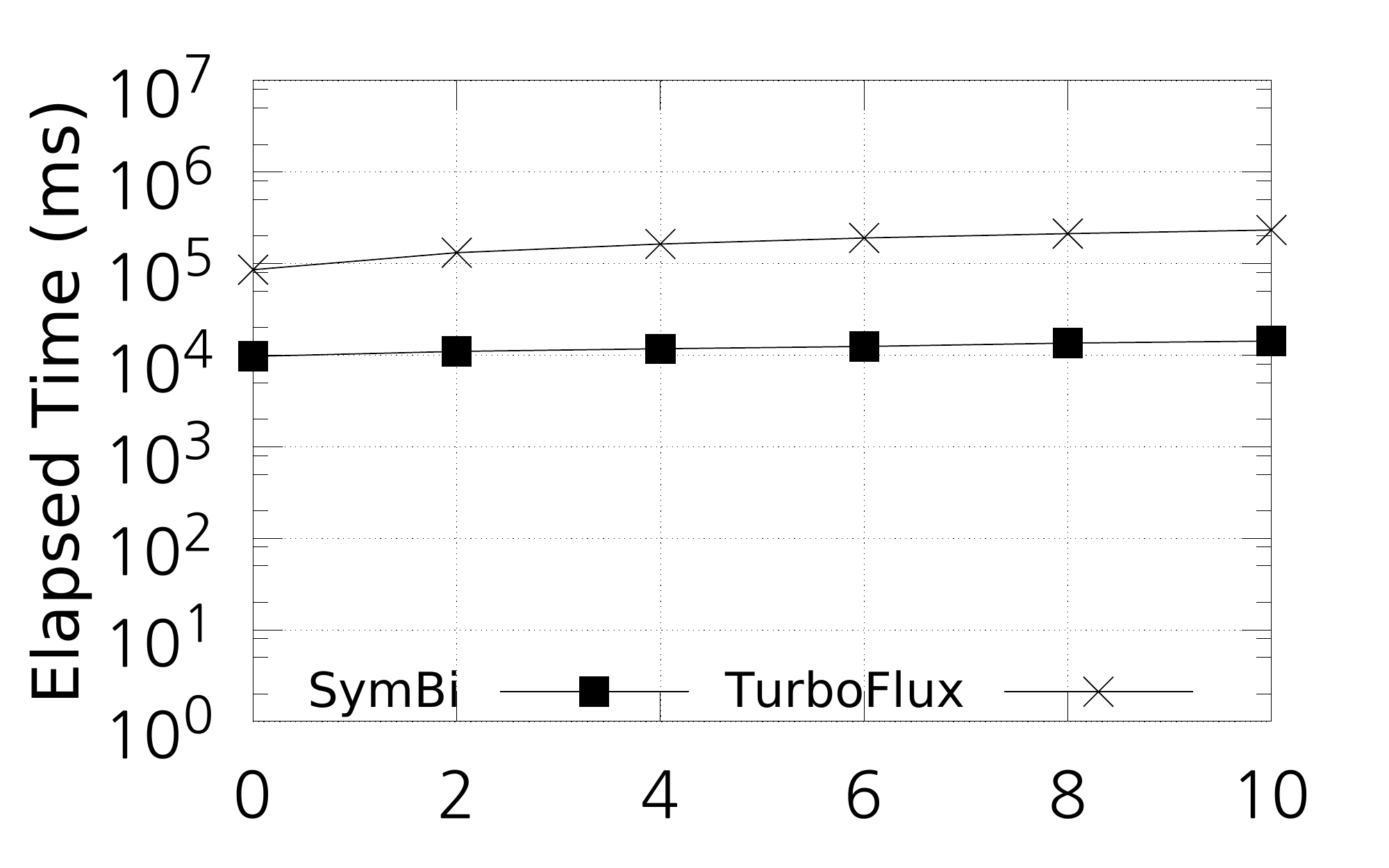}
    }
    \subcaptionbox{Solved queries\label{exp:vary_deletion_rate_lsbench_solved}}{
        \includegraphics[width=0.47\linewidth,scale=0.2]{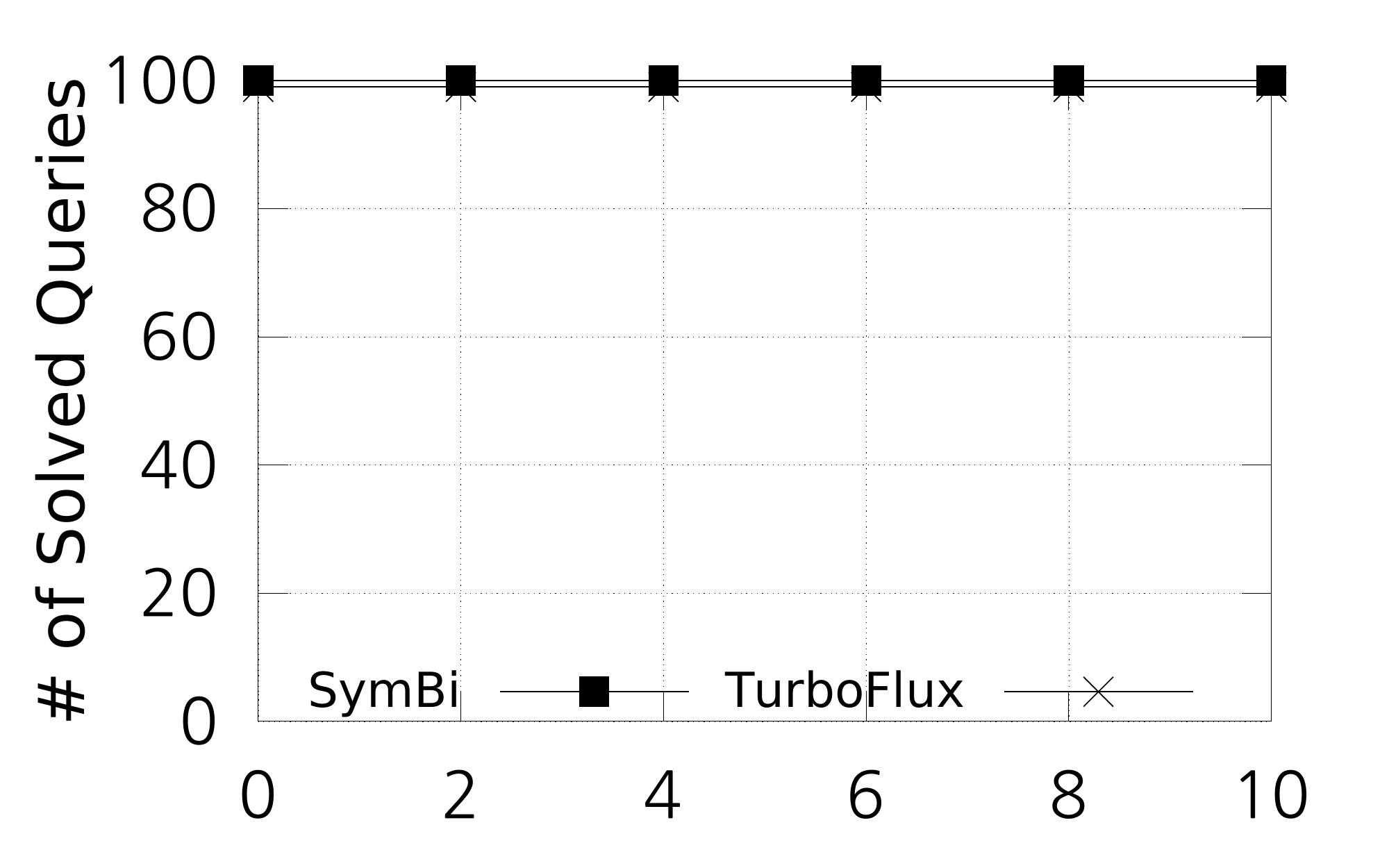}
    }
    \vspace*{-3mm}
    \caption{Varying deletion rate for LSBench}
    \label{exp:vary_deletion_rate_lsbench}
\end{figure}

%\vspace{-2mm}

The performance results for \textsf{LSBench} are shown in Figure \ref{exp:vary_deletion_rate_lsbench}. Figure \ref{exp:vary_deletion_rate_lsbench_solved} shows that \textsf{SymBi} solves all queries while \textsf{TurboFlux} solves 99 queries for all deletion rates. Figure \ref{exp:vary_deletion_rate_lsbench_time} shows that as the deletion rate increases, the performance ratio between two algorithms increases (8.84 to 16.30 times). Similarly to the previous one, considering only the 99 queries that both algorithms solve, \textsf{SymBi} only increases 1.52 times when the deletion rate is 10\% compared to 0\%, but \textsf{TurboFlux} increases 11.70 times.  This shows that \textsf{SymBi} handles the edge deletion case better than \textsf{TurboFlux}.

In order to further analyze why \textsf{SymBi} processes queries better than \textsf{TurboFlux} as the deletion rate increases, we divide the elapsed time when the deletion rate is 10\% into four types: update/backtracking time for edge insertion, and update/backtracking time for edge deletion. Since the number of insertion operations and that of deletion operations are different, we measure the elapsed time per operation by dividing the elapsed time by the number of operations. Table \ref{tab:update_backtracking} shows the results for \textsf{Netflow} and \textsf{LSBench}. It is noteworthy that 
the update time of \textsf{TurboFlux} for edge deletion is much slower than that for edge insertion, while those of \textsf{SymBi} are quite similar.
As noted in Section \ref{sec:introduction}, this happens because the \textsf{DCG} update process of \textsf{TurboFlux} is more complex for edge deletion than for edge insertion.

% \vspace{-2mm}
\begin{table}[h]
    \centering
    \caption{Average update and backtracking time per operation in microseconds (top: \textsf{Netflow}, bottom: \textsf{LSBench})}
    \vspace{-2mm}
    \begin{tabular}{ccccc}
    \toprule
    & \multicolumn{2}{c}{\textbf{TurboFlux}} & \multicolumn{2}{c}{\textbf{SymBi}} \\
    & \textbf{Update}       & \textbf{Backtracking}      & \textbf{Update}     & \textbf{Backtracking}    \\
    \midrule
    \textbf{Ins} & 6.44 & 202.48 & 0.93 & 0.41 \\
    \textbf{Del} & 1867.39 & 2086.18 & 1.68 & 4.20 \\
    \midrule
    \textbf{Ins} & 1.13 & 4.32 & 0.47 & 3.44 \\
    \textbf{Del} & 599.82 & 21.72 & 0.68 & 17.32 \\
    \bottomrule
    \end{tabular}
    
    \label{tab:update_backtracking}
\end{table}

\vspace*{1mm}
\noindent\textbf{Varying the insertion rate.} To test the effect of the insertion rate, we use the G10 query set and vary the insertion rate from 2\% to 10\% in 2\% increments. Note that the size of the initial graph does not change from 90\% of the original dataset.

\begin{figure}[h!]
    \centering
    \subcaptionbox{Average elapsed time (in milliseconds)\label{exp:vary_insertion_rate_netflow_time}}{
        \includegraphics[width=0.47\linewidth,scale=0.2]{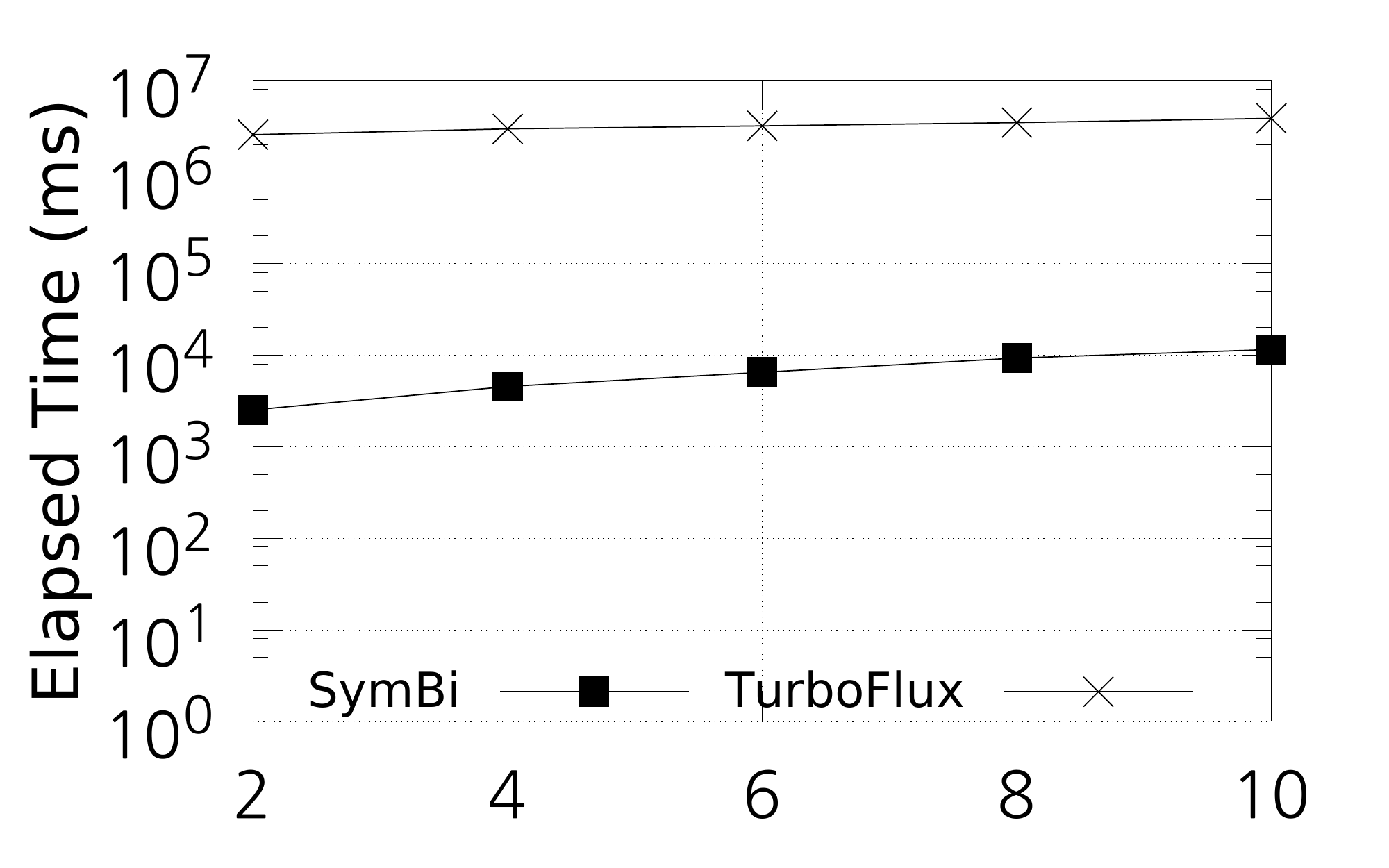}
    }
    \subcaptionbox{Solved queries\label{exp:vary_insertion_rate_netflow_solved}}{
        \includegraphics[width=0.47\linewidth,scale=0.2]{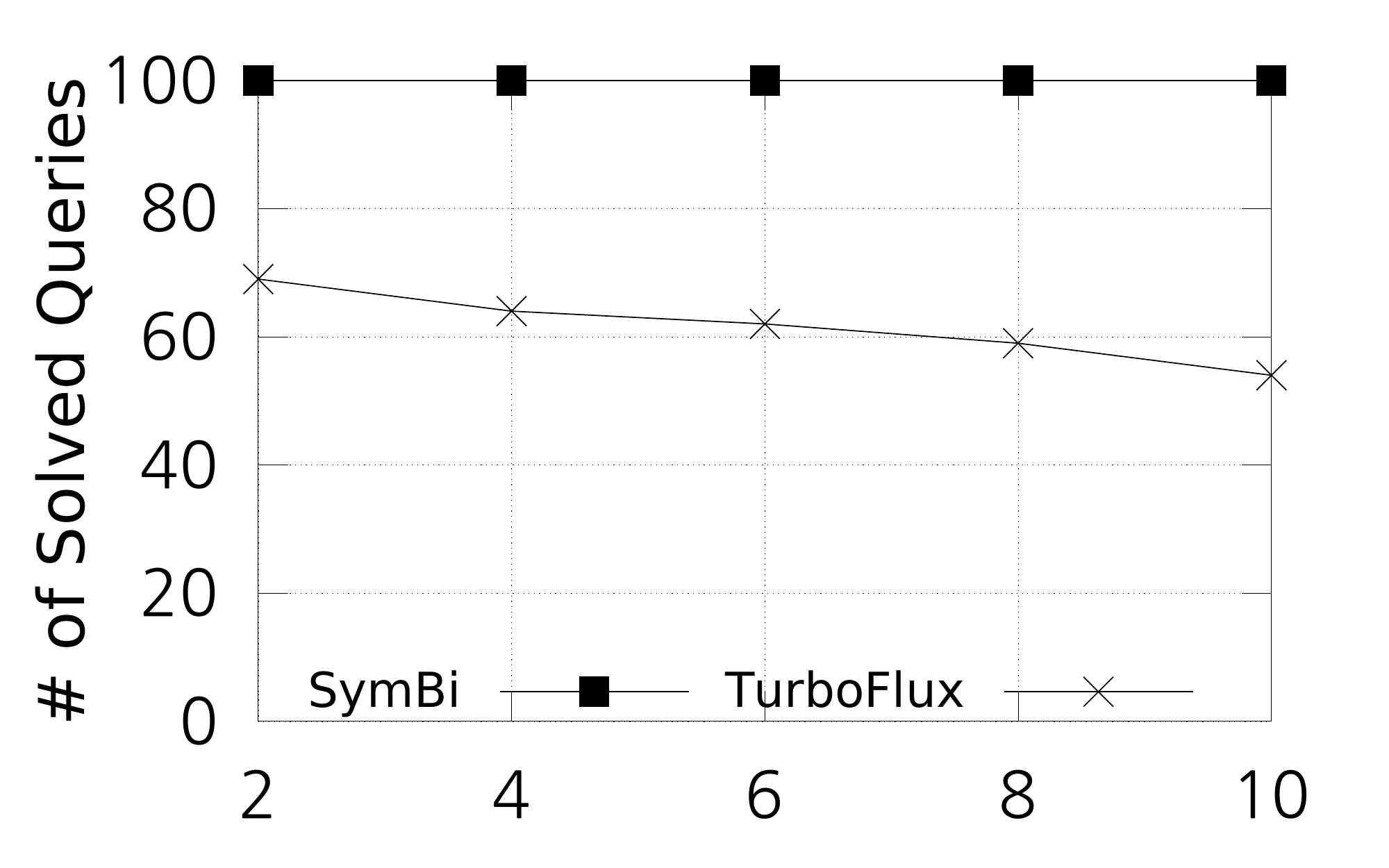}
    }
    \vspace*{-3mm}
    \caption{Varying insertion rate for Netflow}
    \label{exp:vary_insertion_rate_netflow}
\end{figure}

% Figure \ref{exp:vary_insertion_rate_netflow} represents the performance results for \textsf{Netflow}. 
Figure \ref{exp:vary_insertion_rate_netflow} represents the results using \textsf{Netflow} for varying insertion rates. Figure \ref{exp:vary_insertion_rate_netflow_solved} shows that \textsf{SymBi} solves all queries for all insertion rates, while the number of solved queries of \textsf{TurboFlux} decreases from 69 to 54 as the insertion rate increases. In Figure \ref{exp:vary_insertion_rate_netflow_time}, \textsf{SymBi} outperforms \textsf{TurboFlux} regardless of the insertion rate. However, as before, the performance gap between two algorithms decreases as the insertion rate increases due to the queries that \textsf{TurboFlux} cannot solve. 
% As in Figure \ref{exp:sorting_deletion_rate}, Figure \ref{exp:sorting_insertion_rate} represents the elapsed time of all queries for each algorithm in ascending order based on the elapsed time of \textsf{TurboFlux} when the insertion rate is 10\%. 
As in Figure \ref{exp:sorting_deletion_rate}, when we measure the average elapsed time with 54 queries that both algorithms solve within the time limit in all insertion rates, the performance ratio increases from 95.32 times to 276.60 times as the insertion rate increases from 2\% to 10\%.

\vspace{-2mm}

% \begin{figure}[h]
%     \centering
%     \subcaptionbox{Insertion rate 2\%\label{exp:sorting_insertion_rate0}}{
%         \includegraphics[width=0.47\linewidth,scale=0.2]{figures/sorting_insertion_rate0.pdf}
%     }
%     \subcaptionbox{Insertion rate 10\%\label{exp:sorting_insertion_rate10}}{
%         \includegraphics[width=0.47\linewidth,scale=0.2]{figures/sorting_insertion_rate10.pdf}
%     }
%     \vspace{-1mm}
%     \caption{Elapsed time of all queries for each algorithm with insertion rate 2\% and 10\%}
%     \label{exp:sorting_insertion_rate}
% \end{figure}

\begin{figure}[htp]
    \centering
    \subcaptionbox{Average elapsed time (in milliseconds)\label{exp:vary_insertion_rate_lsbench_time}}{
        \includegraphics[width=0.47\linewidth,scale=0.2]{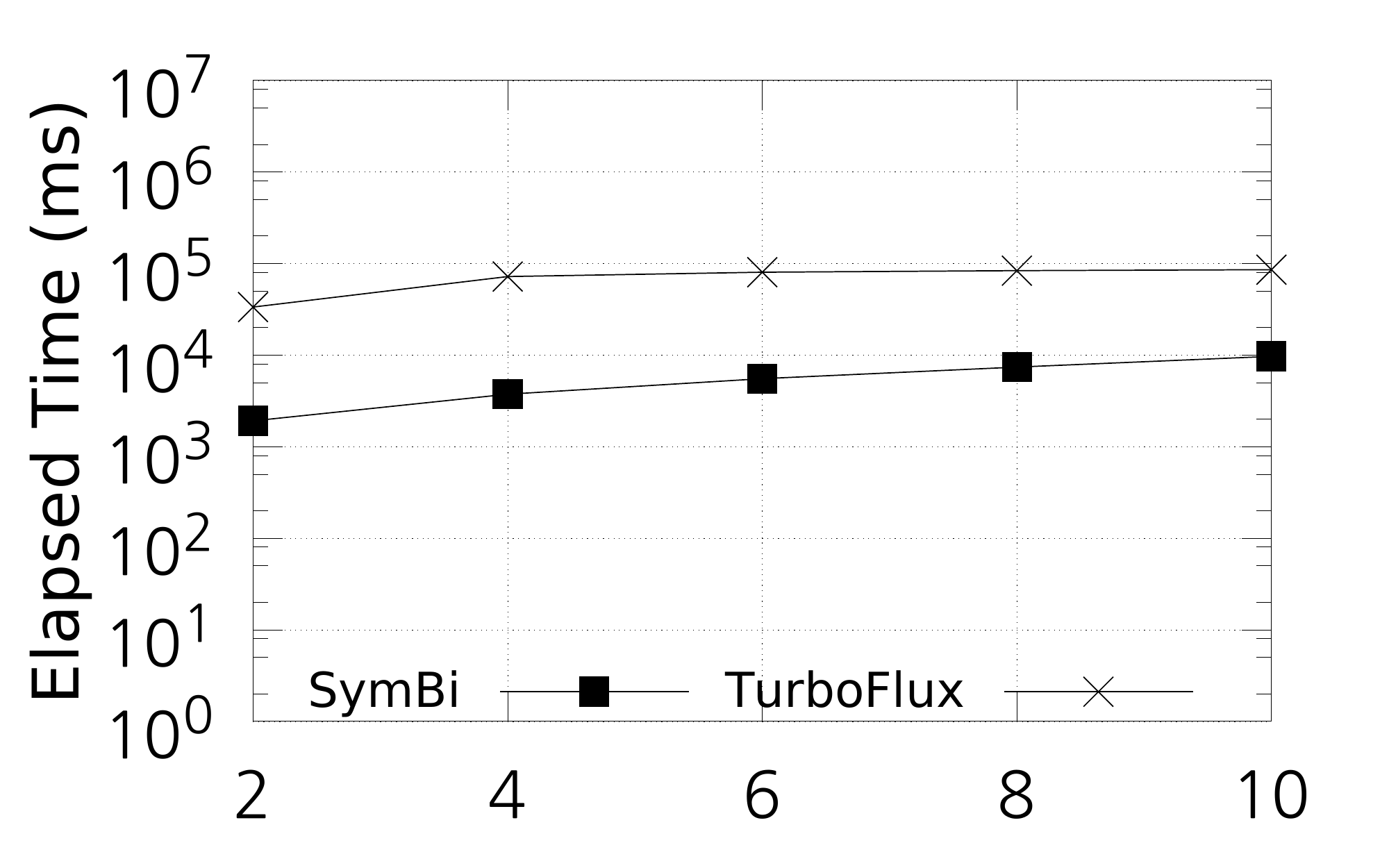}
    }
    \subcaptionbox{Solved queries\label{exp:vary_insertion_rate_lsbench_solved}}{
        \includegraphics[width=0.47\linewidth,scale=0.2]{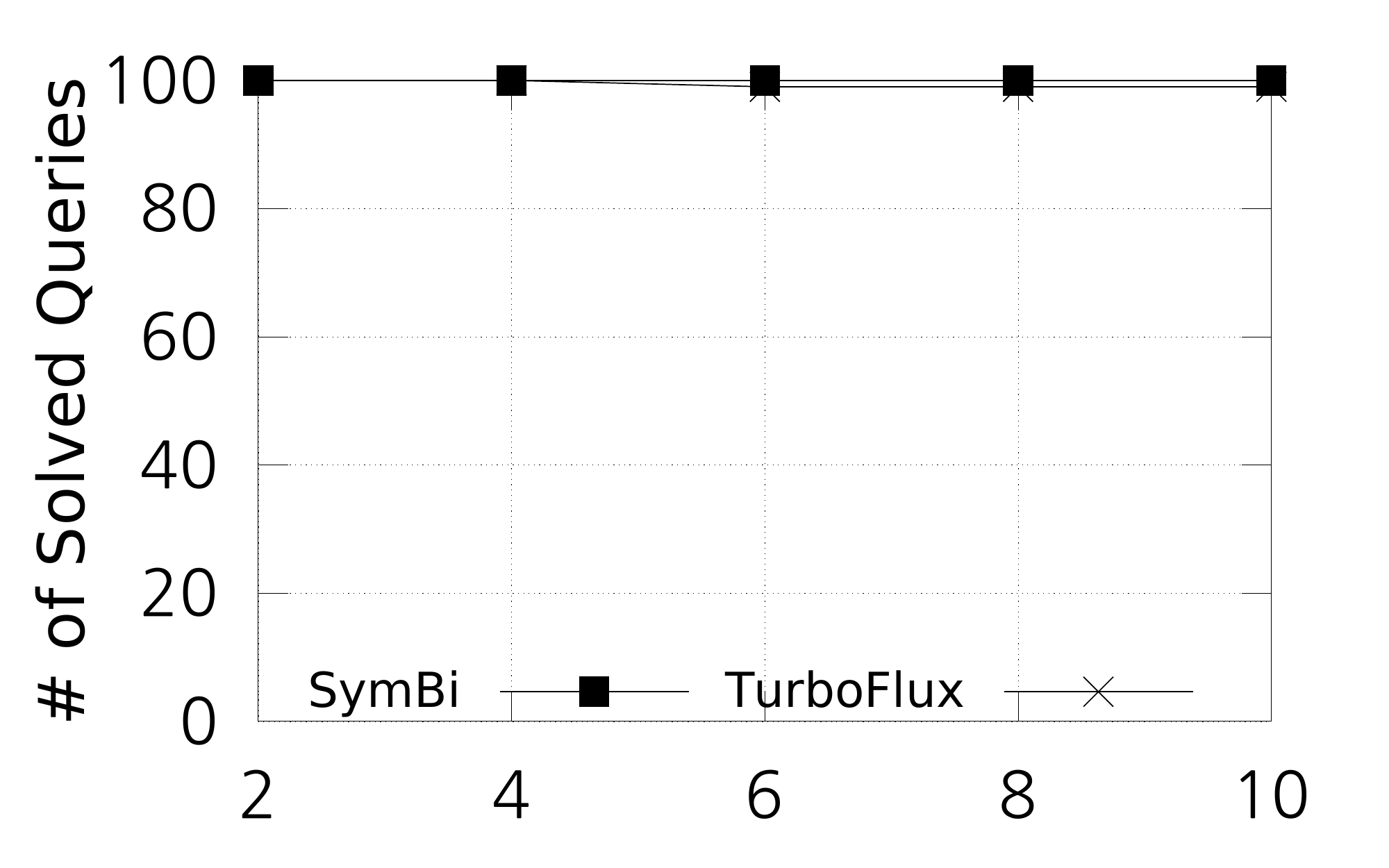}
    }
    \vspace*{-3mm}
    \caption{Varying insertion rate for LSBench}
    \label{exp:vary_insertion_rate_lsbench}
\end{figure}

Figure \ref{exp:vary_insertion_rate_lsbench} shows the results for \textsf{LSBench}. The performance ratio between two algorithms is the largest at 19.30 times when the insertion rate is 4\%. As one query reaches the time limit for \textsf{TurboFlux} at 6\% insertion rate, the performance gap starts to decrease from 6\% insertion rate. Nevertheless, \textsf{SymBi} is 8.84 times faster than \textsf{TurboFlux} at 10\% insertion rate.

\vspace{-2mm}

\begin{figure}[htp]
    \centering
    \subcaptionbox{Average elapsed time (in milliseconds)\label{exp:vary_dataset_size_lsbench_time}}{
        \includegraphics[width=0.47\linewidth,scale=0.2]{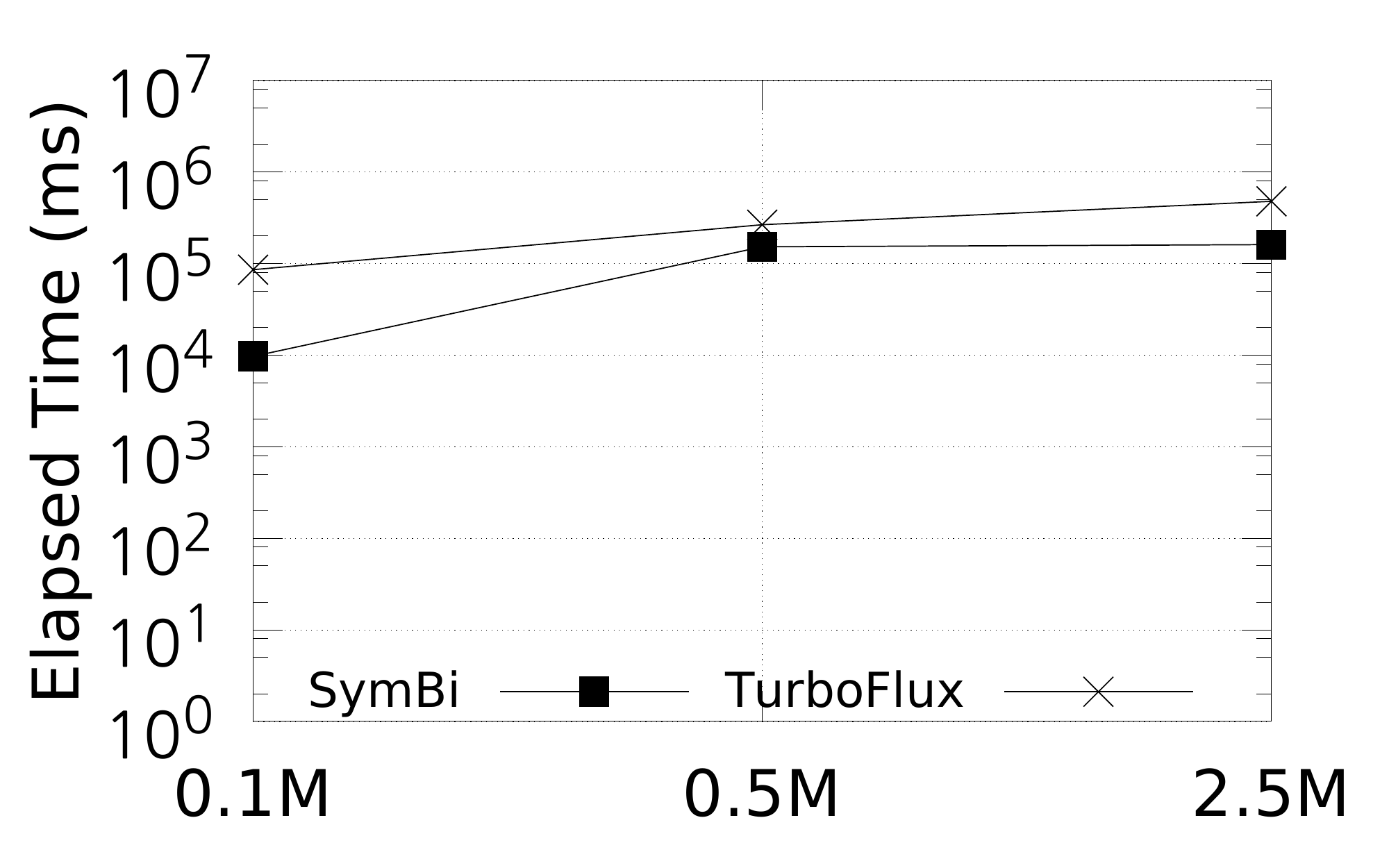}
    }
    \subcaptionbox{Solved queries\label{exp:vary_dataset_size_lsbench_solved}}{
        \includegraphics[width=0.47\linewidth,scale=0.2]{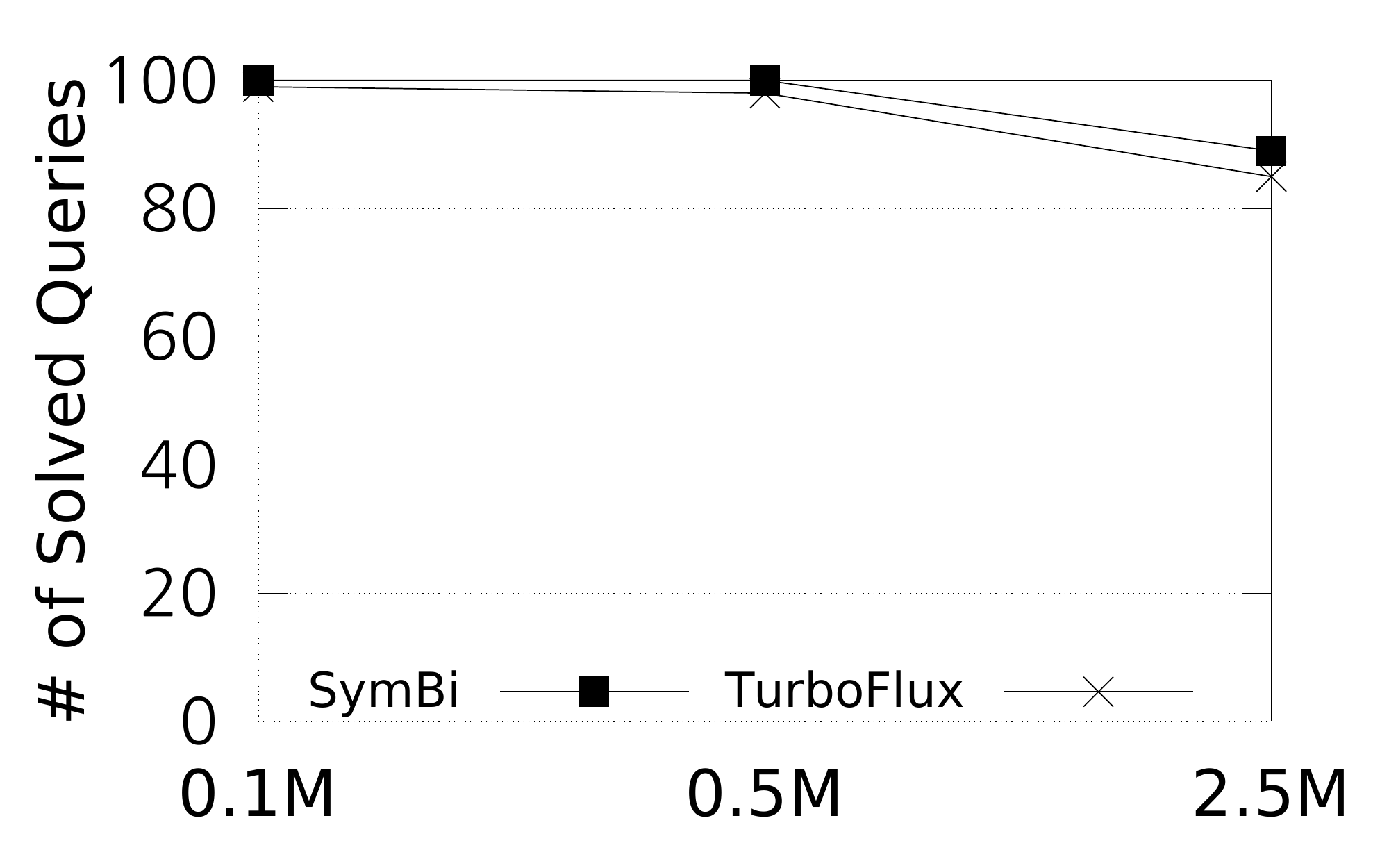}
    }
    \vspace*{-3mm}
    \caption{Varying dataset size}
    \label{exp:vary_dataset_size_lsbench}
\end{figure}

\noindent\textbf{Varying the dataset size.} We measure the performance for different \textsf{LSBench} dataset sizes: 0.1, 0.5, and 2.5 million users. The size of the initial data graph increases from 20,988,361 triples (0.1M users) to 525,446,784 triples (2.5M users). As shown in the experiment of varying the insertion rate, the number of triples in the graph update stream affects the elapsed time. To test only the effect of the dataset size, we set the same number of triples in the three graph update streams. We fix the number of triples in the graph update streams as 10\% of the triples of the first dataset (0.1M users). In Figure \ref{exp:vary_dataset_size_lsbench}, as the dataset size increases, the elapsed time of both algorithm generally increases and the number of solved queries decreases. \textsf{SymBi} is consistently faster and solves more queries than \textsf{TurboFlux} regardless of the dataset sizes.

\vspace{-2mm}

\begin{figure}[htp]
    \centering
    \includegraphics[scale=0.21]{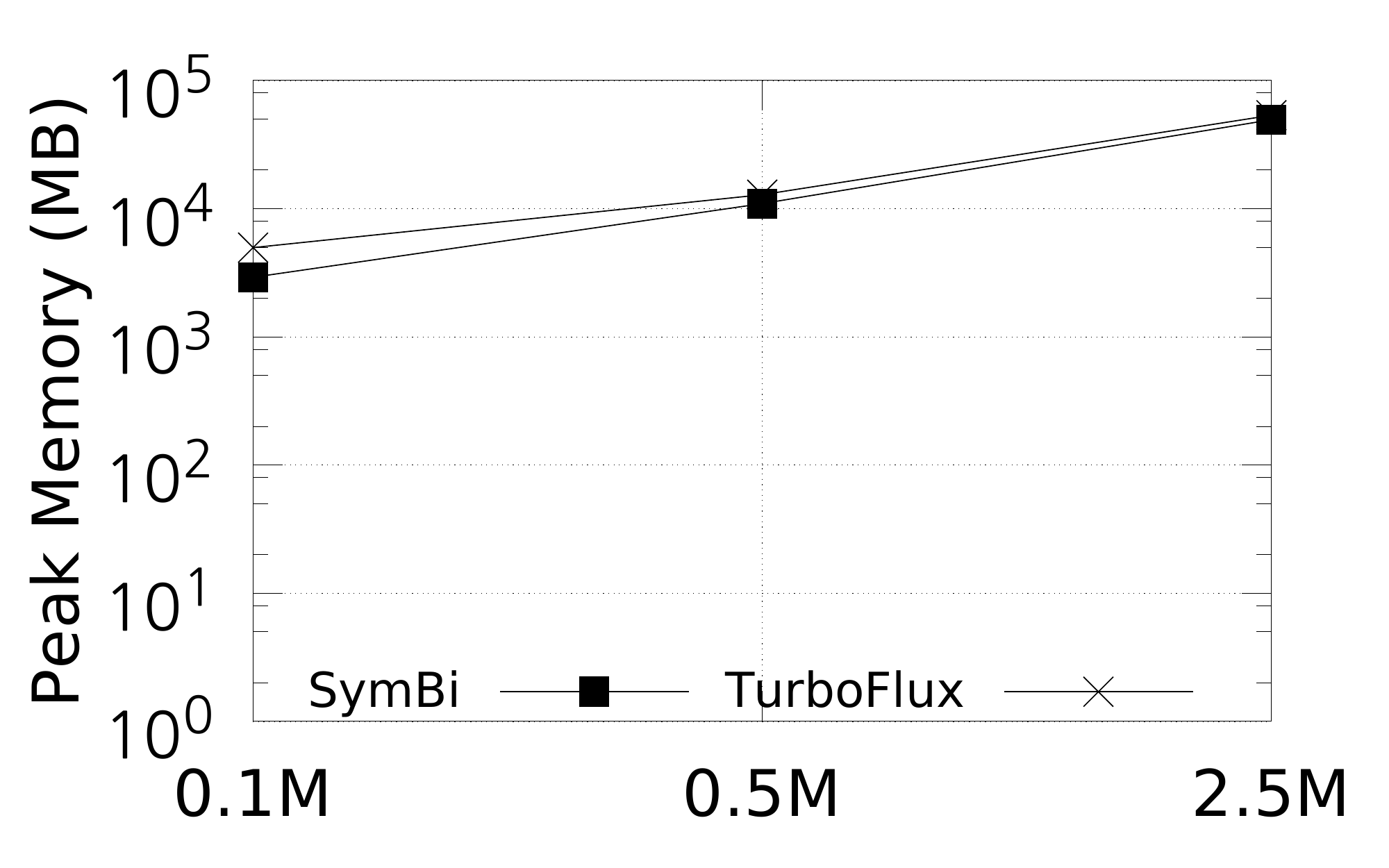}
    \vspace*{-2mm}
    \caption{Average peak memory (in MB)} 
    \label{exp:vary_dataset_size_lsbench_memory}
\end{figure}

\noindent\textbf{Memory usage.} Figure \ref{exp:vary_dataset_size_lsbench_memory} demonstrates the average peak memory of each program for varying the dataset size (the results for the other experiments are similar). Here, peak memory is defined as the maximum of the virtual set size (VSZ) in the ``ps'' utility output. This shows that \textsf{SymBi} uses a slightly less memory than \textsf{TurboFlux} regardless of the dataset sizes.

\vspace{-2mm}

\section{Conclusion}\label{sec:conclusion}
In this paper, we have studied continuous subgraph matching, and proposed an auxiliary data structure called dynamic candidate space (\textsf{DCS}) which stores the intermediate results of bidirectional dynamic programming between a query graph and a dynamic data graph. We further proposed an efficient algorithm to update \textsf{DCS} for each graph update operation. We then presented a matching algorithm that uses \textsf{DCS} to find all positive/negative matches. Extensive experiments on real and synthetic datasets show that \textsf{SymBi} outperforms the state-of-the-art algorithm by up to several orders of magnitude. Parallelizing our algorithm including both intra-query parallelism and inter-query parallelism is an interesting future work.

\section{Acknowledgments}
\noindent S. Min, S. G. Park, and K. Park were supported by Institute of Information communications Technology Planning Evaluation (IITP) grant funded by the Korea government (MSIT) (No. 2018-0-00551, Framework of Practical Algorithms for NP-hard Graph Problems). W.-S. Han was supported by Institute of Information communications Technology Planning Evaluation(IITP) grant funded by the Korea government(MSIT) (No. 2018-0-01398, Development of a Conversational, Self-tuning DBMS). G. F. Italiano was partially supported by MIUR, the Italian Ministry for Education, University and
Research, under PRIN Project AHeAD (Efficient Algorithms for HArnessing Networked Data).

\clearpage

\balance
\bibliographystyle{ACM-Reference-Format}
\bibliography{references}

\appendix
\section{Appendix}

\subsection{Proofs of Lemmas}\label{subsec:proofs_of_lemmas}
\noindent\textbf{Proof of Lemma \ref{lma:construction_correct}.}
We prove the lemma for $D_1$ by induction in a top-down fashion. For each $\langle u, v \rangle$ in \textsf{DCS}, we divide the cases whether $u$ is the root of DAG $\hat{q}$ or not.

When $u$ is the root of DAG $\hat{q}$ (i.e., base cases), the lemma holds trivially since $v \in C(u)$ and thus $u$ and $v$ both have the same label.
    
When $u$ is not the root of DAG $\hat{q}$ (i.e., inductive case), let's assume that $D_1[u_p, v_p]$ is correctly computed for every parent $u_p$ of $u$ and $v_p \in C(u_p)$. Now we show that Recurrence (\ref{eqn}) correctly computes $D_1[u, v]$, i.e., there exists a weak embedding of $\hat{q}_u^{-1}$ at $v$ if and only if $\exists v_p \in C(u_p)$ adjacent to $v$ such that $D_1[u_p,v_p]=1$ for every parent $u_p$ of $u$ in $\hat{q}$.
    
First we show the `only if' part. Let's assume that there exists a weak embedding $M'$ of $\hat{q}^{-1}_u$ at $v$. For each parent $u_p$ of $u$ in $\hat{q}$ (which is also a child of $u$ in $\hat{q}^{-1}$), we can get a weak embedding of $\hat{q}^{-1}_{u_p}$ at $M'(u_p)$ by removing the nodes not in $\hat{q}^{-1}_{u_p}$ from $M'$. Therefore, $D_1[u_p, M'(u_p)] = 1$ holds by inductive hypothesis. Furthermore, $M'(u_p) \in C(u_p)$ and $M'(u_p)$ is adjacent to $M'(u) = v$ because $M'$ is a weak embedding of $\hat{q}^{-1}_u$ and $u_p$ is adjacent to $u$. Therefore we proved the statement.
    
Now we show the converse. If there exist $\exists v_p \in C(u_p)$ adjacent to $v$ such that $D_1[u_p,v_p]=1$ for every parent $u_p$ of $u$ in $\hat{q}$, there is a weak embedding $M'_{u_p}$ of $\hat{q}^{-1}_{u_p}$ at $v_p$ for each $u_p$ by inductive hypothesis. Now we can build a weak embedding $M'$ of $\hat{q}^{-1}_u$ at $v$, by building a tree which has $v$ as a root, and $M'_{u_p}$'s as subtrees under $v$. Therefore we proved the statement.

We can similarly prove that $D_2$ is also correctly computed by Recurrence (\ref{eqn2}).

\noindent\textbf{Proof of Lemma \ref{lma:construction_complexity}.}
We need $O(|V(q)| \times |V(g)|)$ space to store $\langle u,v\rangle$'s in the \textsf{DCS} structure, $O(|V(q)| \times |V(g)|)$ space to store $D_1$ and $D_2$, and $O(|E(q)| \times |E(g)|)$ space to store edges. Hence we need $O(|E(q)| \times |E(g)|)$ space for the \textsf{DCS} structure in total.

We can build the vertices and edges in the \textsf{DCS} structure in $O(|E(q)|\times |E(g)|)$ time by traversing through the vertices and edges in $q$ and $g$. Now we consider $D_1$ and $D_2$. In order to compute $D_1[u,v]$, we have to traverse through all parents $\langle u_p, v_p \rangle$ of $\langle u, v \rangle$. Since we traverse through all edges once, we need $O(|E(q)|\times |E(g)|)$ time to compute $D_1$. Similarly, we can compute $D_2$ with the same time complexity.

\noindent\textbf{Proof of Lemma \ref{lma:update_correct}.}
We first consider $D_1$ here, since we can deal with the case of $D_2$ similarly to $D_1$.

In order to prove that $D_1[u, v]$ is correctly updated after the insertion for every $\langle u,v\rangle$, we need only prove that $N^1_{u,v}[u_p]$'s are correctly updated for all parents $u_p$ of $u$. It is because every time $N^1_{u,v}[u_p]$ is updated in Line 9 in Algorithm \ref{alg:insertion_topdown}, we update $N^1_P[u, v]$ and $D_1[u, v]$ following their definitions in Lines 1-8 in Algorithm \ref{alg:insertion_topdown}.

Now we prove that $N^1_{u,v}[u_p]$'s are correctly updated by induction on $u$, in a topological order on $\hat{q}$.

If $u$ is the root of $\hat{q}$ (i.e., base case), the statement is trivial since since there are no parents $u_p$ of the root $u$.

If $u$ is not the root of $\hat{q}$ (i.e., inductive cases), let's assume that $D_1[u_p, v_p]$'s are correctly updated for every parents of $u_p$ and $v_p \in C(u_p)$. Now we show that $N^1_{u,v}[u_p]$'s are correctly updated. If $N^1_{u, v}[u_p]$ has to be updated from $0$ to $1$, it means that a new edge $(\langle u, v \rangle, \langle u_p, v_p \rangle)$ is inserted to \textsf{DCS} for some $v_p$, or $D_1[u_p, v_p]$ was updated from $0$ to $1$ for some $v_p$. In the former case, $N^1_{u,v}[u']$ is updated properly in Line 6 in Algorithm \ref{alg:DCS_insertion_update}. In the latter case, since $D_1[u_p, v_p]$ was properly updated from $0$ to $1$ by inductive hypothesis, Line 4 in Algorithm \ref{alg:insertion_topdown} was called with $D_1[u_p, v_p]$, and thus $\langle u, v \rangle$ is pushed to $Q_1$ in Line 5. Thus, $N^1_{u,v}[u']$ is updated properly in Line 14 of Algorithm \ref{alg:DCS_insertion_update}, when $\langle u_p, v_p \rangle$ is popped from $Q_1$. Therefore, Algorithm \ref{alg:DCS_insertion_update} updates $N^1_{u,v}[u']$ every time it's necessary, and thus it has the correct value after Algorithm \ref{alg:DCS_insertion_update} is finished.

\noindent\textbf{Proof of Lemma \ref{lma:update_complexity}.} The \textsf{DCS} update process for edge deletion is similar to the \textsf{DCS} update process for edge insertion, so we will only show the time complexity of the update process for edge insertion (Algorithm \ref{alg:DCS_insertion_update}). First, Lines 3-10 of Algorithm \ref{alg:DCS_insertion_update} are executed $|E_{DCS}|$ times. Since \textsc{InsertionTopDown} (Algorithm \ref{alg:insertion_topdown}) and \textsc{InsertionBottomUp} (Algorithm \ref{alg:insertion_bottomup}) take a constant time, the total execution time of Lines 3-10 is $O(|E_{DCS}|)$. Next, the while loop of Lines 11-14 (or Lines 15-20) takes a time proportional to the number of children of $\langle u,v\rangle$ (or the number of parent of $\langle u,v\rangle$ plus the number of children of $\langle u,v\rangle$), which is equal to or less than the number of connected edges of $\langle u,v\rangle$. Since the while loop is executed for $\langle u,v\rangle$ where $D_1[u,v]$ or $D_2[u,v]$ changes, the total execution time of Lines 11-20 is proportional to the sum of the number of edges connected to $\langle u,v\rangle$ where $D_1[u,v]$ or $D_2[u,v]$ changes. Hence, the time complexity of the \textsf{DCS} update is $O(\sum_{p\in P}\deg(p)+|E_{DCS}|)$.
% We update $D_1[u,v]$ and corresponding arrays in time proportional to the number of updated parents of $\langle u,v\rangle$ which is less than or equal to the number of parents of $\langle u,v\rangle$. Thus, the time complexity of the $D_1$ update is proportional to the number of edges in \textsf{DCS} (i.e., $O(|E(q)\times |E(g)|)$. Similarly, the time complexity of the $D_2$ update is $O(|E(q)\times |E(g)|)$. Hence, the time complexity of the \textsf{DCS} update is $O(|E(q)\times |E(g)|)$. 

We need $O(|E(q)|\times |V(g)|)$ space to store $N^1_{u,v}[u_p]$'s because there are $|E(q)|$ edges for $(u,u_p)$ and $|V(g)|$ vertices for $v$. Also, we need $O(|V(q)|\times |V(g)|)$ space to store $N^1_P[u,v]$'s. Similarly, we need $O(|E(q)|\times |V(g)|)$ space for $N^2_{u,v}[u']$ and $O(|V(q)|\times |V(g)|)$ for $N^2_C[u,v]$. Hence the space complexity of the \textsf{DCS} update excluding \textsf{DCS} itself is $O(|E(q)|\times |V(g)|)$.

\subsection{Extensions of Our Algorithm}\label{subsec:extensions_of_algorithm}
In this section, we explain how to extend \textsf{SymBi} to handle edge-labeled graphs and directed graphs.

\noindent\textbf{Edge-labeled Graph.} To deal with edge-labeled graphs, our algorithm needs to be modified as follows. First, when constructing \textsf{DCS}, edge labels should be considered. Specifically, for the edge $(\langle u,v\rangle ,\langle u',v'\rangle )$ in \textsf{DCS} to exist, not only edges $(u,u')$ and $(v,v')$ must exist, but the edge labels of both must also be the same. Also, when computing $D_1[u,v]$ (or $D_2[u,v]$) through recurrences, it is necessary to verify that the label of $(u,u_p)$ (or $(u,u_c)$) and the label of $(v,v_p)$ (or $(v,v_c)$) should be the same. Next, when computing $E_{DCS}$ in \textsc{DCSChangedEdge}, only edges of the query graph with the same edge label as $(v,v')$ should be included. Finally, when computing $S_{u_{min}}$ in Algorithm \ref{alg:computing_Cmu}, we must include vertex $v$ such that the label of $(u,u_{min})$ and the label of $(v,M(u_{min}))$ are the same. Similarly, the edge label must be considered when computing $C_M(u)$.

\noindent\textbf{Directed Graph.} Suppose that we are given a directed data graph and a directed query graph. To create a parent-child relationship of query vertices required in \textsf{DCS}, we regard the directed query graph as an undirected graph and build a rooted DAG as in the paper. The query DAG created in this way is independent of the actual direction of the edge of the query graph. We use the query DAG to order the query vertices when constructing or updating \textsf{DCS} in a top-down or bottom-up fashion. When we map an edge $(u,u')$ of the query graph and an edge $(v,v')$ of the data graph, we need to check that the directions of the two edges match, as in the edge-labeled graph.

\subsection{Distribution of the Number of Matches}\label{subsec:distribution_number_of_matches}
\begin{figure}[H]
    \centering
    \subcaptionbox{Netflow query sets\label{exp:distribution_netflow}}{
        \includegraphics[width=\linewidth,scale=0.1]{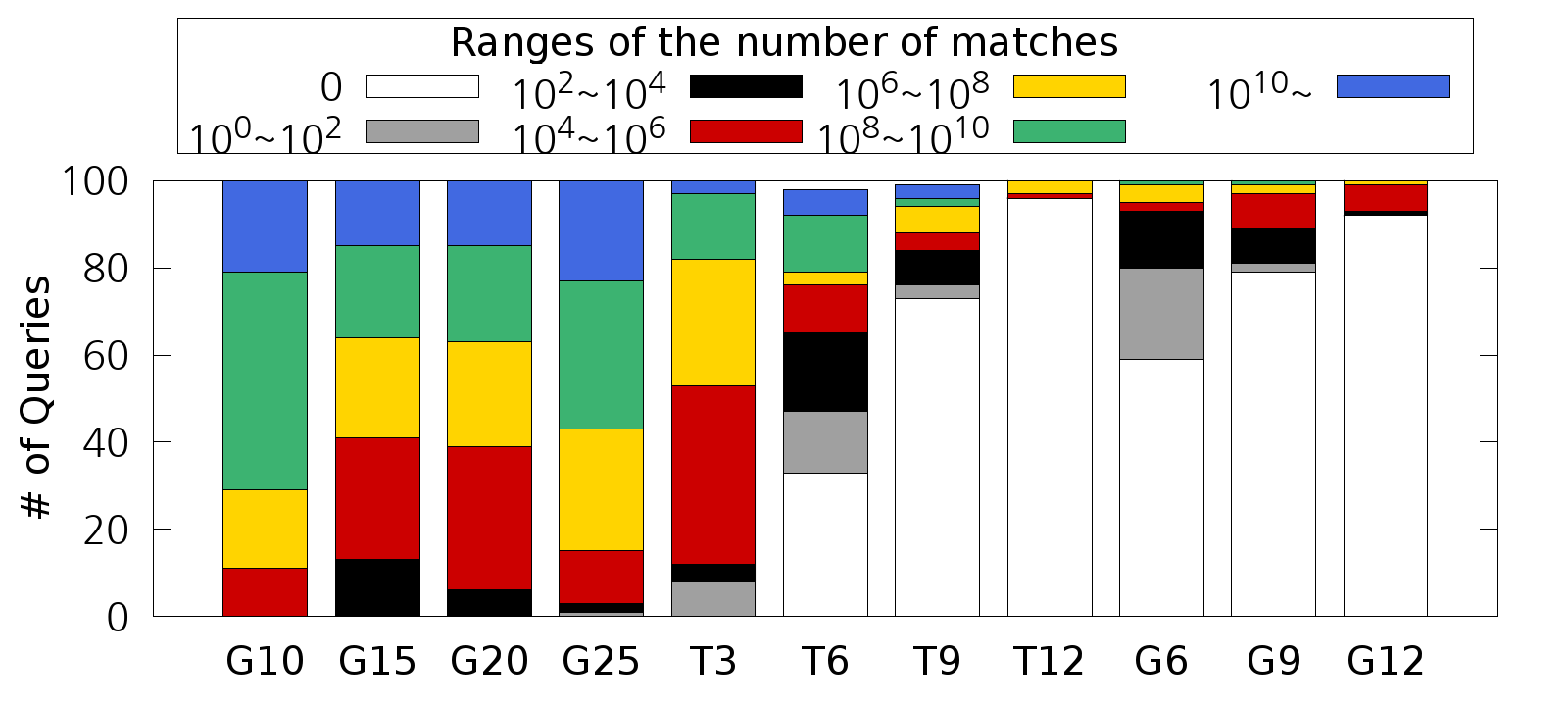}
    }
    \newline
    \vskip 1mm
    \subcaptionbox{LSBench query sets\label{exp:distribution_lsbench}}{
        \includegraphics[width=\linewidth,scale=0.1]{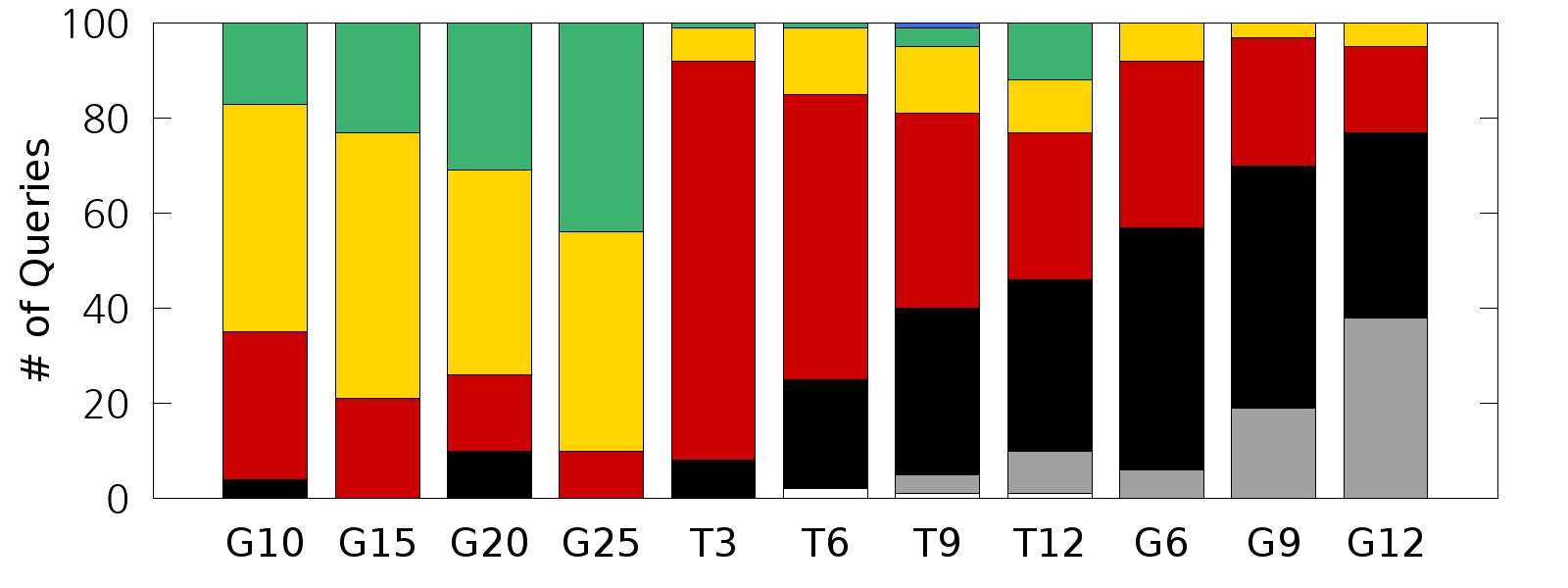}
    }
    \vspace{-1mm}
    \caption{Distribution of the number of matches for queries}
    \label{exp:distribution}
\end{figure}

Stacked bar charts in Figure \ref{exp:distribution} represent the distribution of the number of matches for queries we test. The bar for each query set is made up of seven sub-bars. 
The color of a sub-bar represents the range of the number of matches as shown in the legend of Figure \ref{exp:distribution}, and the size of a sub-bar represents the number of queries for which the number of matches is in that range.
Sub-bars are stacked in order from the smallest number of matches to the largest number of matches from the bottom. For example, the bar for the G10 query set shows that there are 11 queries with $10^4\sim 10^6$ matches, 18 queries with $10^6\sim 10^8$ matches, 50 queries with $10^8\sim 10^{10}$ matches, and 21 queries with more than $10^{10}$ matches. In Figure \ref{exp:distribution_netflow}, two queries in the T6 query set bar and one in the T9 query set bar are missing because neither algorithm can solve them within the time limit. Also, the stacked bars in Figure \ref{exp:distribution} show different results from the stacked bars of \cite{kim2018turboflux}, since \cite{kim2018turboflux} uses graph homomorphism as matching semantics while we use graph isomorphism as matching semantics.

Figure \ref{exp:distribution} shows that our generated queries have a larger number of matches than queries from \cite{kim2018turboflux}.  Also, Figure \ref{exp:distribution} shows that the \textsf{Netflow} queries have more matches than the \textsf{LSBench} queries. This supports the fact that the \textsf{LSBench} queries are easier to solve than the \textsf{Netflow} queries as mentioned above.

\subsection{Effect of Our Techniques}\label{subsec:effect_of_our_techniques}
% In this subsection, we evaluate the practical impact of each of the proposed techniques introduced in this paper.

\noindent\textbf{Effect of DCS Update.} To see the effect of the proposed \textsf{DCS} update method, we compare the elapsed time of the proposed method with the elapsed time of recomputing from scratch. We also measure the number of updated vertices and the number of visited edges in \textsf{DCS} during the \textsf{DCS} update, which are presented in Section \ref{subsec:experimental_resutls}. Since recomputing from scratch takes a long time, we limit the number of update operations to 1000.

\begin{figure}[H]
    \centering
     \subcaptionbox{Netflow query sets\label{exp:update_recompute_netflow}}{
        \includegraphics[width=\linewidth,scale=0.2]{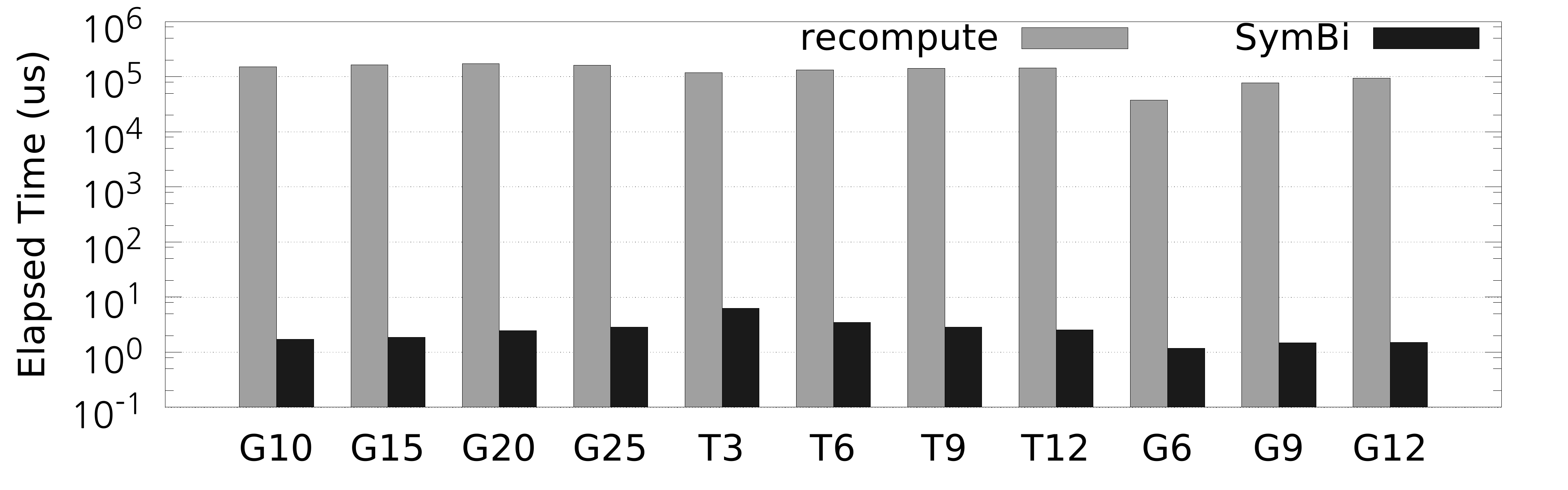}
    }
    \newline
    \vskip 1mm 
    \subcaptionbox{LSBench query sets\label{exp:update_recompute_lsbench}}{
        \includegraphics[width=\linewidth,scale=0.2]{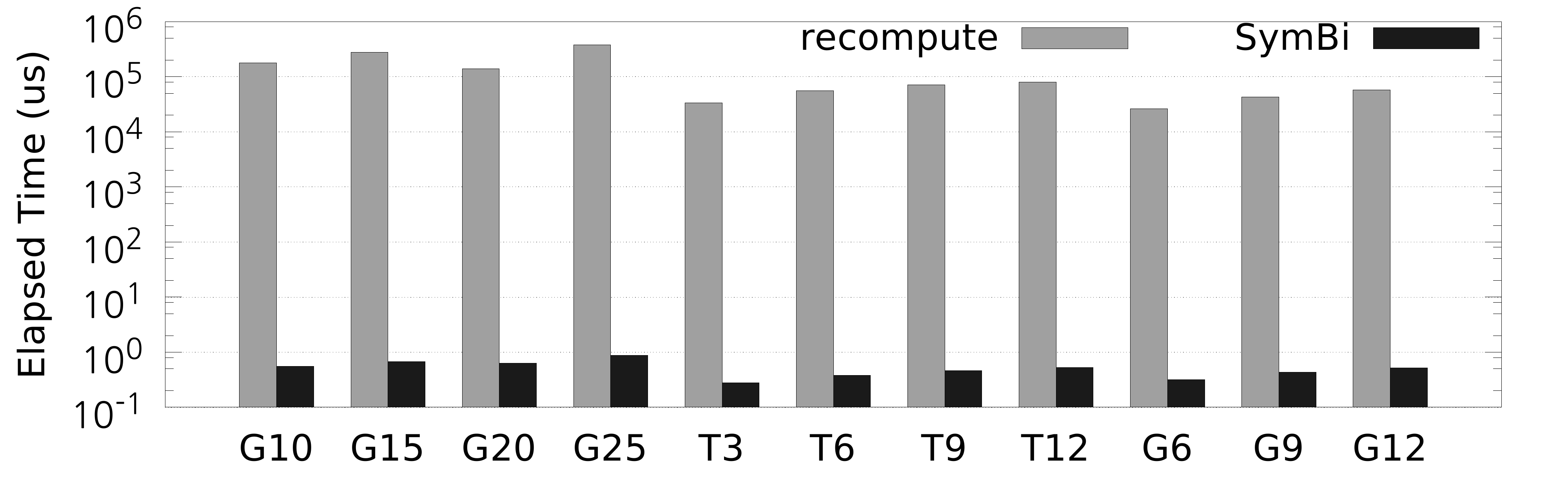}
    }
    \vspace{-3mm}
    \caption{Average elapsed time per update operation (in microseconds)}
    \label{exp:update_recompute}
\end{figure}

Figure \ref{exp:update_recompute} shows the average \textsf{DCS} update time per update operation (i.e., average \textsf{DCS} update time / 1000) for \textsf{Netflow} and \textsf{LSBench}. 
% Note that the elapsed time is in microseconds. 
Our proposed update method is faster than recomputing from scratch by up to four orders of magnitude for \textsf{Netflow} and five orders of magnitude for \textsf{LSBench}.

\noindent\textbf{Effect of Estimated Size.} To evaluate the effect of the estimated size of extendable candidates, we compare the estimated size of extendable candidates and the exact size of extendable candidates through two scores. Also, we compare the elapsed time when using the estimated size and the elapsed time when using the exact size.

\vspace{-3mm}
To define the two scores $S_1$ and $S_2$ for a query graph and a dataset, we consider the query vertex $u$ selected to match according to the estimated candidate size order for each partial embedding $M$ in the search process. To show how similar the estimated size and the exact size are, we define the first score $S_1$ as the sum of $|C_M(u)|$ of the vertices that we consider divided by the sum of $E(u)$ of the same vertices. Next, the second score $S_2$ is defined as the number of partial embeddings in which the estimated candidate size order and the exact candidate size order select the same query vertex $u$ (i.e., both $E(u)$ and $|C_M(u)|$ are the smallest among extendable vertices) divided by the number of partial embeddings in the search process.

Table \ref{tab:score} shows the average scores of 100 queries in a query set for \textsf{Netflow} and \textsf{LSBench}. We exclude tree-shaped query sets because extendable vertices in a tree-shaped query have only one matched neighbor. The average $S_1$ score for the 7 query sets is 0.434 for \textsf{Netflow} and 0.874 for \textsf{LSBench}. The average $S_2$ score is 0.830 for \textsf{Netflow} and 0.966 for \textsf{LSBench}. The $S_1$ score indicates that depending on the dataset, there may be differences between the estimated size and the exact size. However, the $S_2$ score shows that in most cases of both datasets, the estimated candidate size order chooses the same vertex as the exact candidate size order. Although the computational overhead of the estimated size of extendable candidates is negligible, the estimated candidate size order works almost like the exact candidate size order.

\vspace{-1mm}
\begin{table}[h]
    \centering
    \caption{Estimated size $E(u)$ vs. exact size $|C_M(u)|$. $S_1$: $\sum{|C_M(u)|} / \sum{E(u)}$, $S_2$: proportion that the estimated candidate size order is the same as the exact candidate size order. (top: \textsf{Netflow}, bottom: \textsf{LSBench})}
    \vspace*{-1mm}
    \begin{tabular}{cccccccc}
    \toprule
    \textbf{Score} & \textbf{G10} & \textbf{G15} & \textbf{G20} & \textbf{G25} & \textbf{G6} & \textbf{G9} & \textbf{G12}\\ 
    \midrule
    $S_1$ & 0.567 & 0.535 & 0.445 & 0.509 & 0.308 & 0.332 & 0.341\\
    $S_2$ & 0.869 & 0.831 & 0.811 & 0.863 & 0.772 & 0.842 & 0.822\\
    \midrule
    $S_1$ & 0.939 & 0.950 & 0.891 & 0.955 & 0.754 & 0.807 & 0.823\\
    $S_2$ & 0.998 & 0.991 & 0.899 & 0.975 & 0.969 & 0.973 & 0.959\\
    \bottomrule
    \end{tabular}

    \label{tab:score}
\end{table}

\clearpage

\begin{figure}[H]
    \centering
    \subcaptionbox{Average elapsed time (in milliseconds) for Netflow query sets\label{exp:estimated_exact_netflow_time}}{
        \includegraphics[width=\linewidth,scale=0.2]{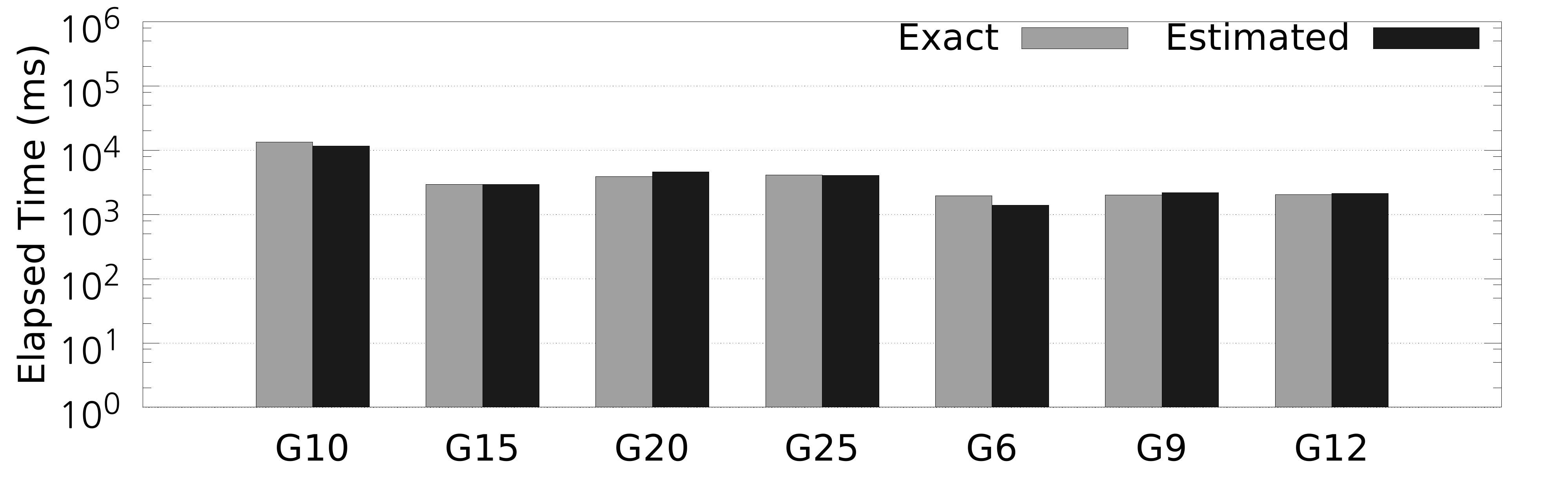}
    }
    \newline
    \vskip 1mm
    \subcaptionbox{Average elapsed time (in milliseconds) for LSBench query sets\label{exp:estimated_exact_lsbench_time}}{
        \includegraphics[width=\linewidth,scale=0.2]{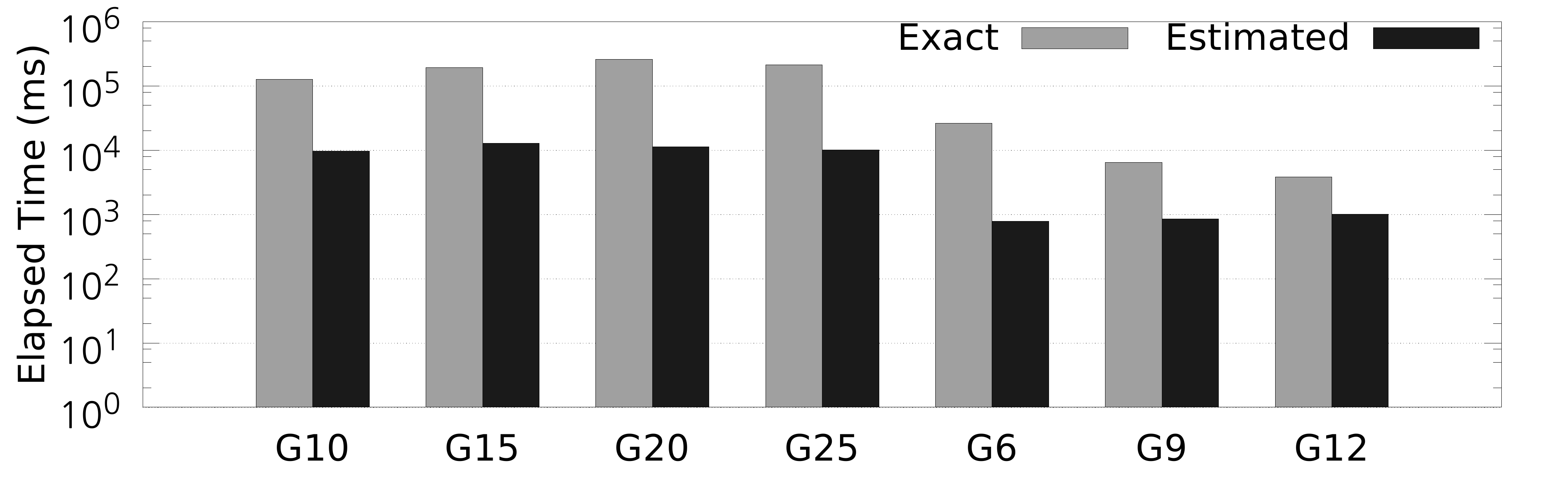}
    }
    \vspace{-1mm}
    \caption{Estimated size vs. exact size}
    \label{exp:estimated_exact}
\end{figure}

Figure \ref{exp:estimated_exact} represents the average elapsed time when using the estimated size and the exact size. In Figure \ref{exp:estimated_exact_lsbench_time}, the algorithm using the estimated size outperforms the algorithm using the exact size for all query sets in \textsf{LSBench}. In particular, the algorithm using the exact size failed to solve 1 and 2 queries in G10 and G20, respectively, while the algorithm using the estimated size solves all queries. This is because the estimated size in \textsf{LSBench} works almost the same as the exact size, but there is no overhead for updating $C_M(u)$.

\begin{figure}[H]
    \centering
    \subcaptionbox{Average elapsed time (in milliseconds) for Netflow query sets\label{exp:isolated_netflow}}{
        \includegraphics[width=\linewidth,scale=0.2]{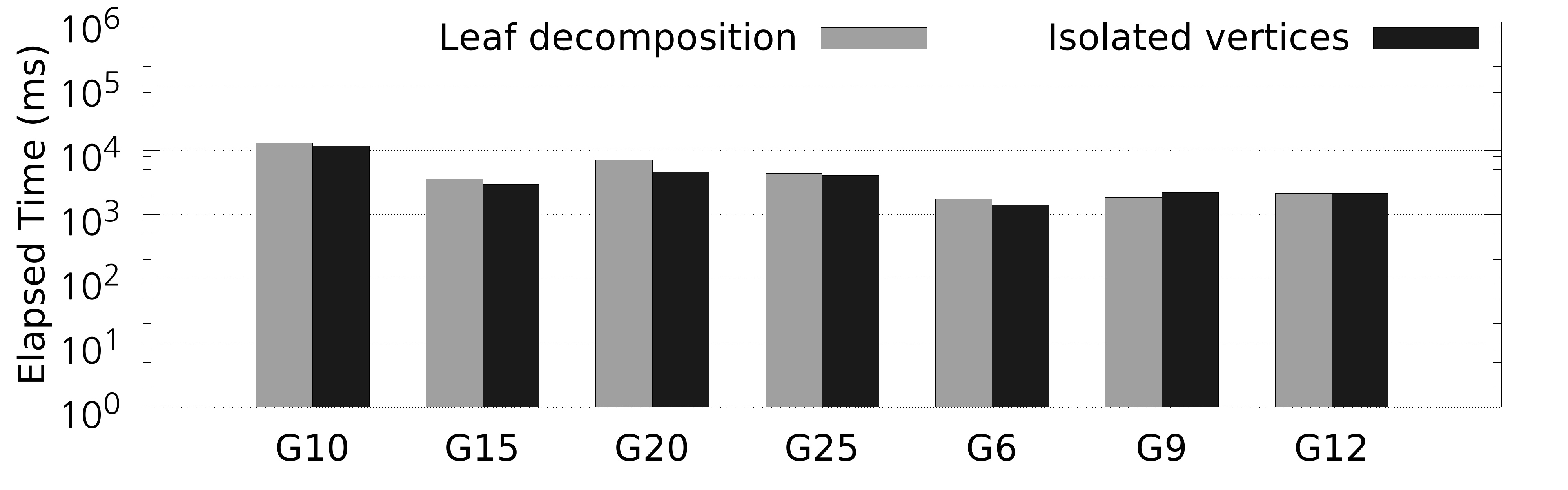}
    }
    \newline
    \vskip 1mm
    \subcaptionbox{Average elapsed time (in milliseconds) for LSBench query sets\label{exp:isolated_lsbench}}{
        \includegraphics[width=\linewidth,scale=0.2]{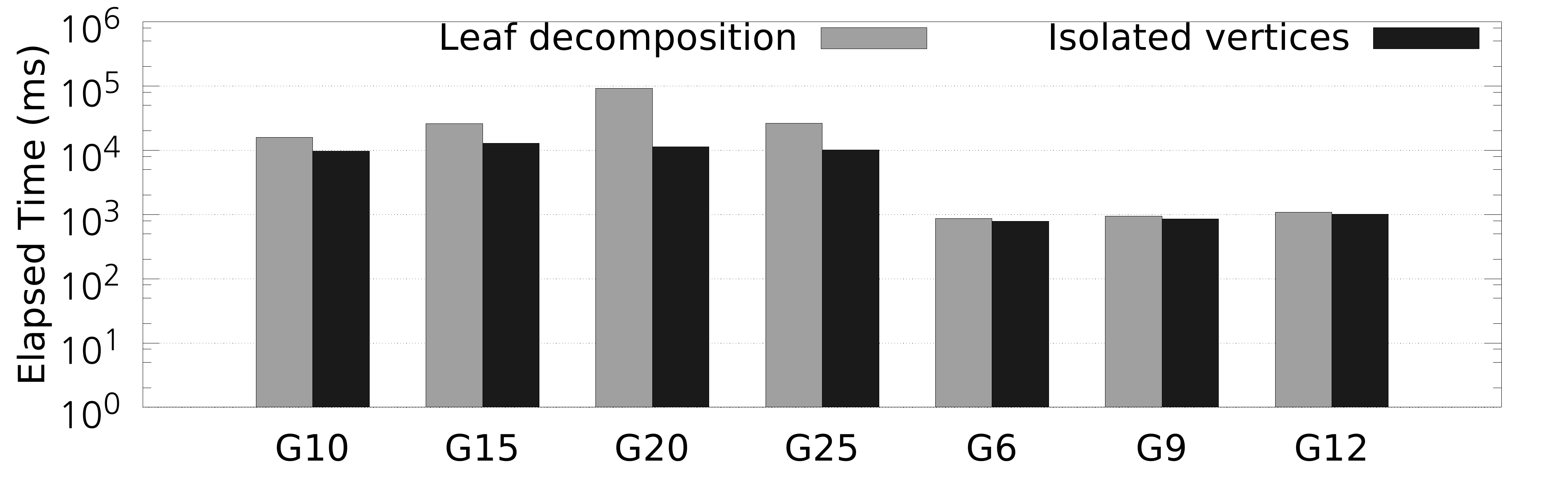}
    }
    \vspace{-1mm}
    \caption{Isolated vertex vs. leaf decomposition}
    \label{exp:isolated_vs_leafdecomp}
\end{figure}

\noindent\textbf{Effect of Isolated Vertex.} To test the effect of isolated vertices, we compare the elapsed time when using the isolated vertex technique and the elapsed time when not using it (i.e., just using the leaf decomposition technique in \cite{bi2016efficient}). Figure \ref{exp:isolated_vs_leafdecomp} shows the results. Since the isolated vertex technique and the leaf decomposition technique are the same in a tree-shape query, we exclude tree-shaped query sets. Due to the characteristics of the datasets, the generated queries are sparse. Because of this, there are not many situations where a query vertex becomes an isolated vertex without being a leaf, so the performance of the two techniques is similar in many query sets. Nevertheless, when using the isolated vertex technique, it is 1.63, 2.01, 8.07, and 2.59 times faster on query sets G10, G15, G20, and G25 of \textsf{LSBench}, respectively.

\end{document}